\documentclass[a4paper,12pt]{article}

\usepackage{graphics,epsfig}
\usepackage{amsmath}
\usepackage{amssymb}

\begin{document}

\begin{flushright}
SHEP-07-14, LPT-Orsay-07-129\\
\today
\end{flushright}
\vspace*{1.0truecm}

\begin{center}
{\Large\bf
Exploring the Di-Photon Decay of a Light Higgs Boson \\[0.25cm]
in the MSSM With Explicit CP Violation
}

\vspace*{1.0truecm}
{\large
S. Hesselbach\footnote{s.hesselbach@hep.phys.soton.ac.uk}$^a$,
S. Moretti\footnote{stefano@hep.phys.soton.ac.uk}$^{a,b}$,
S. Munir\footnote{shobig@hep.phys.soton.ac.uk}$^a$,
P. Poulose\footnote{poulose@hep.phys.soton.ac.uk; poulose@iitg.ernet.in}}$^{a,c}$

\vspace*{0.5truecm}
$^a${\it
School of Physics \& Astronomy,\\
University of Southampton,  Highfield, Southampton SO17 1BJ, UK}\\[3mm]
$^b${\it
Laboratoire de Physique Th\'eorique,\\
Universit\'e Paris--Sud, F--91405 Orsay Cedex, France}\\[3mm]
$^c${\it
Physics Department, IIT Guwahati, Assam, INDIA - 781039}
\end{center}
 
\date{\today}

\begin{abstract}
The di-photon decay channel of the lightest Higgs boson is
considerd as a probe to explore CP violation in the 
Minimal Supersymmetric Standard Model (MSSM).
The scalar/pseudo-scalar mixing is considered along with
CP violation entering through the Higgs-sfermion-sfermion
couplings, with and without light sparticles. The impact
of a light stop on the decay width and Branching Ratio (BR)
is established through a detailed study of
the amplitude of the process $H_1\rightarrow \gamma\gamma$.
The other sparticles
have little influence even when they are light.
With a suitable combination of other MSSM parameters,
a light stop can change the BR by more than 50\% with a
CP-violating phase $\phi_\mu\sim 90^\circ$, while the change
is almost nil with a heavy stop. 
\end{abstract}

\newcommand{\ra}{\rightarrow}
\newcommand{\lra}{\longrightarrow}
\newcommand{\ee}{e^+e^-}
\newcommand{\gam}{\gamma \gamma}
\newcommand{\tb}{\tan \beta}
\newcommand{\s}{\smallskip}
\newcommand{\nn}{\noindent}
\newcommand{\non}{\nonumber}
\newcommand{\beq}{\begin{eqnarray}}
\newcommand{\eeq}{\end{eqnarray}}

\providecommand{\UI}{}
\renewcommand{\UI}{\ensuremath{\mathrm{U}(1)}}
\providecommand{\smgroup}{}
\renewcommand{\smgroup}{\ensuremath{\mathrm{SU}(3) \otimes \mathrm{SU}(2)_L \otimes \mathrm{U}(1)_Y } }
\providecommand{\Tr}{}
\renewcommand{\Tr}{\ensuremath{\mathrm{Tr}}}
\providecommand{\identity}{}
\renewcommand{\identity}{\ensuremath{\mathrm{I}}}
\providecommand{\Kahler}{}
\renewcommand{\Kahler}{K\"ahler}
\providecommand{\Fbar}{}
\renewcommand{\Fbar}{\overline{F}}

\section{Introduction}

Despite its success to describe the physics of elementary particles
there are strong hints that the Standard Model (SM) 
is only  an effective theory valid up to the TeV range, and new
physics is needed to explain particle dynamics (much) beyond this
energy scale. Among many others
Supersymmetry (SUSY) is one of the most favoured scenarios and will
be explored for in all possible ways at the upcoming Large 
Hadron Collider (LHC) at CERN.
Furthermore, one of the main tasks of the LHC is 
the search for Higgs bosons, i.e.\ the determination of the underlying
mechanism of Electro-Weak Symmetry Breaking (EWSB).
Precision measurements at CERN LEP and SLAC SLC prefer
a light Higgs particle, which is indeed  predicted in SUSY models.
The scalar potential of the MSSM 
conserves CP at tree level \cite{Higgs-hunter}, because
SUSY imposes an additional (holomorphic) 
symmetry on the Higgs sector of a general two-Higgs doublet model, that 
enforces flavour conservation in tree-level neutral currents and absence of 
CP-violating scalar/pseudo-scalar mixing in the Born approximation.
Beyond tree-level, the CP invariance of the Higgs potential may in
principle be spontaneously broken by radiative corrections when the Vacuum
Expectation Values (VEVs) of the two Higgs doublets develop a 
relative phase \cite{Maekawa,Pilaftsis-PLB}.
According to the Georgi-Pais theorem \cite{Georgi-Pais} though,
this type of CP violation requires a very light Higgs 
state, which is essentially ruled out by experiment \cite{Pamoral}.

On the other hand many new MSSM parameters can well be complex and
thus explicitly break CP invariance.
Beyond the Born approximation these new CP-violating phases induce CP
violation also in the Higgs sector \cite{Pilaftsis-PLB,PlWagner}.
The possibly complex parameters include: (i) the 
higgsino mass term $\mu$, (ii) the soft SUSY-breaking gaugino masses 
$M_a~(a = 1,2,3)$, (iii) the soft bi-linear term $B\mu$ 
and (iv) the soft trilinear 
Yukawa couplings $A_f$ of the Higgs particles to scalar fermions of flavour
$f$. 
In general, all of these new phases are independent.
However, when imposing universality conditions at a unification scale
$M_\mathrm{GUT}$ all the three gaugino masses $M_a$ have a common phase as
well as all the trilinear couplings $A_f$ have another common phase,
i.e., four independent phases remain: those of $\mu$, $B\mu$, $M_a$ and $A_f$.
Furthermore,
the two $U(1)$ symmetries of the conformal-invariant part of the
MSSM may be employed to re-phase one of the Higgs doublet fields and the
gaugino fields such that $M_a~\rm{and}~\textit{B}\mu$ are real 
\cite{PlWagner,Dugan}. 
In this paper we will work within this setup with two independent
physical phases, which we take to be $\arg(\mu)=\phi_\mu$ and
$\arg(A_f)=\phi_{A_f}$.

The new CP-violating phases in the MSSM are severely constrained by
bounds on the Electric Dipole Moments (EDMs) of electron, neutron 
and the Hg atom.
In order to avoid problems with phases associated with the sfermions
of the first and second generations one may deviate from exact universality and
consider $A_f$ to be diagonal in flavour space with vanishing first
and second generation couplings \cite{dchang}.
In general, the constraints are rather model dependent and 
there have been several suggestions \cite{EDM1}--\cite{Bartl:2003ju}
to evade these
constraints allowing large CP-violating phases of $\mathcal{O}(1)$.
One possibility is to 
arrange for partial cancellations among various contributions to the 
EDMs \cite{EDM3}.
In this scenario,  it has recently 
been pointed out that for large trilinear scalar
couplings $A_f$, phases $\phi_\mu\sim \mathcal {O}(1)$ can be compatible with
the EDM bounds \cite{YaserAyazi:2006zw}.
Another option is to make the first two generations 
of scalar fermions rather heavy, of order a few TeV, so that the one-loop 
EDM constraints are automatically evaded,
however, two-loop contributions of third generation scalar fermions
may still be large \cite{dchang}.
(A detailed analysis of the so-called CPX
scenario with very heavy
first and second generations squarks is available in Ref.~\cite{ap-edm}.)
As a matter of fact, one can consider
so-called effective SUSY models \cite{EDM2} where decoupling of the first 
and second generation sfermions are invoked to solve the SUSY Flavour Changing 
Neutral Current (FCNC) and CP 
problems without spoiling the naturalness condition.
Furthermore, the restrictions on the phases may also disappear if
lepton flavour violating terms in the MSSM Lagrangian are included
\cite{Bartl:2003ju}.
In conclusion, this means that
large phases cannot be ruled out and therefore we analyse the full range
$0^\circ \leq (\phi_\mu,\phi_A) \leq 180^\circ$ in the following.

The CP-violating SUSY phases can in principle be determined
directly in SUSY particles production and decay
at high energy colliders \cite{PlWagner},
\cite{ChoiSik}--\cite{ToddILC} or indirectly via their radiative effects
on the Higgs sector \cite{PlWagner,ChoiKao}.
Here we focus on the di-photon decay mode, $H_1\rightarrow \gamma\gamma$,
of the lightest neutral Higgs boson $H_1$, which involves
direct, i.e\ leading, effects of the SUSY phases through couplings of
the $H_1$ to SUSY particles in the loops (see Fig.~\ref{fig:HiggsPho})
as well as indirect, i.e.\ sub-leading, effects through the
scalar/pseudo-scalar mixing yielding the Higgs mass-eigenstate $H_1$.
The di-photon decay mode is important for the study of CP-violating
effects in the MSSM Higgs sector for two reasons.
On the one hand, it is 
the most promising channel for the discovery of a light neutral Higgs state
with mass between 80 and 130 GeV at the LHC \cite{ATLASTDR2,CMS}.
On the other hand, the coupling strength of the dominant CP-violating
terms 
of the di-photon decay width
which depend on $\mu$ and $A_f$ (with $f=b,t,\tau$, hereafter)
is of the same order, 
${\cal O}(\alpha^3)$, as that of the CP-conserving ones.

The entire 
$gg/qq\ra H_1\ra\gamma\gamma$ process can be factorised exactly
into three parts:
the production process, the Higgs propagator and the decay channel.
(Herein, we will adopt this factorisation in Narrow Width
Approximation,
thereby neglecting small corrections of $~{\cal O}(\Gamma_{H_1}/M_{H_1})$, 
as $\Gamma_{H_1}\ll M_{H_1}$
for a light Higgs state.)
In this process, effects of CP violation can occur through the
aforementioned couplings in the production, through a possible
mixing of Higgs states at one-loop and above in the propagator and through
the same couplings in the decay. 
CP violation in the production of a Higgs state in the gluon-gluon 
fusion process at hadron colliders was studied first by \cite{dedes}, 
choosing a parameter space region which is not sensitive to the CP mixing of 
the Higgs states, and later by \cite{ChoiSik2,ChoiKaoJae}, including
the presence of CP mixing of the Higgs states\footnote{
CP violation in vector boson associated production ($VH_i$) is studied
by \cite{dilipVH}.}.  Effects of CP mixing in the propagator are 
discussed separately but in great detail in \cite{EllisPLee}. 
A detailed study of the other MSSM Higgs decay channels in presence
of CP violation can be found in 
\cite{ChoiSik}--\cite{EllisPLee}, \cite{CarEllis1}--\cite{dilip}\footnote{Where all these
decay modes have been studied inclusively. A discussion of
CP violation in exclusive four lepton final states via gauge boson decays can be found in 
Ref.~\cite{4leptons}.}.

Results of a random parameter space scan to understand the general
behaviour of the $\mathrm{BR}(H_1\rightarrow \gamma\gamma)$ for non-zero
$\phi_\mu$ values are 
reported in \cite{paper1}. It has been seen that about 50\%
 deviations are possible for
$M_{H_1}$ around 104 GeV for $\phi_\mu=100^\circ$, and an average of 30\%
deviation occurs over
 the mass range 90--130 GeV. Masses around and below 110
GeV show a decrease in the BR for non-zero $\phi_\mu$ values, 
while masses above this value 
show an increase. Certain individual parameter space points were also
discussed in \cite{paper1}. Specifically, $|A_f|=1.5$ TeV, $|\mu|=1 $ TeV,
$\tan\beta=20$ was considered as a benchmark scenario. Then, by choosing 
(a) $M_{\tilde U_3}=1$
TeV and (b) $M_{\tilde U_3}= 250$ GeV, it was demonstrated that the 
light stop $\tilde t_1$ has a strong impact on our decay mode, through
the $\mu$ and the trilinear couplings $A_f$, which
is quite different from the effects due only to the (one-loop)
change of the $H_1W^+W^-$ coupling (discussed in \cite{ChoiKaoJae}). In Ref.~\cite{paper1},
it was also noticed that the effect of a light $\tilde t_1$ is in the 
opposite direction as compared to that due to  modifications of the
$H_1$ coupling to the SM
particles.

In this article we will consolidate the results of \cite{paper1} by
a detailed discussion of the following points:
\begin{itemize}
\item
The Higgs mixing matrix elements are discussed in detail showing the changes
in mixing as the phase $\phi_\mu$ is varied.

\item
The decay amplitudes due to individual (s)particles in the loops are presented
in detail and the real and imaginary parts of the scalar and pseudo-scalar
contributions are given separately. We will see that this unambiguously shows
that a light stop contribution is comparable to the SM one, 
while the other sparticles have negligible 
effects even in scenarios when they are light.

\item
Compared to \cite{paper1}, where only one particular value was
considered for $\tan\beta$, $|\mu|$ and $|A_f|$,  
we present here results with four different
$\tan\beta = 2, 5, 20, 50$, two different $|\mu| = 0.5, 1$~TeV and
two different $|A_f| = 0.5, 1.5$~TeV.

\end{itemize}
{We postpone the full analysis of $gg/qq\ra H_1\ra\gamma\gamma$
to a forthcoming publication \cite{preparation}.}
The outline of this paper is as follows.
In Sect.~2 the CP mixing in the Higgs sector is explained in more detail.
In Sect.~3 we analyse the phase dependence of $H_1\rightarrow \gamma\gamma$.
We conclude in Sect.~4.

\section{Higgs Mixing in the CP-violating MSSM}

In the Higgs sector of the
MSSM with explicit CP violation the CP-violating phases 
introduce non-vanishing
off-diagonal mixing terms in the neutral Higgs mass matrix, which in the
weak basis $(\phi_1,\phi_2,a)$, where $\phi_{1,2}$ are the  CP-even states 
and $a$ is the CP-odd state, may schematically be written as
\cite{PlWagner,ChoiKao,CarEllis1,heinmeyer}
\beq
\mathcal{M}_H^2 = \left(%
\begin{array}{cc}
  \mathcal{M}_S^2 & \mathcal{M}_{SP}^2 \\
  \mathcal{M}_{PS}^2 & \mathcal{M}_P^2 \\
\end{array} \right).
\eeq
Here, ${\cal M}_S^2$ is a $2\times 2$ matrix describing the transition between 
the CP-even states, ${\cal M}_P^2$ gives the mass of the CP-odd
state while ${\cal M}_{PS}^2=({\cal M}_{SP}^2)^T$ (a $1\times 2$ matrix) 
describes the mixing between the CP-even and CP-odd states. The mixing matrix 
elements are typically proportional to 
\beq
{\cal M}_{SP}^2\propto \mathcal{I}m (\mu A_f)\,
\label{Eq.MSP}
\eeq
and dominated by loops involving the top squarks. 
As a result, the neutral Higgs bosons of the MSSM no longer carry any definite 
CP-parities. Rotation from the EW states to the mass eigenvalues, 
\beq
(\phi_1,\phi_2,a)^T=O\;(H_1,H_2,H_3)^T, \non
\eeq
is now carried out by a $3\times 3$
real orthogonal matrix $O$, such that
\beq 
O^T\mathcal{M}_H^2O = {\rm{diag}}(M^2_{H_1},M^2_{H_2},M^2_{H_3}) 
\eeq
with $M_{H_1}\leq M_{H_2}\leq M_{H_3}$. 
As a consequence, it is now appropriate to parametrise the Higgs sector of the CP-violating MSSM in terms of
the mass of the charged Higgs boson, $M_{H^\pm}$, as the latter remains basically unaffected. (For a detailed 
formulation of the MSSM Higgs sector with explicit CP violation, see Refs. \cite{PlWagner,CarEllis1}.)

\section{\boldmath $H_1 \to \gamma\gamma$ in the CP-violating MSSM}

\begin{figure}
\begin{center}
\includegraphics[width=8cm]{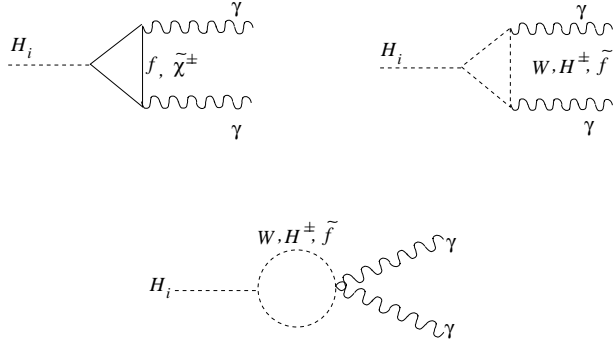}
\end{center}
\caption{\it Diagrams for Higgs decay into $\gamma\gamma$ pairs in the
  CP-violating MSSM: 
$f\equiv t,\: b,\:\tau;\;\;\tilde f\equiv \tilde t_{1,2}, \tilde b_{1,2},
\tilde \tau_{1,2}$.}
\label{fig:HiggsPho}
\end{figure}

A Higgs boson in the MSSM decays at one-loop level into two photons
through loops of fermions, sfermions, $W^\pm$ bosons, charged Higgs
bosons and charginos, see Fig.~\ref{fig:HiggsPho}.
The analytical expressions for the respective amplitude
along with relevant couplings in the CP-violating MSSM
can be found in \cite{CPSuperH} and references therein.
The amplitude has the form
\begin{equation}
{\cal M}_{\gamma\gamma H_i}=-\frac{\alpha\:M^2_{H_i}}{4\pi v}\;
\left\{S^\gamma_i(M_{H_i})\:(\epsilon_{1\perp}^*\cdot\epsilon_{2\perp}^*)-
P^\gamma_i(M_{H_i})\frac{2}{M^2_{H_i}}\:
\left<\epsilon_1^*\epsilon_2^*k_1k_2\right> \right\},
\label{eq:Melement}
\end{equation}
where $k_{1,2}$ are the momenta of the two photons and $\epsilon_{1,2}$ are
their polarisation vectors, which are conveniently written as
$\epsilon_{r\perp}^\mu=\epsilon_r^\mu-2k_r^\mu(k_s\cdot \epsilon_r)/M_{H_i}^2
\;\;\;(r\ne s)$ and where $\left<\epsilon_1\epsilon_2k_1k_2\right>\equiv
\varepsilon_{\mu\nu\rho\sigma}\epsilon_1^\mu\epsilon_2^\nu k_1^\rho k_2^\sigma$.
$S^\gamma_i$ and $P^\gamma_i$ 
are given by (retaining only the dominant loop contributions):
\begin{eqnarray}
S^\gamma_i(M_{H_i})&=&2\sum_{f=t,b,\tau,\tilde\chi_1^\pm,\tilde\chi_2^\pm}
N_C\:Q_f^2\:g_fg^S_{H_if\bar f}\frac{v}{m_f}F_{sf}(\tau_{if})\nonumber\\
&&-\sum_{\tilde f_j=\tilde t_1,\tilde t_2,\tilde b_1,\tilde b_2,
\tilde\tau_1,\tilde\tau_2}N_C\: Q_f^2\:g_{H_i\tilde f_j^*\tilde f_j}
\frac{v^2}{2m_{\tilde f_j}^2}F_0(\tau_{i\tilde f_j})\nonumber\\
&&-g_{H_iWW}F_1(\tau_{iW})-
g_{H_iH^+H^-}\frac{v^2}{2M_{H^\pm}^2}F_0(\tau_{iH^\pm}),\nonumber\\[2mm]
P^\gamma_i(M_{H_i})&=&2\sum_{f=t,b,\tau,\tilde\chi_1^\pm,\tilde\chi_2^\pm}
N_C\:Q_f^2\:g_fg^P_{H_if\bar f}\frac{v}{m_f}F_{sf}(\tau_{if}).\nonumber
\end{eqnarray}
For the 
expressions of the various couplings ($g$'s) and form factors ($F(\tau$)) we
refer to \cite{CPSuperH}.
Then the partial decay width is given by,
\begin{equation}
\Gamma (H_i\rightarrow
\gamma\gamma)=\frac{M^3_{H_i}\alpha^2}{256\pi^3v^2}\:\left[
\left|S^\gamma_i(M_{H_i})\right|^2+
\left|P^\gamma_i(M_{H_i})\right|^2\right].
\label{eq:width}
\end{equation}
The decay mode $H_i \to \gamma\gamma$, $i=1,2,3$, is 
discussed by Ref.~\cite{ChoiKaoJae} along with Higgs production 
through gluon-gluon fusion. However, that study was confined to
MSSM parameter space regions with suitably heavy sparticles
$\tilde{f}$ and $\tilde{\chi}^\pm$
where CP-violating effects are only due to
the changed SM particle (especially $W^\pm$) couplings to the $H_1$
and effects of sparticles in the triangle loops entering the decay amplitude are negligible.
Here we examine the complementary region of MSSM parameter space
with light sparticles so that they contribute substantially to the latter.
In particular,
we will show that, in the presence of non-trivial CP-violating phases,
regions of MSSM parameter space exist where the couplings of the 
Higgs bosons to all sparticles in the decay loops are strongly modified 
with respect to the CP-conserving MSSM, thereby inducing dramatic changes
on the $H_1\to\gamma\gamma$ width and Branching Ratio (BR). 

We have analysed the Higgs decay widths and BRs with the
publicly available \textsc{Fortran} code \textsc{CPSuperH} 
\cite{CPSuperH} version 2,
which is based on the results obtained in
Refs.~\cite{ChoiSik}--\cite{ChoiDrees} and the most recent
renormalisation group improved effective-potential approach, which
includes dominant higher-order logarithmic and threshold corrections,
$b$-quark Yukawa-coupling re-summation effects and Higgs boson
pole-mass shifts \cite{CarEllis1,CarEllis2}.
\textsc{CPSuperH} calculates the mass spectrum and decay widths of the
neutral and charged Higgs bosons
in the general CP-violating MSSM including the phases of $A_f$ and $\mu$.
Furthermore, it computes all the couplings of the 
neutral Higgs bosons $H_{1,2,3}$ and the charged Higgs boson $H^\pm$
to SM particles and their superpartners.

The open non-SM parameters of the model now include:
the higgsino mass $|\mu|$, 
its phase $\phi_{\mu}$, the charged Higgs mass $M_{H^\pm}$, 
the soft gaugino masses $M_a$, 
the soft sfermion masses of the third generation
 $M_{(\tilde Q_3,\tilde U_3, \tilde D_3, \tilde L_3, \tilde E_3)}$,  
the (unified) soft trilinear coupling of the third generation $|A_f|$
and its phase $\phi_{A_f}$. 

\vskip 5mm
\noindent
In our analysis we fix the following SUSY parameters:
\begin{itemize}
\item $M_1=100$ GeV, $M_2=1$ TeV, $M_3=1$ TeV
\item $M_{\tilde Q_3}= M_{\tilde D_3}= M_{\tilde L_3}= 
M_{\tilde E_3}=M_{\rm SUSY}=1$ TeV,
\end{itemize}
whereas the following parameters are varied as given below: 
\begin{itemize}
\item $\tan\beta=2,~5,~20,~50$ 
\item $|A_f|=500$ GeV, $1.5$ TeV 
\item $\phi_{A_f}=0^\circ$
(whereas the CP-violating effects in the sparticle sector depend on
both $\phi_\mu$ and $\phi_{A_f}$,
the leading CP-violating effects on the Higgs sector, as stated above,
are proportional to ${\cal I}m (\mu A_f)$,
and so we opted to fix 
$\phi_{A_f}$ to 0$^\circ$ and varied only $\phi_\mu$)
\item $|\mu|=500$ GeV, $1$ TeV 
\item $\phi_\mu=0^\circ - 180^\circ$ 
\item $M_{H^+}=100 - 300$ GeV
\item $M_{\tilde U_3}=250$ GeV (case with light $\tilde t_1$) and
$M_{\tilde U_3}=1$ TeV (case with no light sfermion)
\end{itemize}

\noindent 
For this analysis, threshold corrections induced by the exchange of gluinos 
and charginos in the Higgs-quark-antiquark vertices \cite{Hemph,Coarasa} were 
not included. While these corrections may change the actual values of the 
width and BR, we expect it to be the same for both CP-conserving and 
CP-violating cases. The situation will be different if $\phi_3$, the phase
of $M_3$, could be non-zero. As mentioned in the introduction, 
we have considered the case of a common phase for the gaugino mass terms, 
which is rotated away making $M_a$ ($a=1,2,3$) real.

\begin{figure}
\includegraphics[width=8cm]{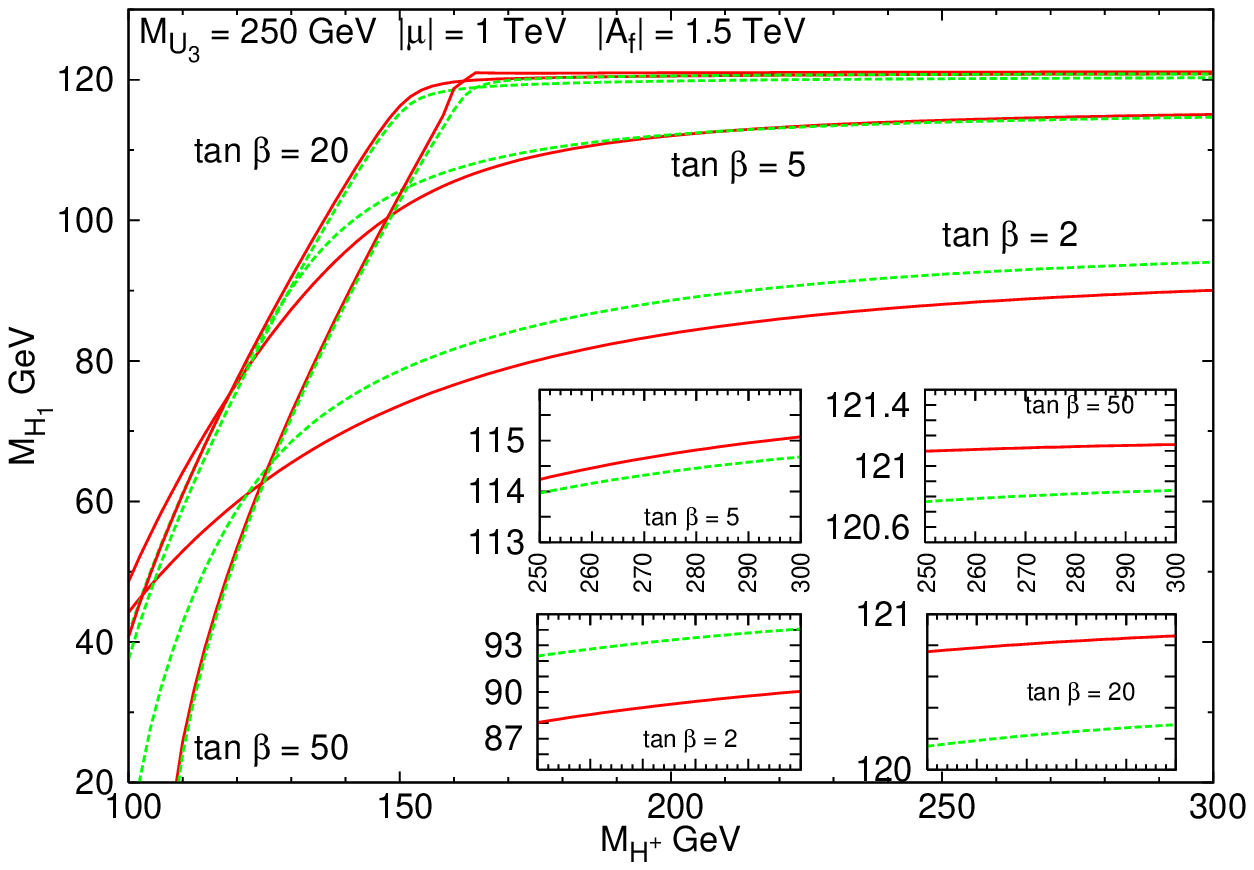}
\includegraphics[width=8cm]{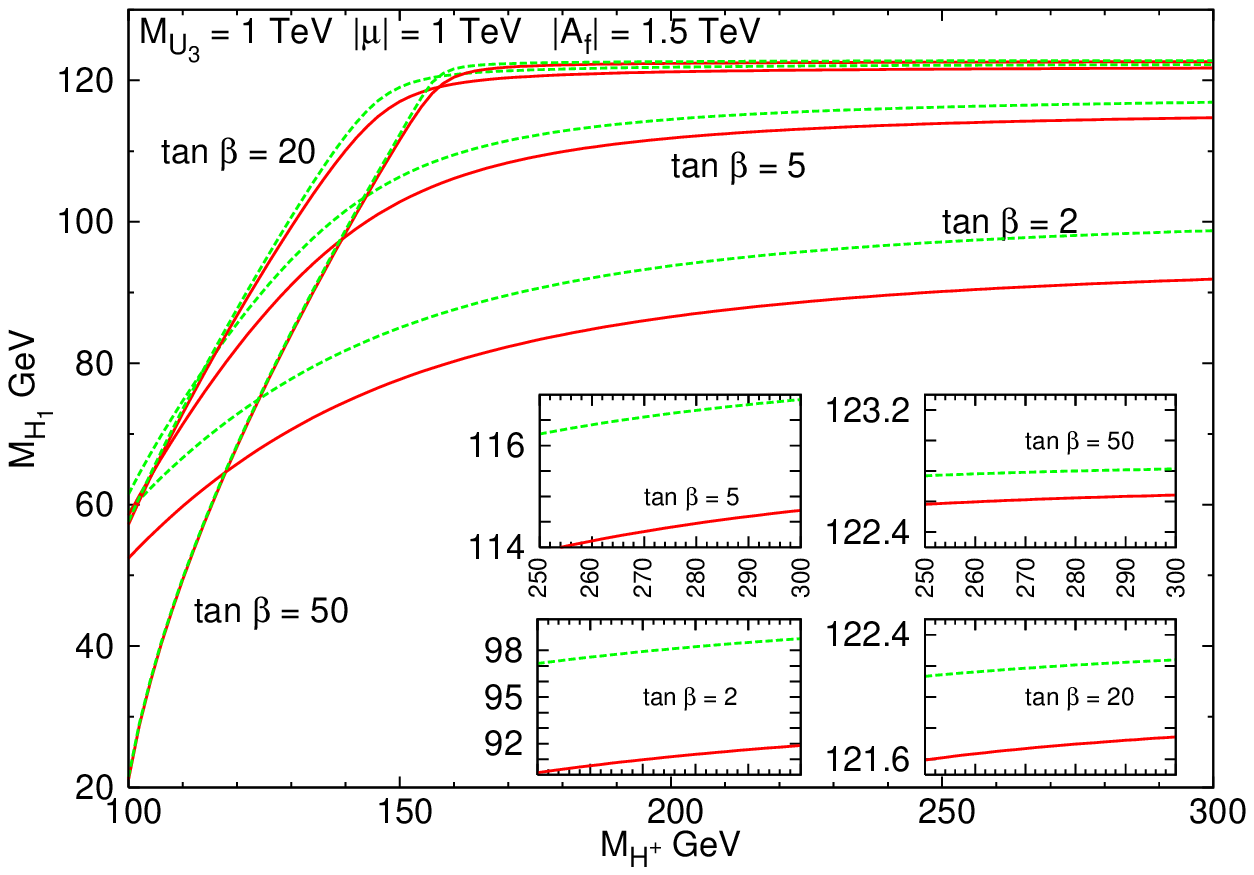}

\includegraphics[width=8cm]{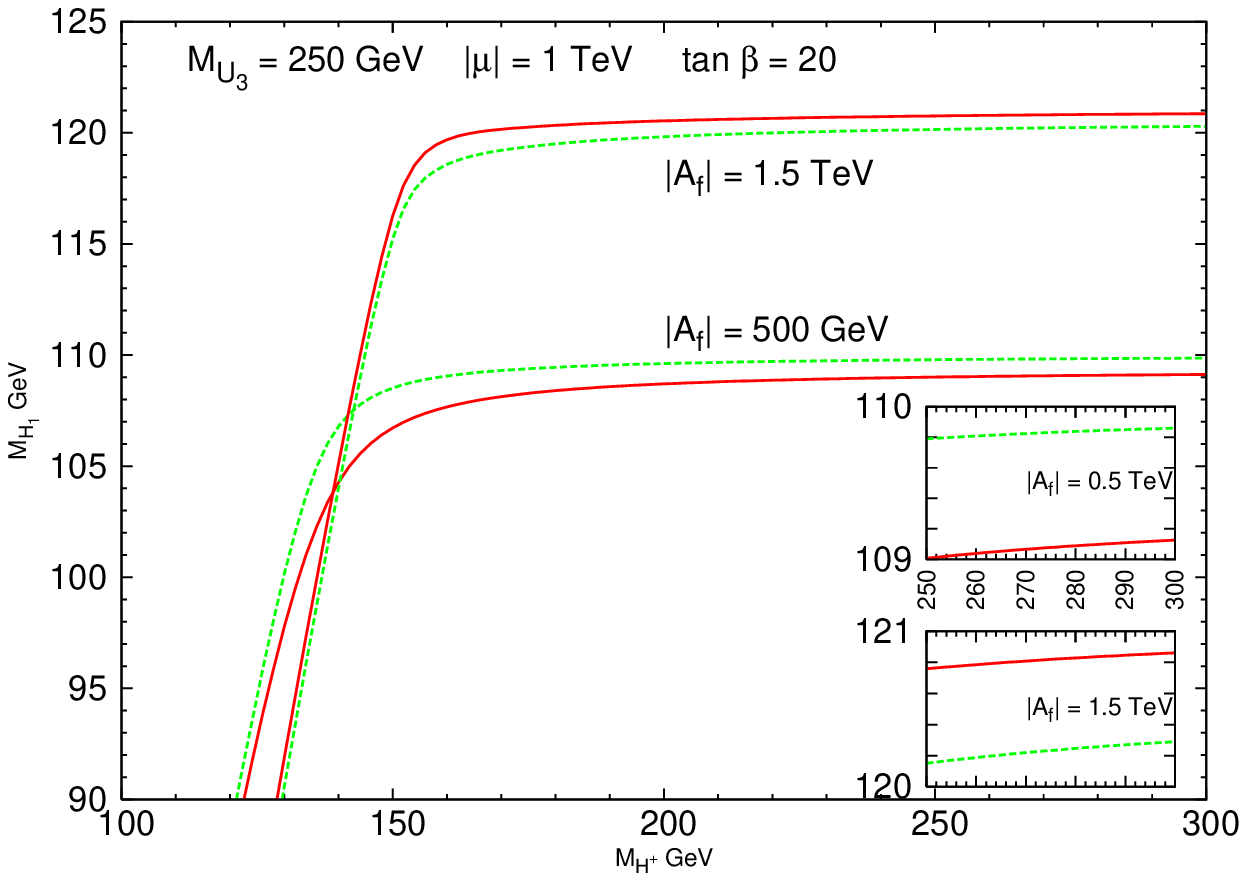}
\includegraphics[width=8cm]{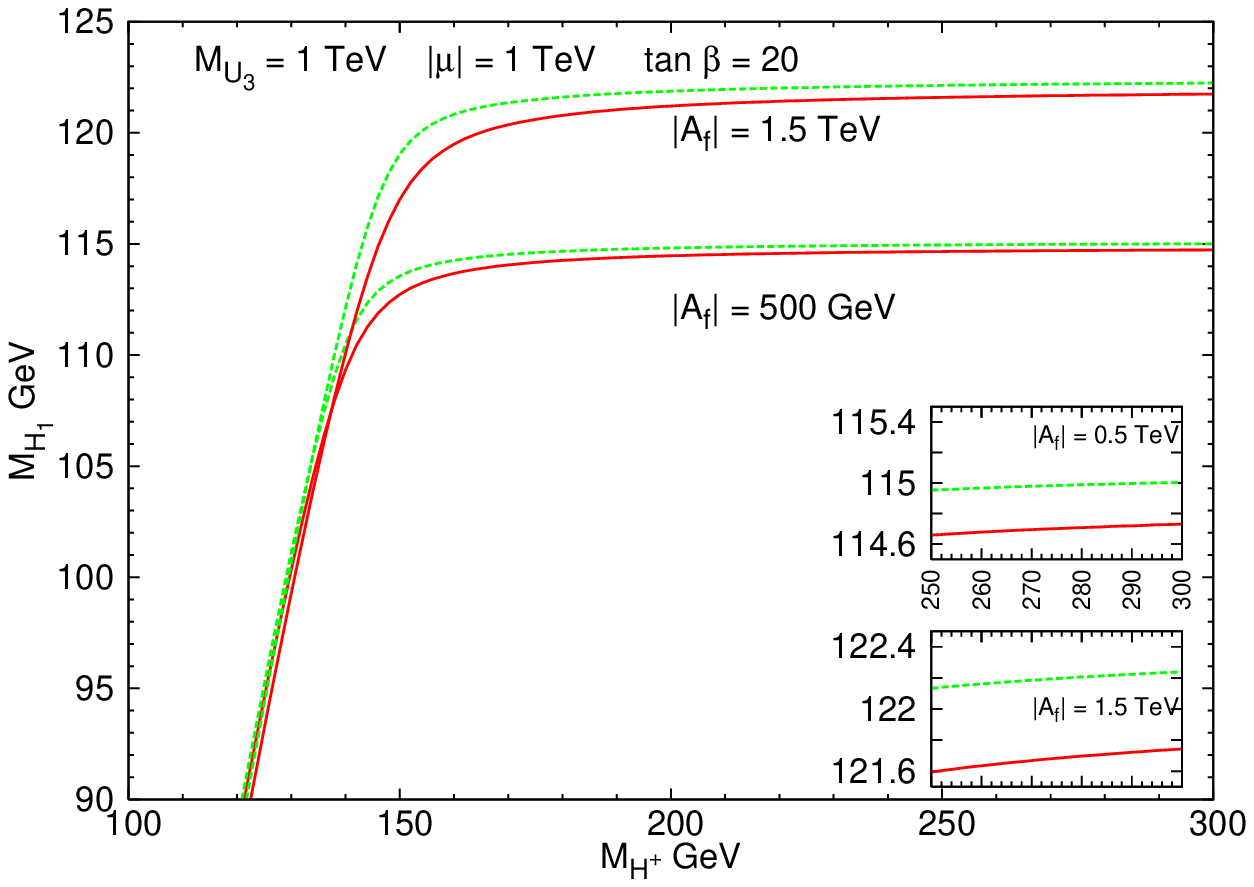}

\caption{\it 
Mass of the lightest neutral Higgs boson $H_1$ against $M_{H^+}$
for $\phi_\mu=0^\circ$ (solid, red line) 
and $\phi_\mu=90^\circ$ (dashed, green line)
with $|A_f|=1.5$ TeV, $|\mu|=1$~TeV and different values of $\tan\beta$
(top row) and
$\tan\beta=20$, $|\mu|=1$~TeV and different values of $|A_f|$
(bottom row), respectively.
In the left column a light stop ($\sim 200$ GeV) is present for
$M_{\tilde U_3}=250$ GeV and $M_{\tilde Q_3}=M_{\rm SUSY}=1$~TeV,
whereas in the right column all sparticles are heavy ($\sim 1$ TeV) for
$M_{\tilde U_3}=M_{\tilde Q_3}=M_{\textrm{SUSY}} = 1$~TeV.
}
\label{fig:MH1}
\end{figure}

The mass of the lightest neutral Higgs boson,
$M_{H_1}$, is sensitive to the value of $\phi_\mu$ chosen.
This dependence on $\phi_\mu$ along with that on other
SUSY parameters is illustrated in Fig. \ref{fig:MH1}, where
$M_{H_1}$ is plotted against the charged Higgs mass for:
 $\tan\beta = 2,~5,~20,~50$; $|A_f|=0.5,~1.5$~TeV 
and $|\mu| = 1$~TeV. Concerning the sparticles in the loop two cases are
considered. The first case comprises
a light $m_{\tilde t_1} \sim 200$ GeV 
(corresponding to $M_{\tilde U_3}=250$ GeV and $M_{\tilde Q_3}=
M_{\textrm{SUSY}}=1$ TeV) while all other sparticles are heavy,
while in 
the other case $m_{\tilde t_1}$ is also taken in the TeV range
with  $M_{\tilde U_3}=M_{\tilde Q_3}= M_{\textrm{SUSY}}=1$ TeV.
The top row in Fig.~\ref{fig:MH1} shows the sensitivity of $M_{H_1}$
to $\tan\beta$ for $|\mu|=1$ TeV and $|A_f|=1.5$~TeV. 
While in the low $\tan\beta$ case the mass shift induced by the change in $\phi_\mu$ from
$0^\circ$ to $90^\circ$
is about 10 \%, in the case of 
$\tan\beta=20$ or above it is about 1 \% or less. 
Notice that the relevant parameters here are $|A_f|$ and $|\mu|$ and
$M_{H_1}$ is found to increase with increasing $|A_f|$ while its 
dependence on  $|\mu|$ is basically negligible. 
In the bottom row of Fig. \ref{fig:MH1} we plot the $M_{H_1}$ dependence
on $M_{H^\pm}$ 
for two representative values of $|A_f|$, 0.5 and 1.5 TeV, by keeping 
$\tan\beta=20$ and $|\mu|=1$ TeV.  

\begin{figure}
\includegraphics[width=8cm]{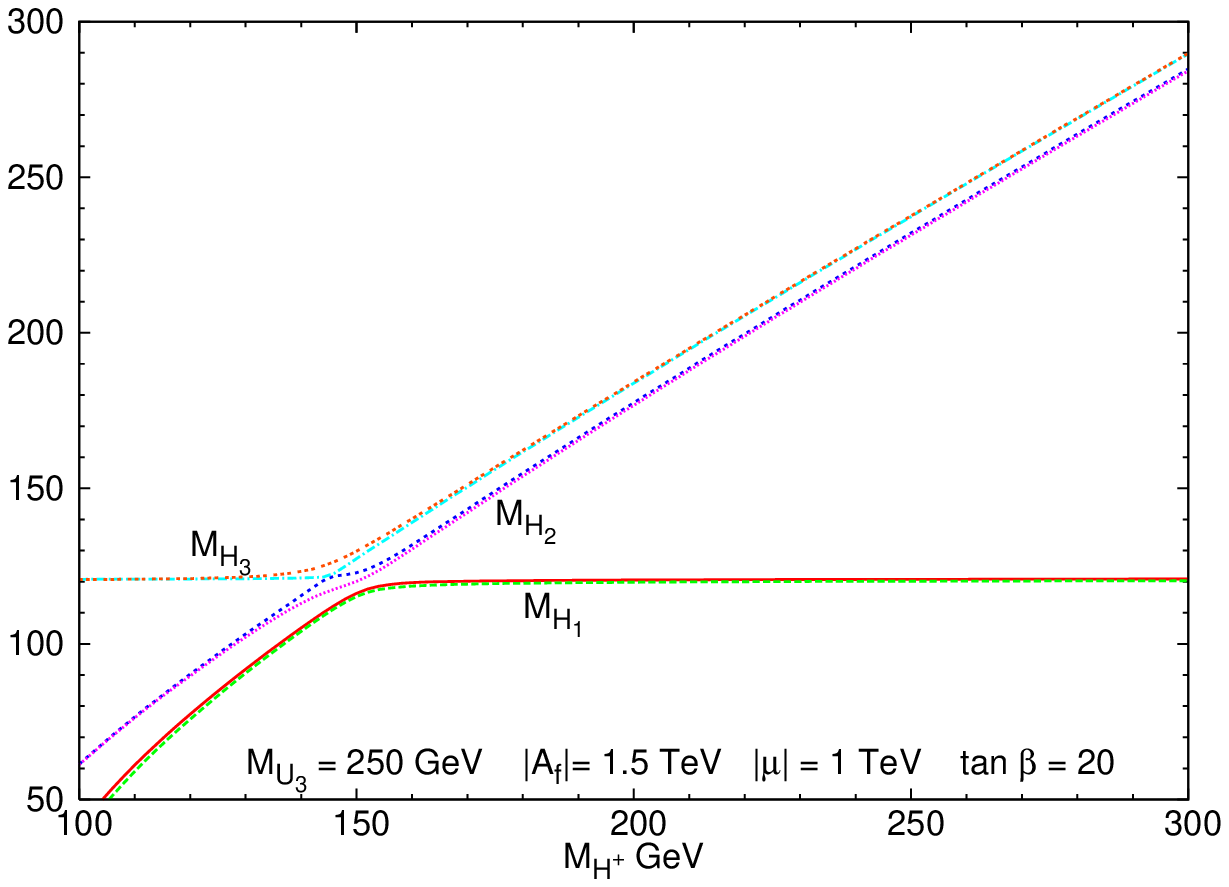}
\includegraphics[width=8cm]{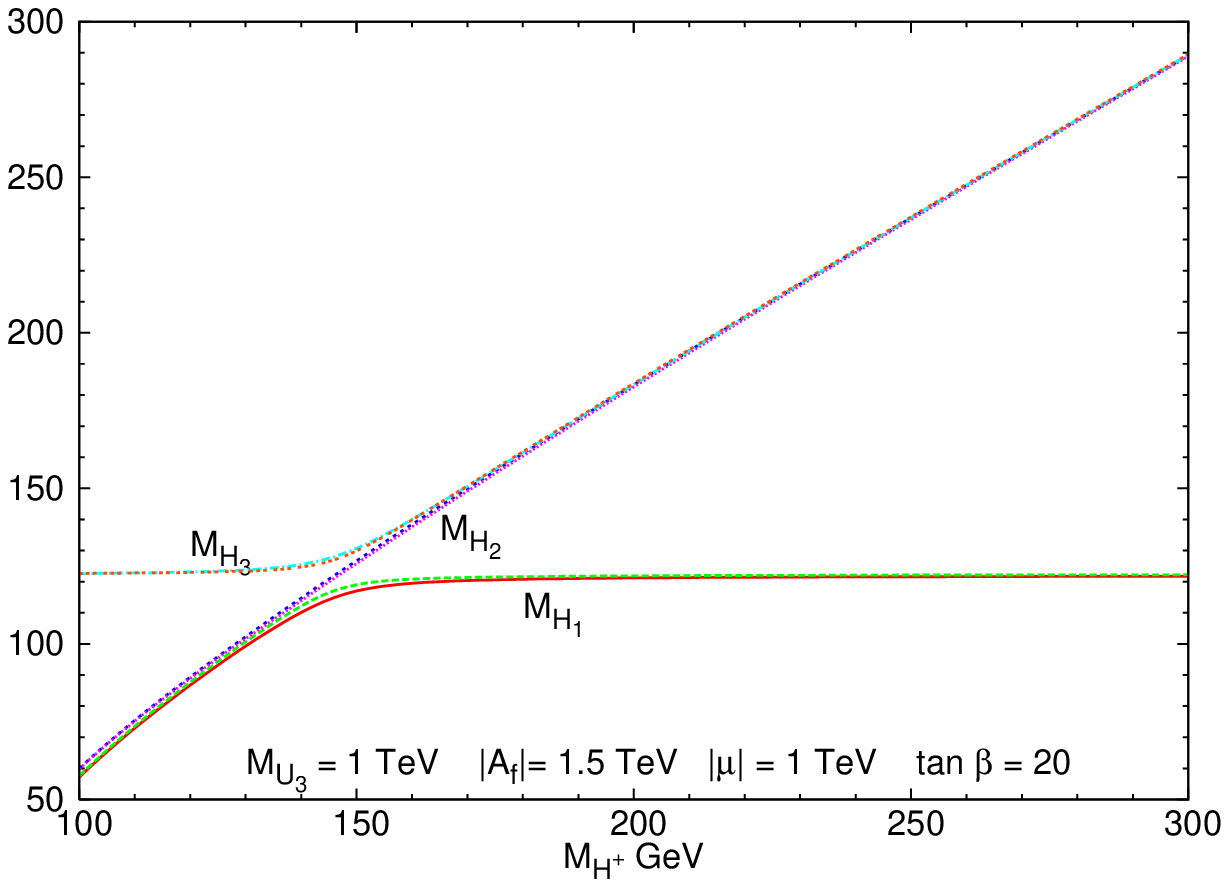}

\caption{\it 
Mass of $H_{1,2,3}$ against $M_{H^+}$ for $\tan\beta = 20$ 
showing the cross over at $M_{H^+}\sim 150$ GeV. Left column is for
$M_{\tilde U_3}=250$ GeV
($m_{\tilde t_1}\sim 200$ GeV), while the right one is 
for $M_{\tilde U_3} = 1$~TeV
(no light  sparticle).
Both plots are with $|A_f|=1.5$ TeV, $|\mu|=1$ TeV.
{Red}, blue and cyan curves represent $\phi_\mu=0^\circ$, while
green, magenta and orange curves correspond to $\phi_\mu=90^\circ$.
}
\label{fig:MHs}
\end{figure}

\begin{table}[h]
\begin{tabular}{|c|c|c|c|c|c|c|}
\hline
&\multicolumn{3}{c|}{$\phi_\mu=0^\circ$}&\multicolumn{3}{c|}{$\phi_\mu=90^\circ$}\\ \cline{2-7}
$M_{H^+}$ (GeV)&
$M_{H_1}$ (GeV)&
$M_{H_2}$ (GeV)&
$M_{H_3}$ (GeV)&
$M_{H_1}$ (GeV)&
$M_{H_2}$ (GeV)&
$M_{H_3}$ (GeV)\\ \cline{1-7}
100&40.7&61.4&120.7&37.6&61.2&120.7\\
120&77.4&90.3&120.8&75.8&89.7&121.0\\ \cline{1-7}
200&120.5&179.8&183.8&119.8&176.7&184.2\\
250&120.8&232.1&237.3&120.2&231.4&237.6\\
300&120.9&284.8&289.6&120.3&284.2&289.8\\
\hline
\end{tabular}
\caption{\it Selected values of $M_{H_i}$ ($i=1,2,3$) for $\phi_\mu=0^\circ$ and
$\phi_\mu=90^\circ$. All SUSY parameters are as in Fig. \ref{fig:MHs} with
$M_{\tilde U_3}=250$ GeV.}
\label{table:MHs1}
\end{table}

\begin{table}[h]
\begin{tabular}{|c|c|c|c|c|c|c|}
\hline
&\multicolumn{3}{c|}{$\phi_\mu=0^\circ$}&\multicolumn{3}{c|}{$\phi_\mu=90^\circ$}\\ \cline{2-7}
$M_{H^+}$ (GeV)&
$M_{H_1}$ (GeV)&
$M_{H_2}$ (GeV)&
$M_{H_3}$ (GeV)&
$M_{H_1}$ (GeV)&
$M_{H_2}$ (GeV)&
$M_{H_3}$ (GeV)\\ \cline{1-7}
100&57.2&60.0&122.6&57.6&59.8&122.6\\
120&86.8&89.3&123.1&87.7&88.9&123.0\\ \cline{1-7}
200&121.2&183.3&183.4&121.9&182.6&183.7\\
250&121.6&236.9&237.0&122.1&236.4&237.2\\
300&121.7&289.2&289.3&122.2&288.9&289.5\\
\hline
\end{tabular}
\caption{\it  Same as Table \ref{table:MHs1}, but with 
$M_{\tilde U_3}=1$ TeV.}
\end{table}

The sudden shift in the dependence of $M_{H_1}$ on $M_{H^+}$
around $M_{H^+}=150$ GeV is understood in terms of the cross over 
in the mass eigenstates at that point. We have illustrated this in
Fig.~\ref{fig:MHs}, where the masses of $H_{1,2,3}$ are plotted against 
$M_{H^+}$ for $\tan\beta=20,~|A_f|=1.5$ TeV and $|\mu|=1$ TeV, again with light
and heavy stops. 
The cross over is a reflection of the
changing compositions of the CP-indefinite mass eigenstates, $H_1,~H_2,~H_3$, with 
eigenvalues $M_{H_1}<M_{H_2}<M_{H_3}$, 
in terms of the CP-definite gauge eigenstates, $\phi_1,~\phi_2$ and $a$.  
To explain this in a little more detail, let us denote the mass eigenvalues 
as $m_1,~m_2,~m_3$ and the corresponding eigenstates as $h_1,~h_2,~h_3$, 
before ordering them from lightest to heaviest
(i.e., $h_1$ need not be the lightest for all values of $M_{H^+}$).
Of the three mass eigenvalues, two ($m_1$ and $m_2$) 
grow linearly with $M_{H^+}$ with almost
the same slope, and lying close to each other. In the CP-conserving case
one of these is a pseudo-scalar, while the other is a scalar. The other 
eigenvalue ($m_3$) corresponds to a scalar eigenstate, 
and is more or less independent of $M_{H^+}$. 
At $M_{H^+}\sim 150$ GeV all  three CP-conserving eigenstates are degenerate with
eigenvalues around 120 GeV.
In the $M_{H^+} \lesssim 150$ GeV region $h_1$ is the lightest, while in the  
$M_{H^+} \gtrsim 150$ GeV region it is $h_3$ which is the lightest. 
When we order such that the lightest is $H_1$, there is a transition
from $H_1=h_1$ to $H_1=h_3$ around $M_{H^+}=150$ GeV.
For other values of $\tan\beta$, $A_f$ and $\mu$ the situation is very similar, 
with small shifts in the actual values of $M_{H^+}$ and the degenerate mass
where the cross over happens. In the CP-violating case with non-zero value of 
$\phi_\mu$, there is mixing between scalar and pseudo-scalar states.
For larger values of $M_{H^+}$ the lightest Higgs state, $H_1$, is almost 
a pure scalar, hence we will not be subject to
 any CP violation through mixing. 
The only possible way to have CP violation here is through the 
$H_1\tilde f\tilde f^*$ coupling, especially that of the stop quark.
We restrict to regions of parameter space with small $M_{H^+}$ where the 
effect of mixing as well as that due to a complex $\phi_{1,2}\tilde f\tilde f^*$
coupling are present. At large $M_{H^+}$ values there will be 
scalar/pseudo-scalar mixing in the heavier Higgs states, $H_2$ and $H_3$. (We will
not discuss the two heavier states in the present article though.) Tabs.~1 and 2 illustrate selected values
of the $H_1, H_2$ and $H_3$ masses for sample choices of $M_{H^\pm}$
when $\phi_\mu=0^\circ$ and $90^\circ$ in the presence
of a light stop and otherwise, respectively. Mass shifts can typically be of a few percent, particularly
for light Higgs masses, in the former case while they are negligible in the latter.

\begin{figure}

\epsfysize=5cm \epsfxsize=5cm
\epsfbox{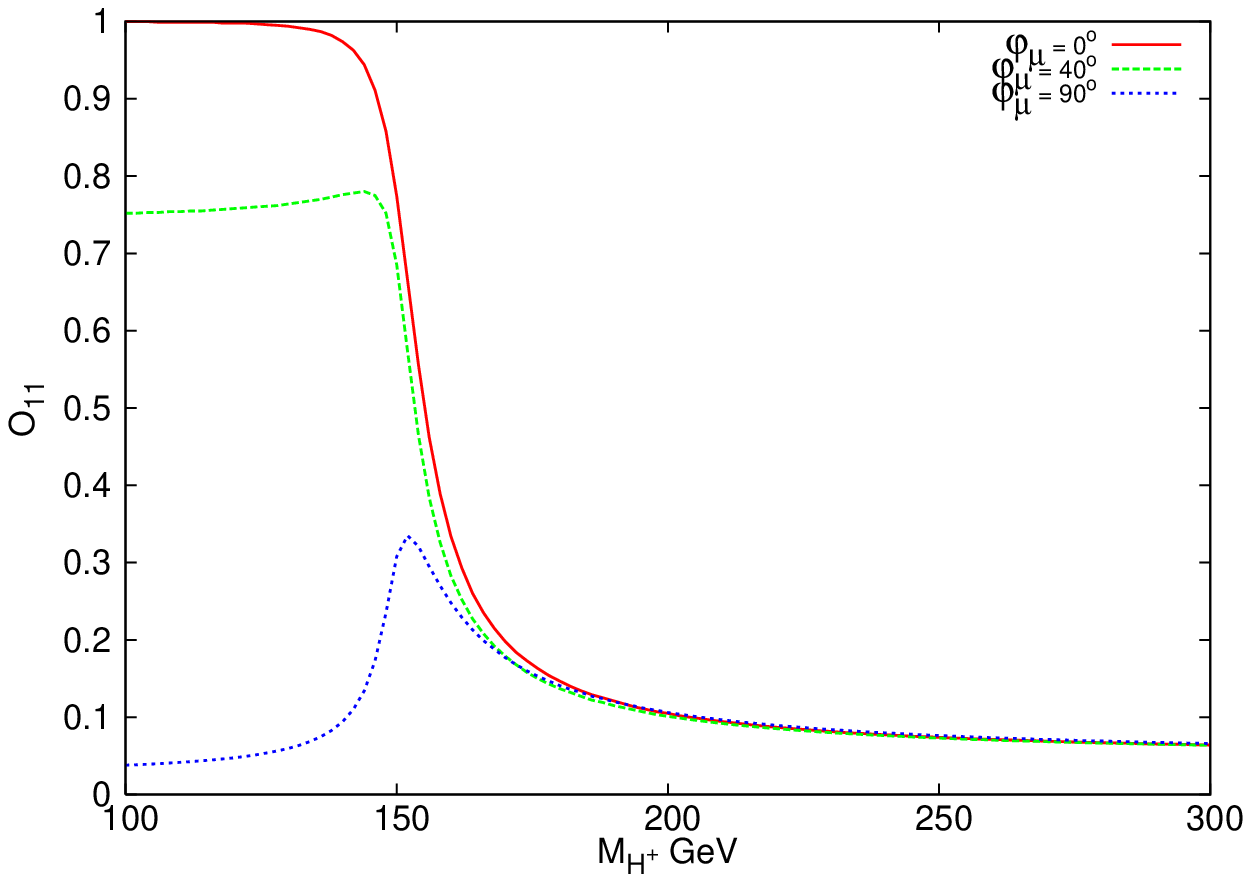}
\epsfysize=5cm \epsfxsize=5cm
\epsfbox{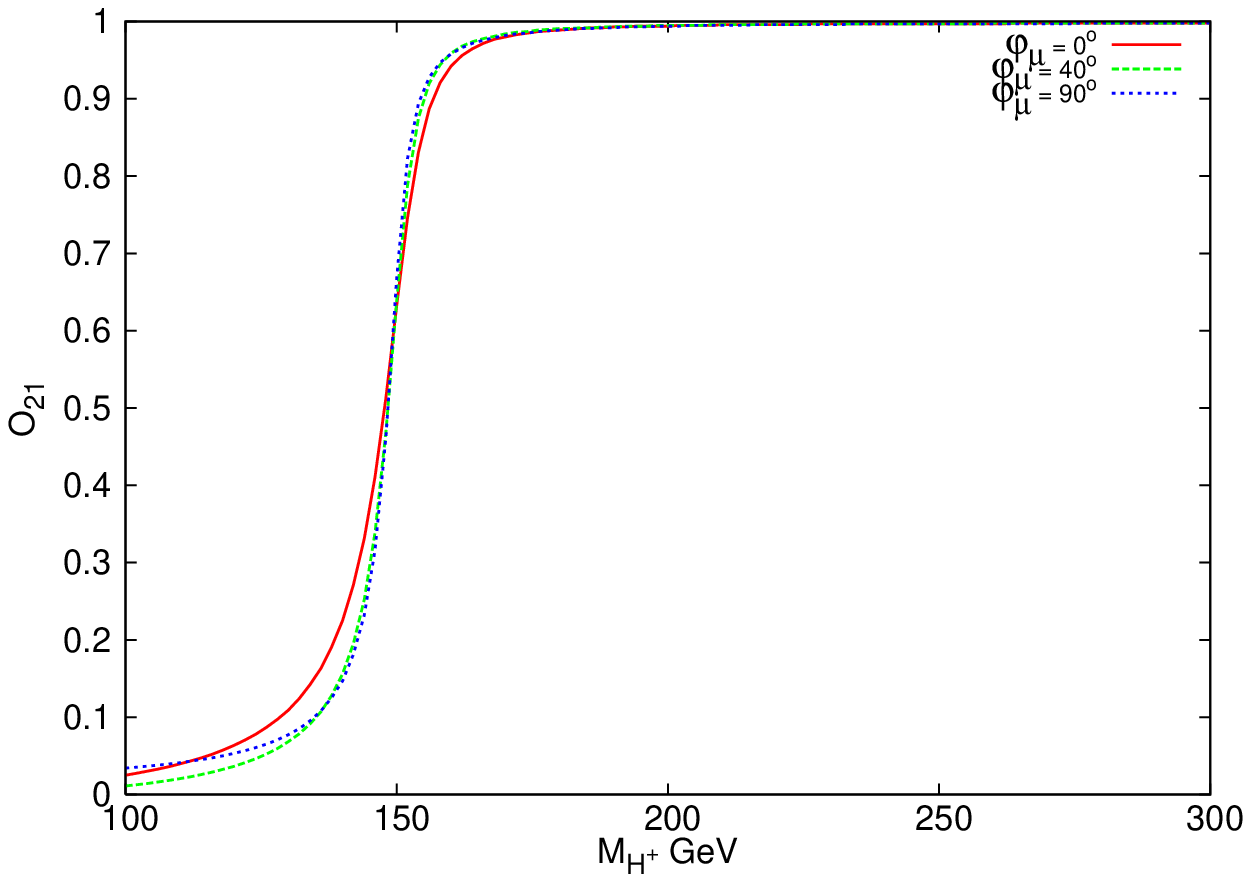}
\epsfysize=5cm \epsfxsize=5cm
\epsfbox{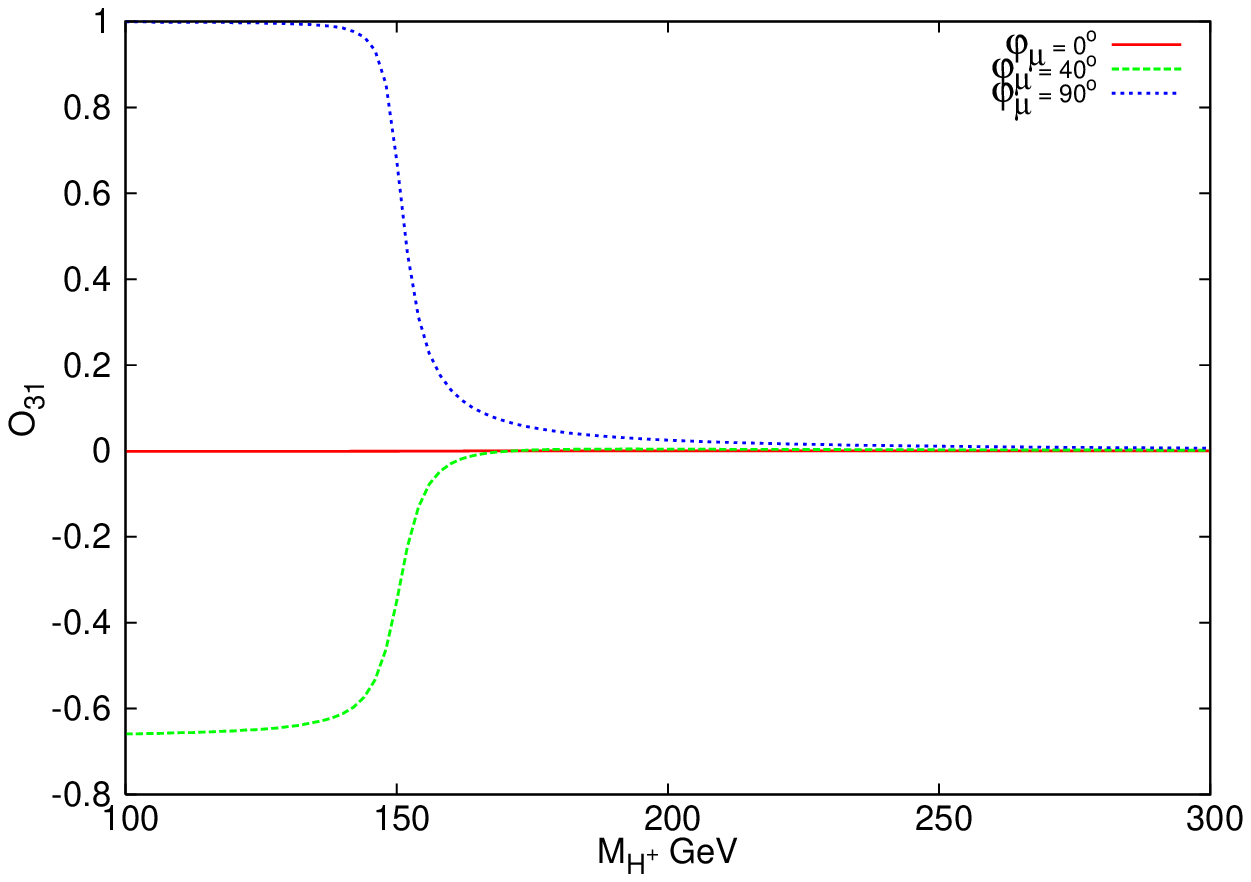}

\epsfysize=5cm \epsfxsize=5cm
\epsfbox{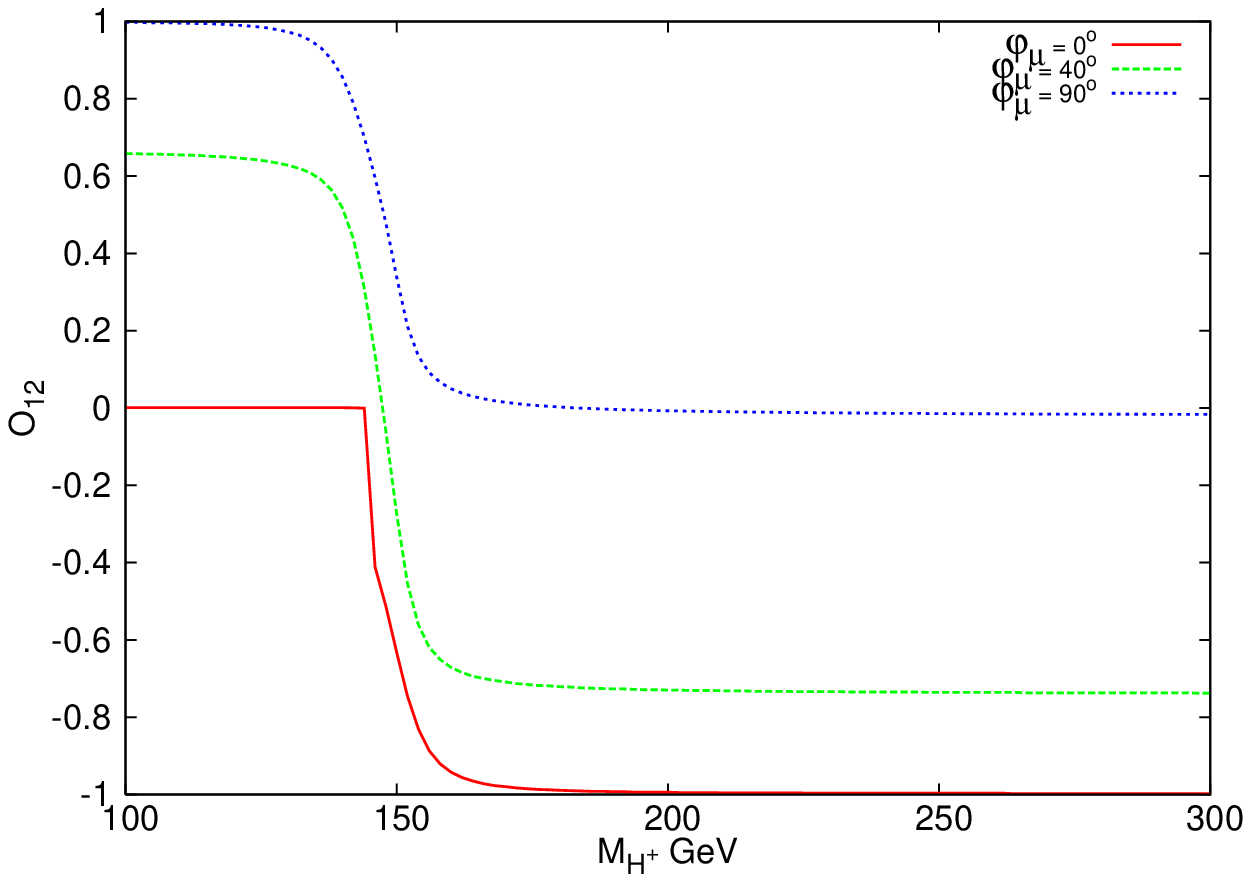}
\epsfysize=5cm \epsfxsize=5cm
\epsfbox{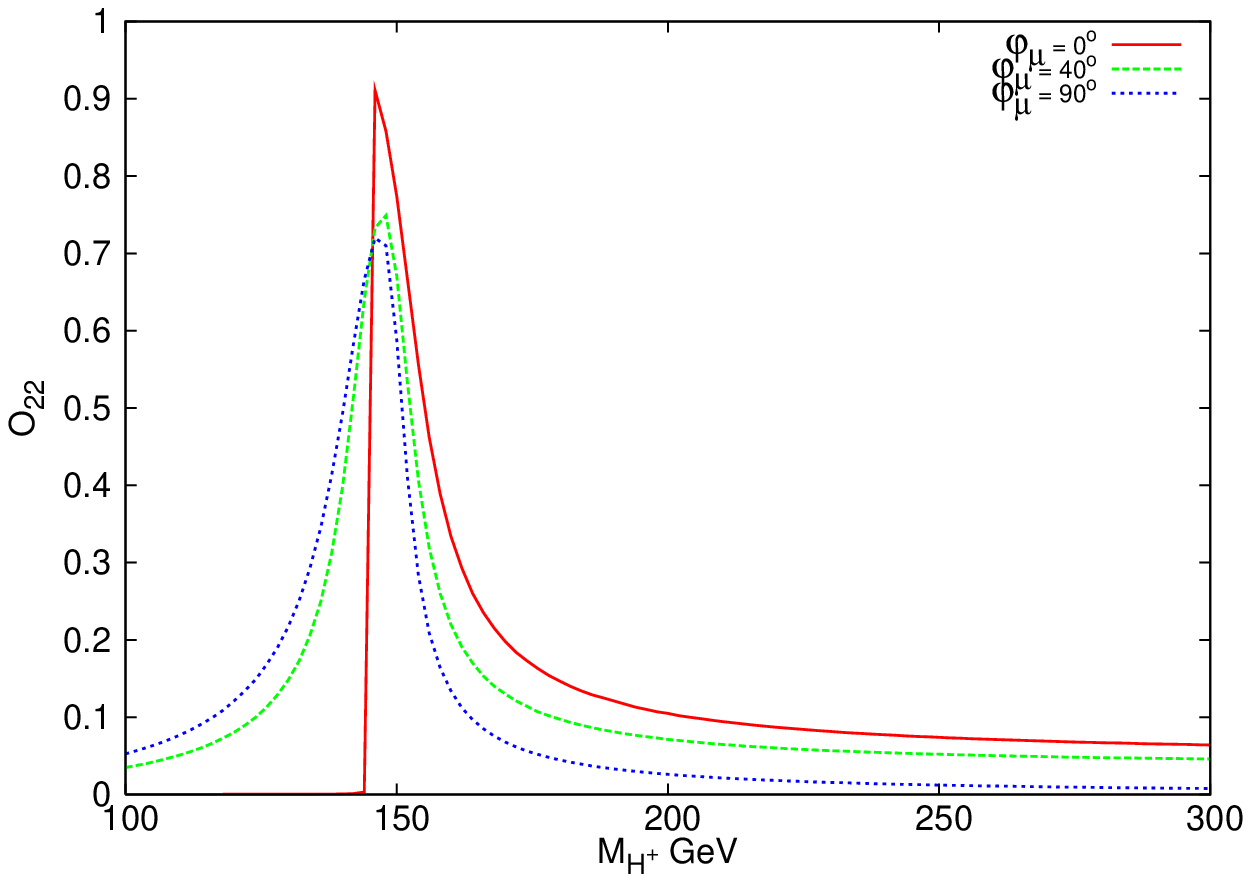}
\epsfysize=5cm \epsfxsize=5cm
\epsfbox{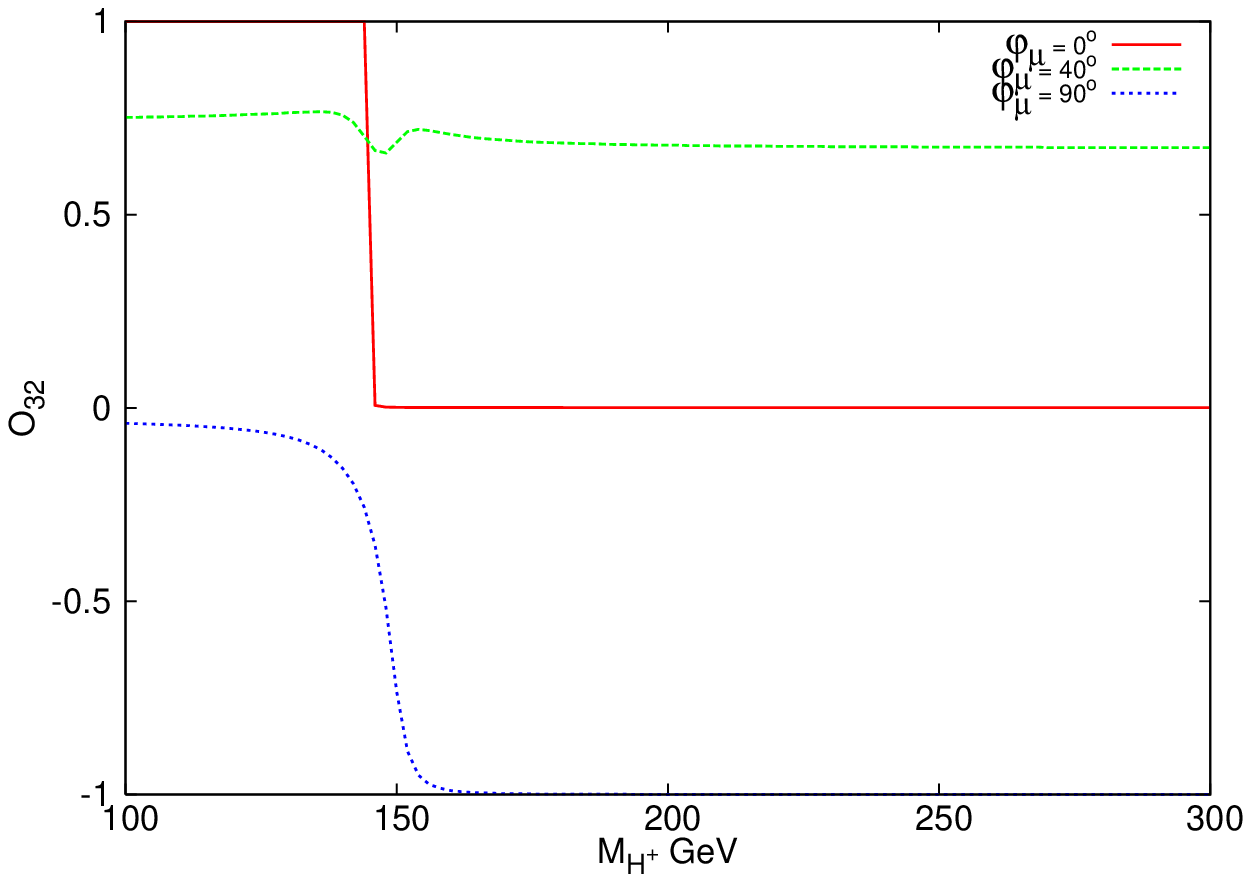}

\epsfysize=5cm \epsfxsize=5cm
\epsfbox{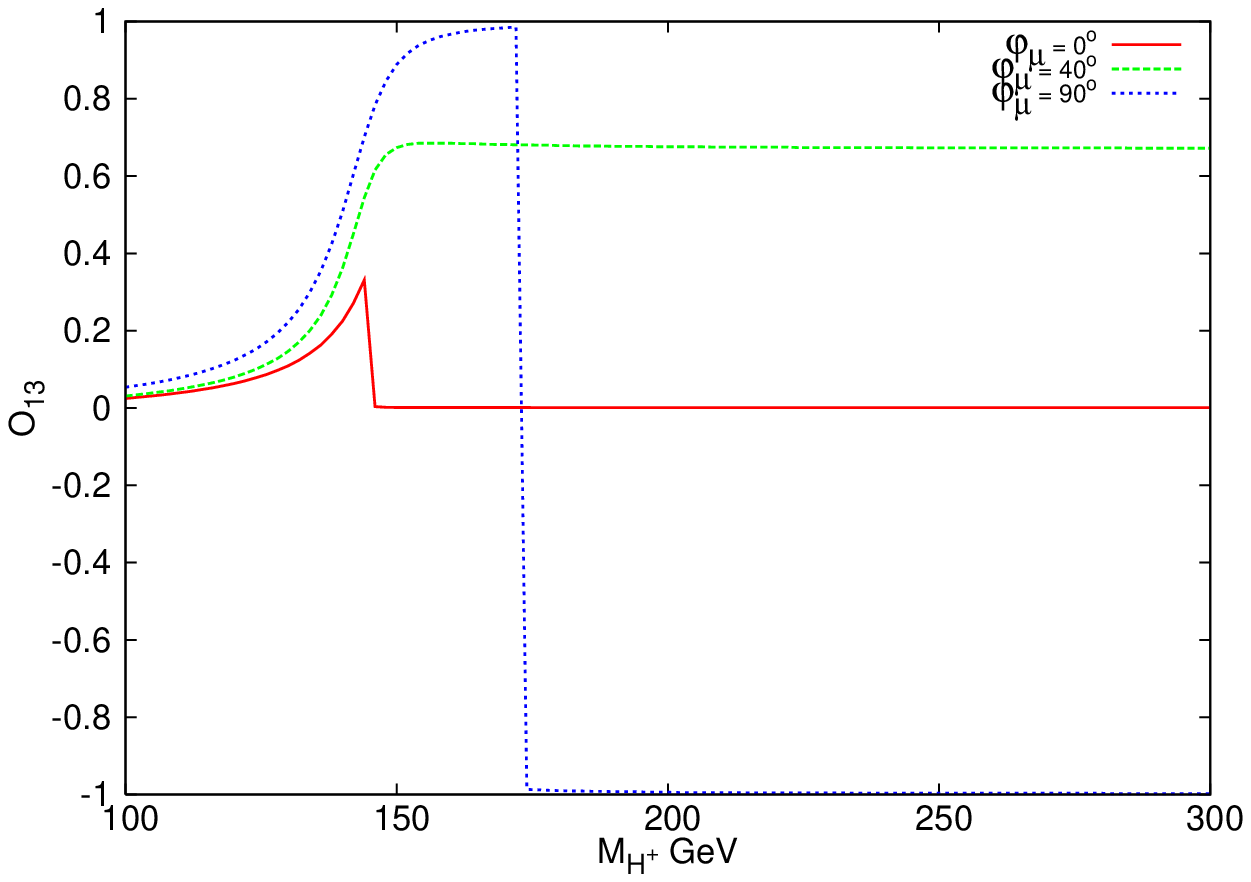}
\epsfysize=5cm \epsfxsize=5cm
\epsfbox{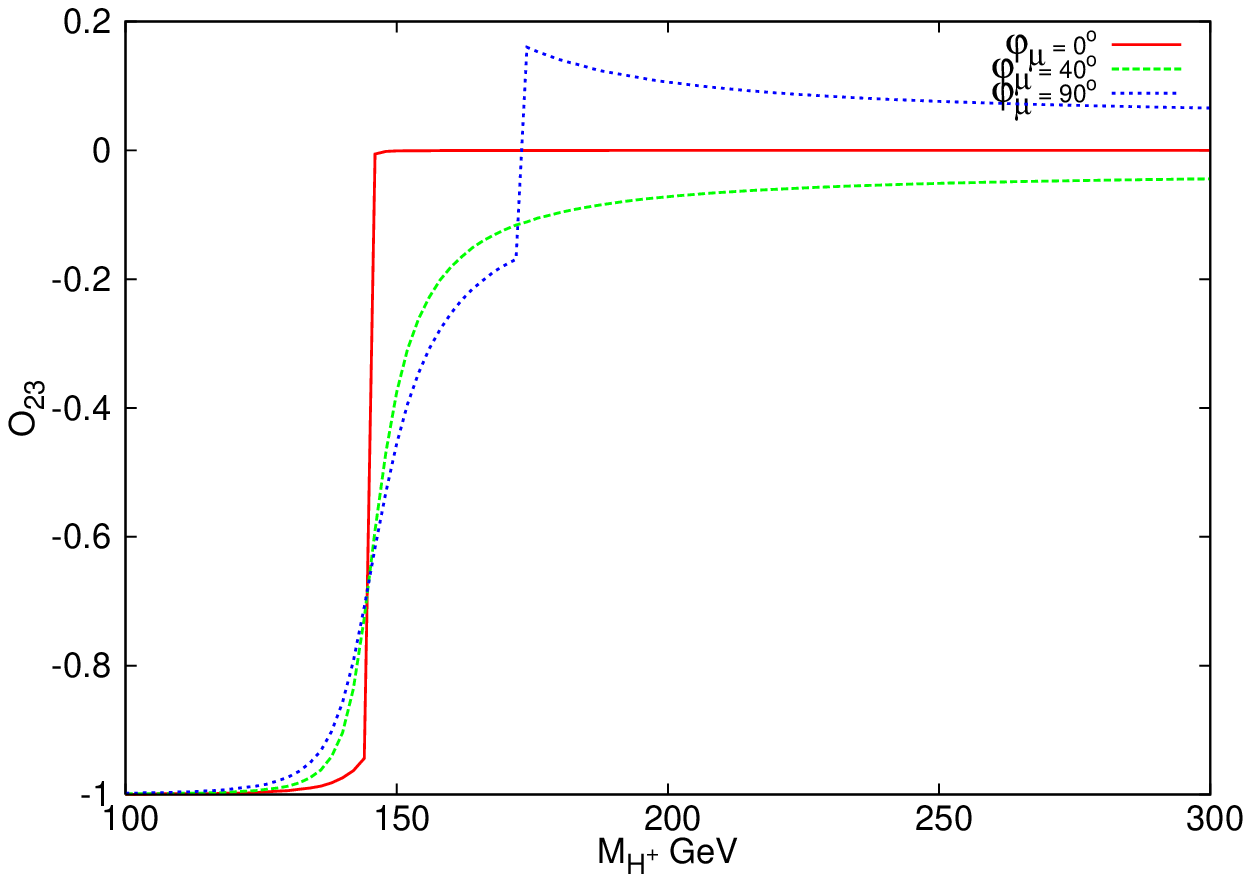}
\epsfysize=5cm \epsfxsize=5cm
\epsfbox{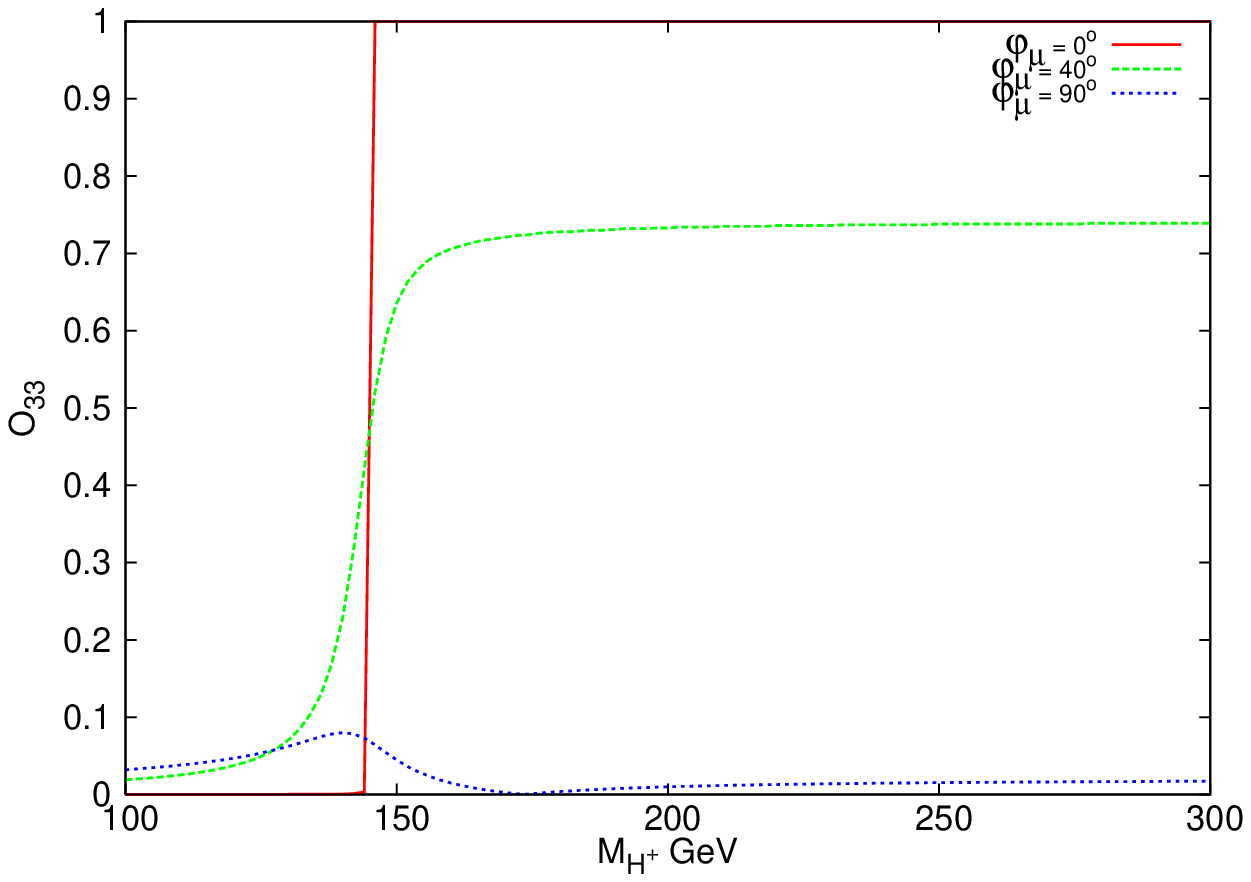}

\caption{\it 
Mixing matrix elements $O_{ij}$ vs. $M_{H^+}$ 
(such that $H_i=O_{1i}~\phi_1+O_{2i}~\phi_2+O_{3i}~a$) with 
$\tan\beta=20$, $|A_f|=1.5$ TeV, $|\mu|=1$ TeV, 
$M_{\tilde Q_3}= M_{\tilde D_3}= M_{\tilde L_3}=
M_{\tilde E_3}= M_{\rm SUSY}= 1$ TeV, 
$M_{\tilde U_3}= 250$ GeV and $\phi_\mu$ as indicated in the plots.
}
\label{fig:mixO1}
\end{figure}

\begin{figure}

\epsfysize=5cm \epsfxsize=5cm
\epsfbox{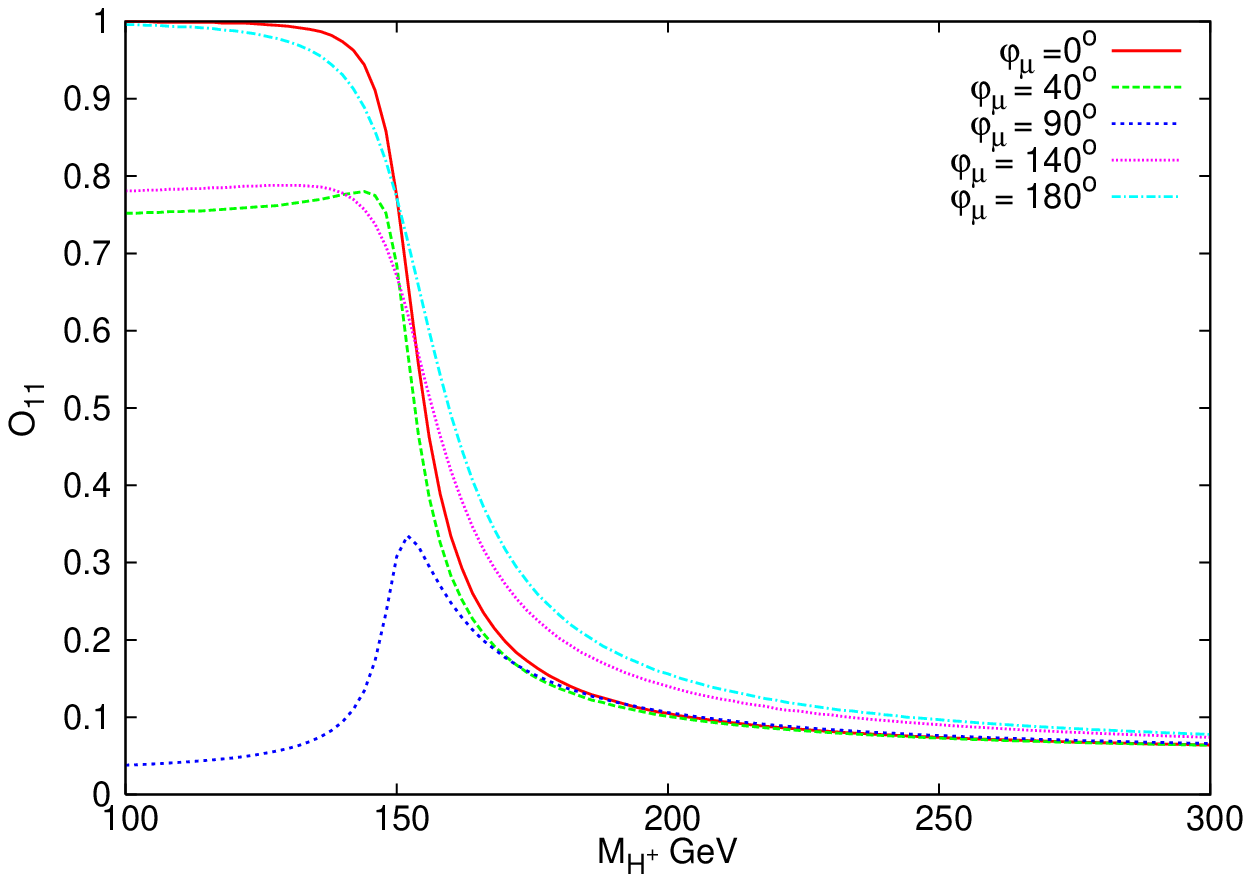}
\epsfysize=5cm \epsfxsize=5cm
\epsfbox{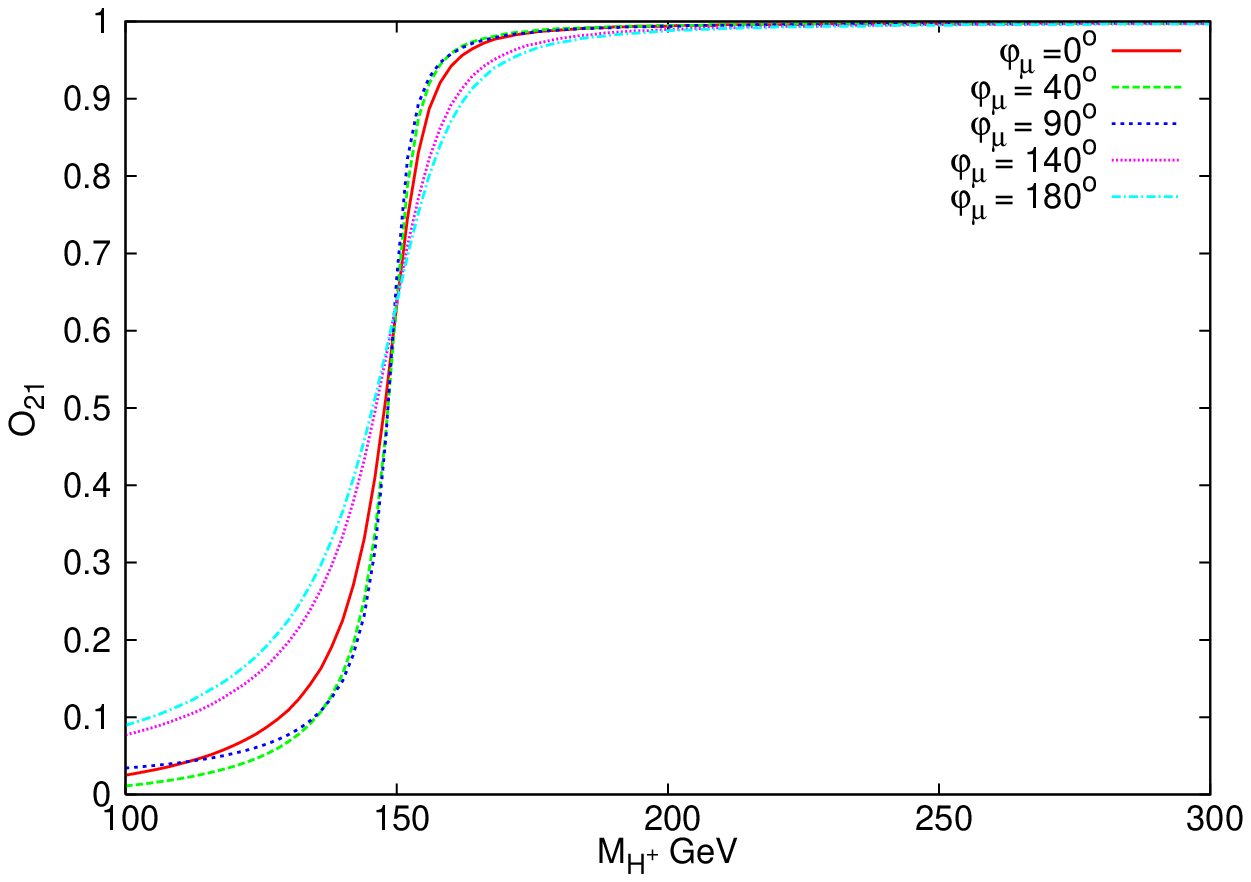}
\epsfysize=5cm \epsfxsize=5cm
\epsfbox{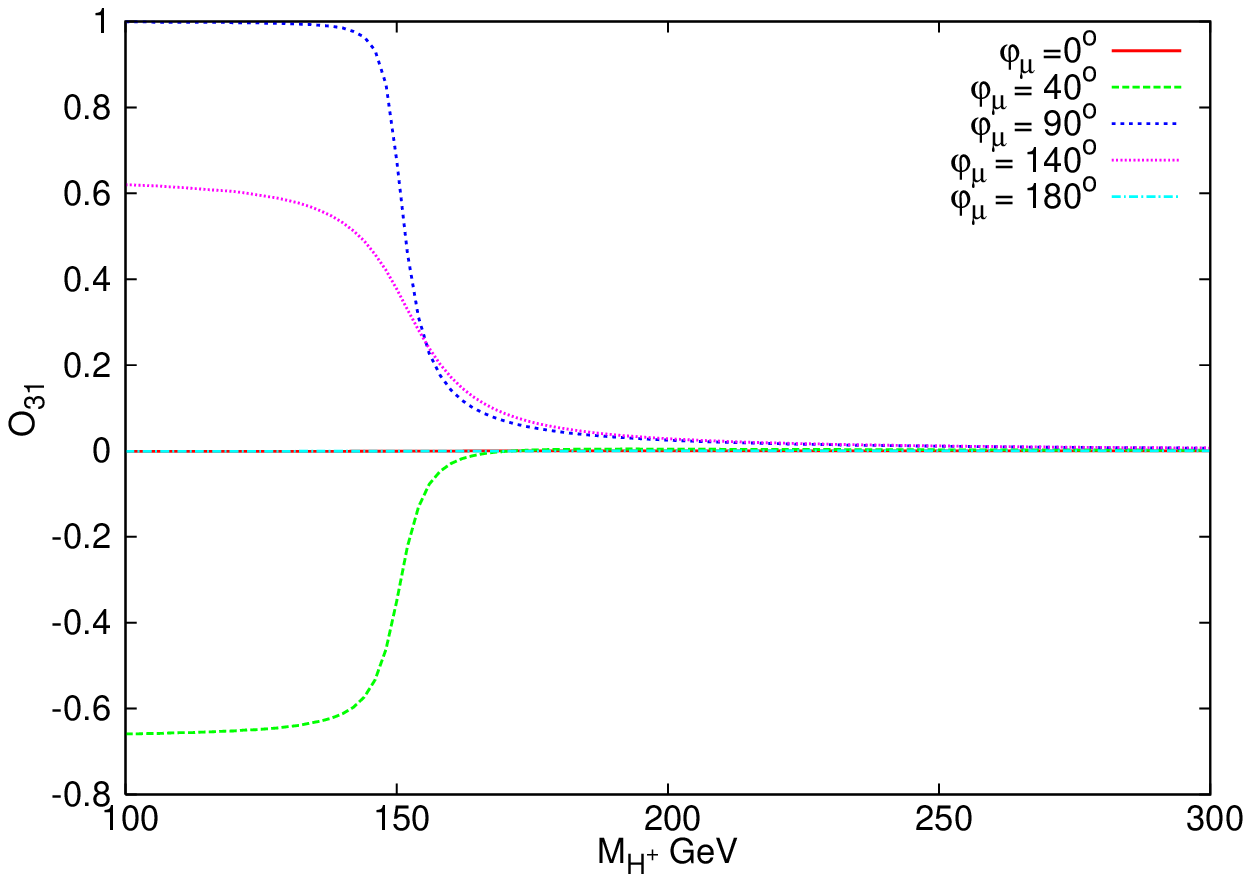}

\epsfysize=5cm \epsfxsize=5cm
\epsfbox{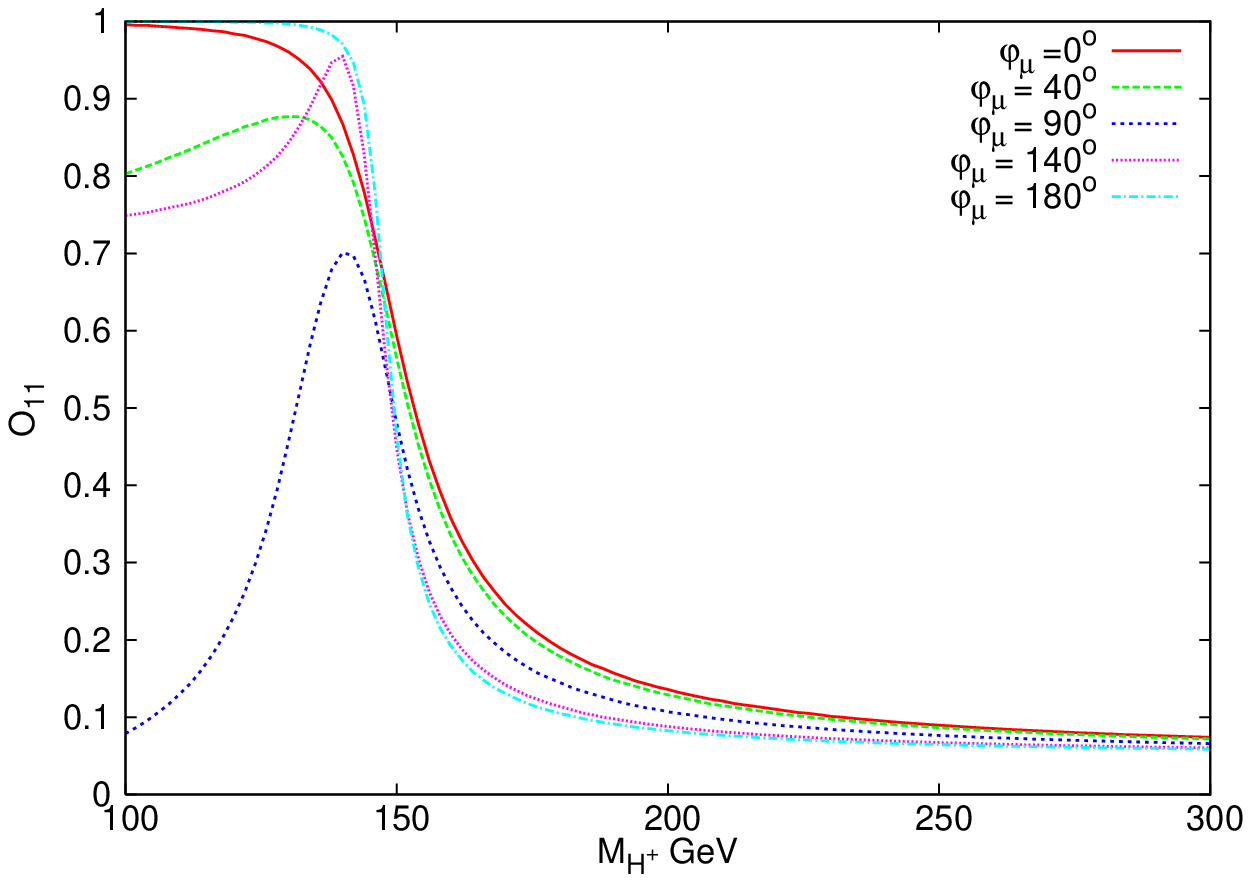}
\epsfysize=5cm \epsfxsize=5cm
\epsfbox{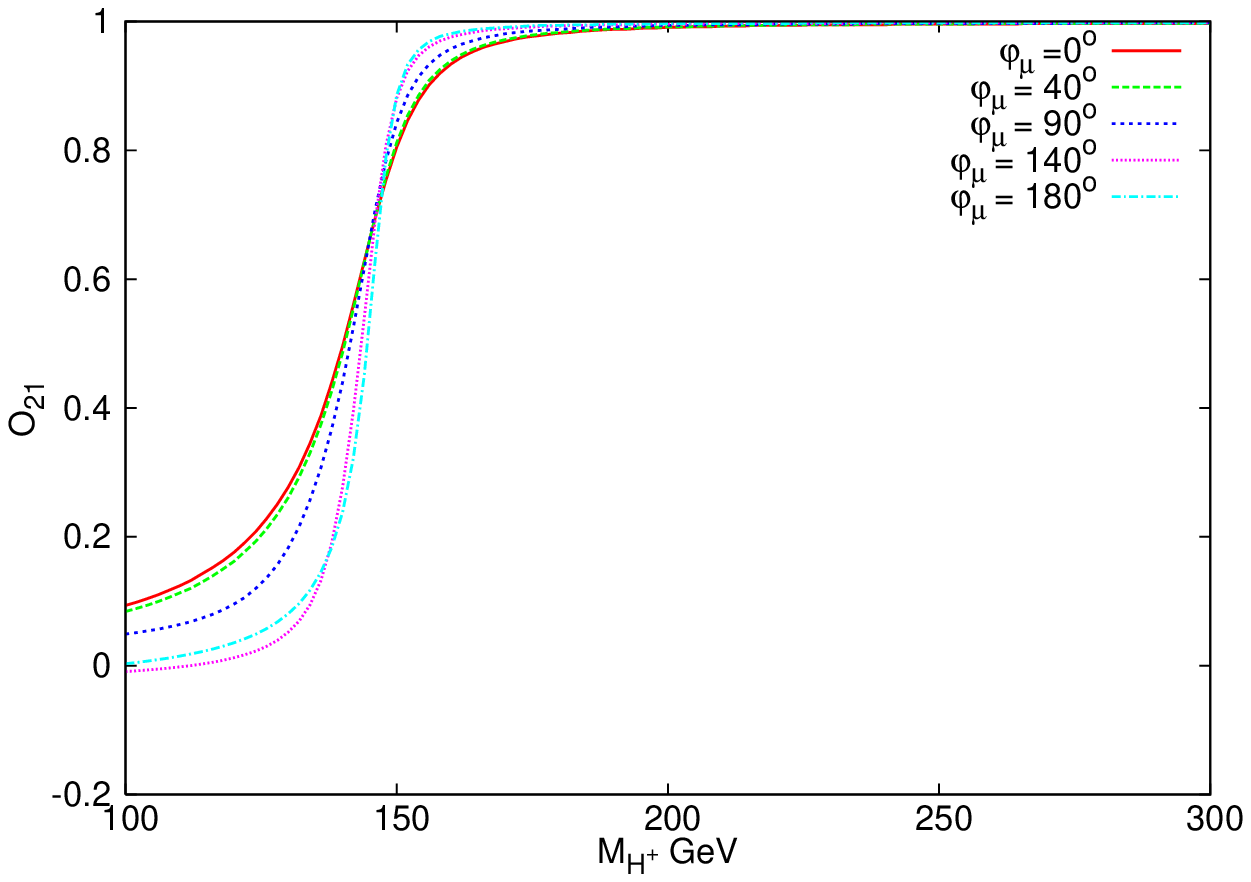}
\epsfysize=5cm \epsfxsize=5cm
\epsfbox{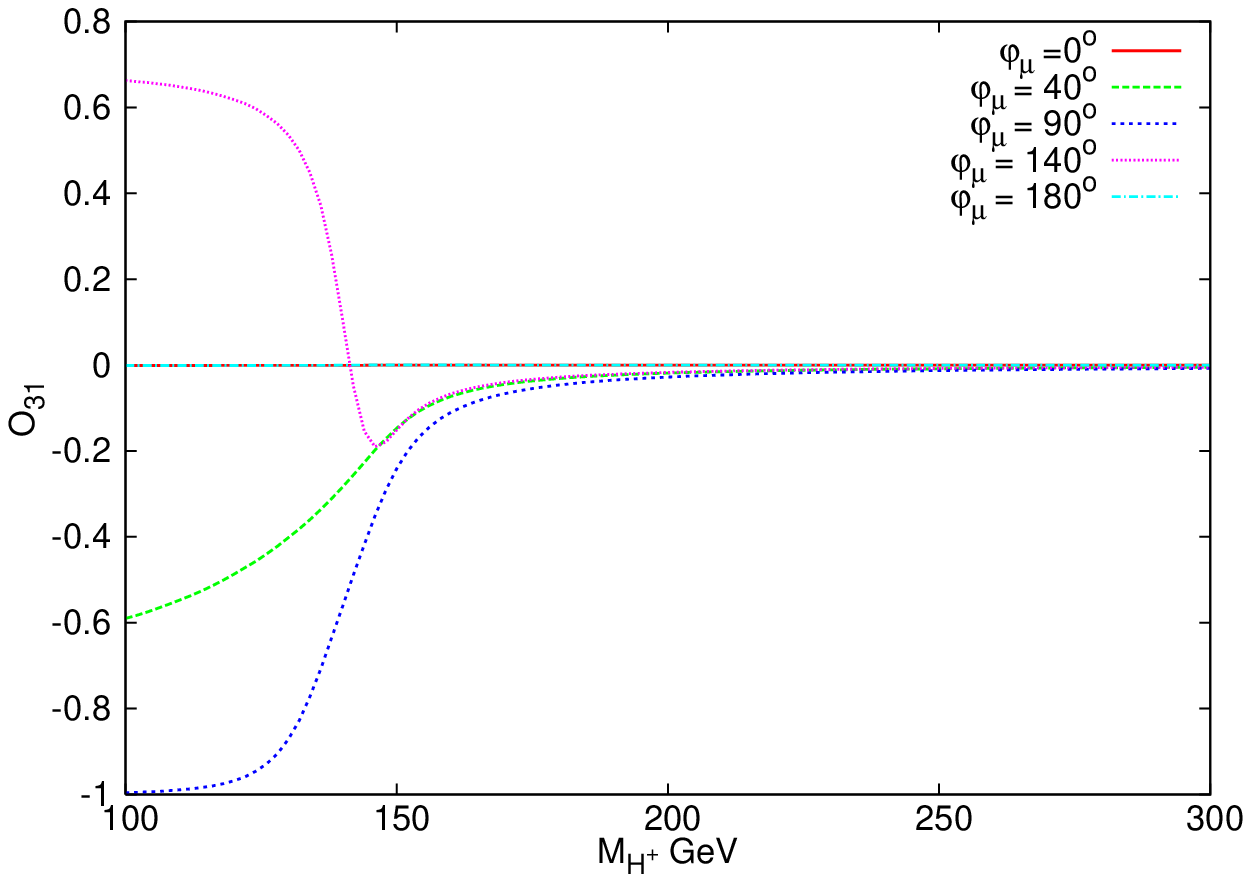}

\caption{\it 
Mixing matrix elements $O_{i1}$ vs. $M_{H^+}$ 
(such that $H_1=O_{11}~\phi_1+O_{21}~\phi_2+O_{31}~a$) for $\tan\beta=20$.
Top row corresponds to $M_{\tilde U_3}= 250$ GeV while 
bottom row corresponds to $M_{\tilde U_3}= 1$ TeV. All other
parameters are as in Fig. \ref{fig:mixO1}.
}
\label{fig:mixO2}
\end{figure}

\vspace*{0.15cm}
Next we analyse the Higgs mixing matrix for the parameters
$\tan\beta=20$, $|A_f|=1.5$ TeV, $|\mu|=1$ TeV, as an example
of a generic pattern over the entire MSSM parameter space. 
We show the mixing matrix elements in Fig.~\ref{fig:mixO1}. 
In the CP-conserving case ($\phi_\mu=0^\circ$) 
$H_1$ is mostly $\phi_1$ below $M_{H^+}\sim 150$ GeV and mostly
$\phi_2$ above it.  $H_2$ is the pseudo-scalar ($a$) below $M_{H^+}\sim 150$ GeV
while above $M_{H^+}\sim 150$ GeV it is mostly $\phi_1$. $H_3$ on the 
other hand is mostly $\phi_2$ below $M_{H^+}\sim 150$ GeV and is the 
pseudo-scalar above this value. Indeed there is some mixing between $\phi_1$
and $\phi_2$ in the transition region.
For the maximum value of $\phi_\mu=90^\circ$ the lightest ($H_1$) is mostly 
the pseudo-scalar below $M_{H^+}\sim 150$ GeV, $H_2$ is mostly
$\phi_1$ and $H_3$ is mostly $\phi_2$.  Above this region $H_1$ is mostly 
$\phi_2$, $H_2$ is mostly $a$ and $H_3$ is mostly $\phi_1$.  There is of course some mixing 
(albeit small) between all the three states (especially in the transition region).
For values of $\phi_\mu$ in between $0^\circ$ and $90^\circ$ mixing could 
be large, as demonstrated by the case of $\phi_\mu=40^\circ$ 
in Fig.~\ref{fig:mixO1}.  
Similar features also happen
between $\phi_\mu=90^\circ$ and $180^\circ$ (the other CP-conserving value).
Concentrating on $H_1$, the lightest eigenstate, we have plotted the 
relevant mixing matrix elements in Fig.~\ref{fig:mixO2} for two cases
with and without the presence of a light $\tilde t_1$, which shows that 
indeed the mixing is affected by the presence of a light stop.

\begin{figure}
\includegraphics[width=5cm]{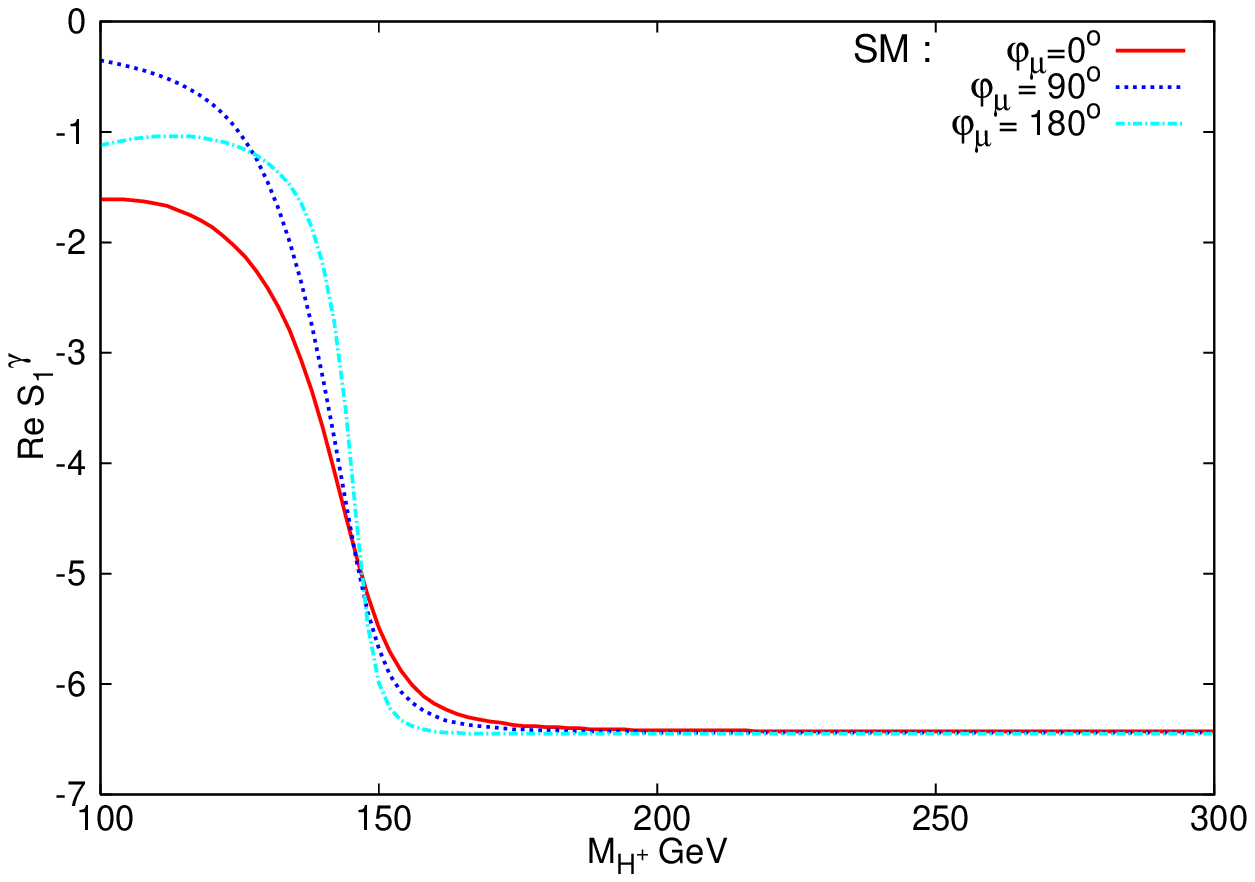}
\includegraphics[width=5cm]{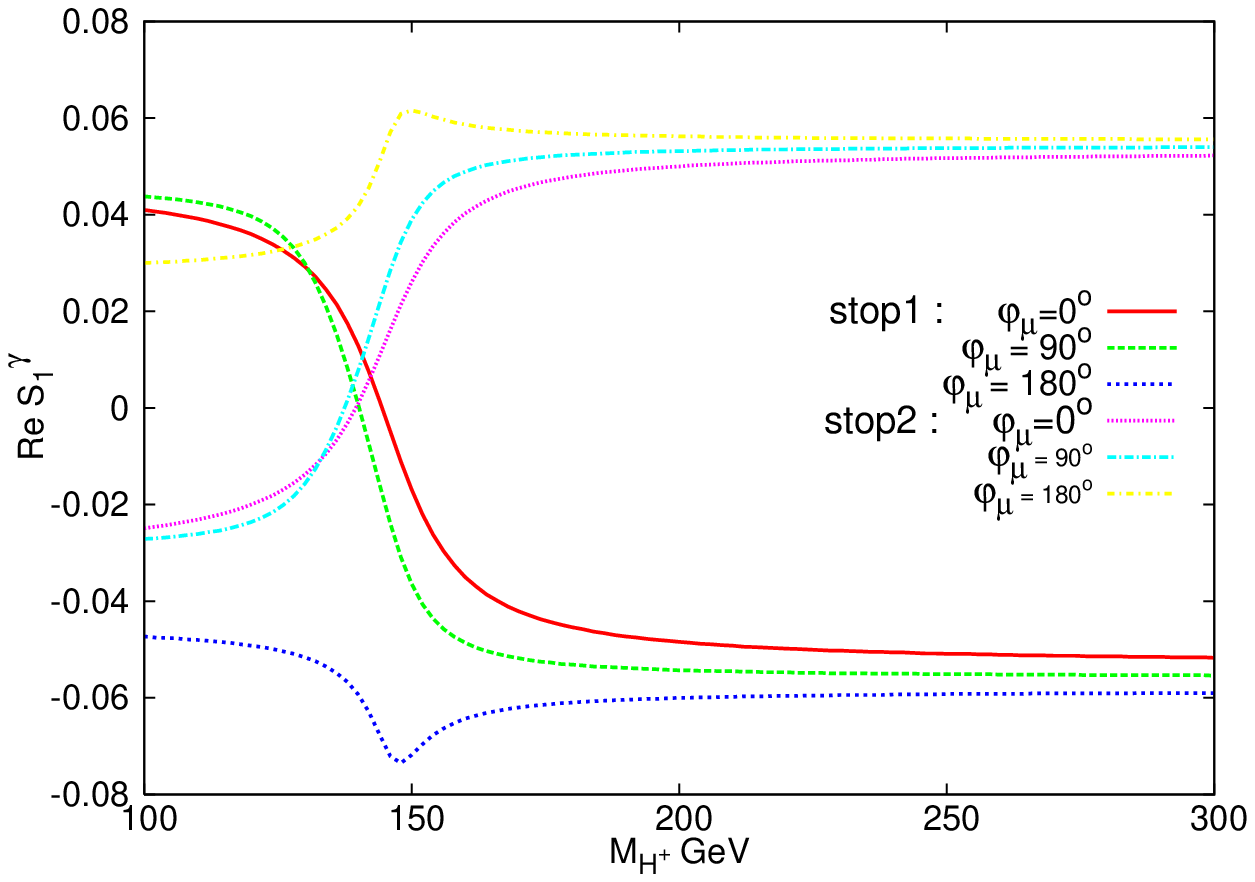}
\includegraphics[width=5cm]{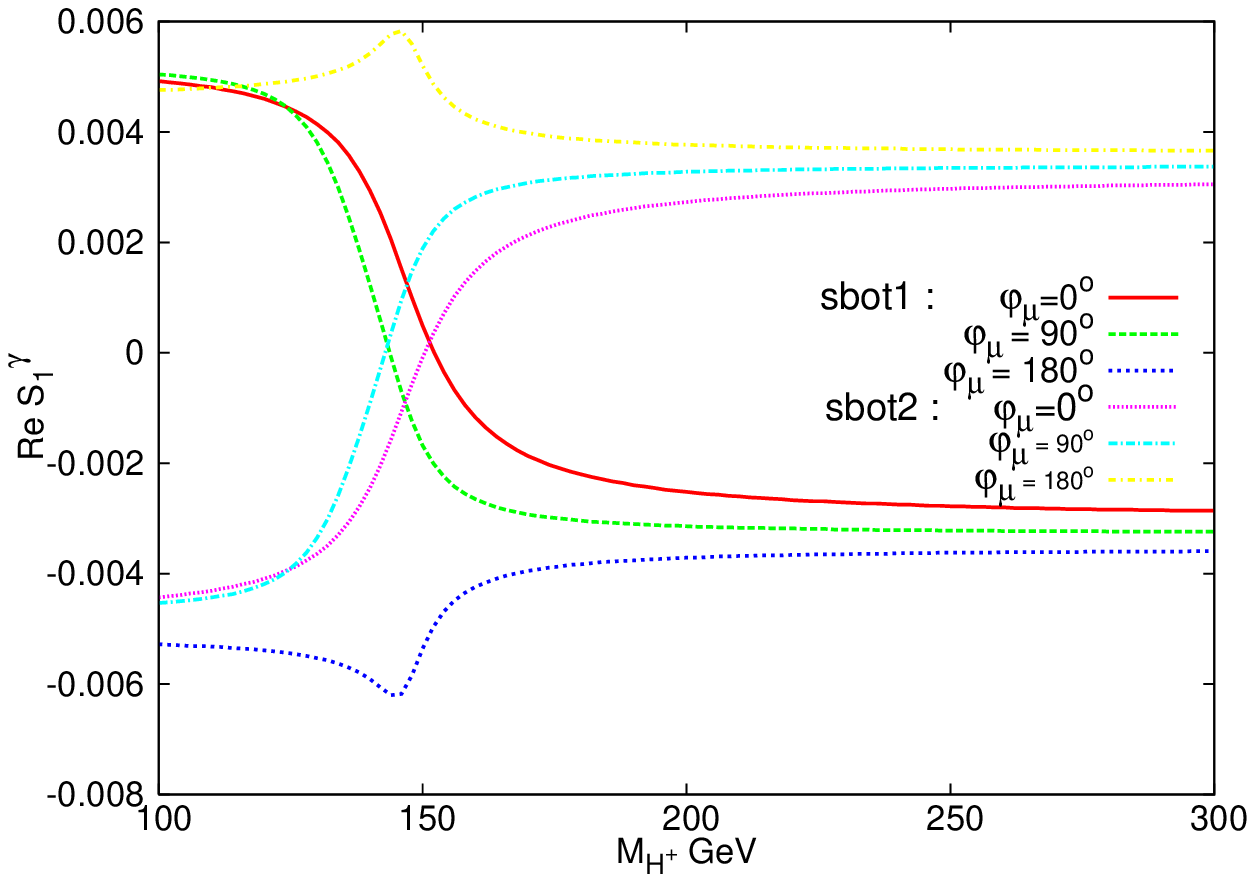}

\includegraphics[width=5cm]{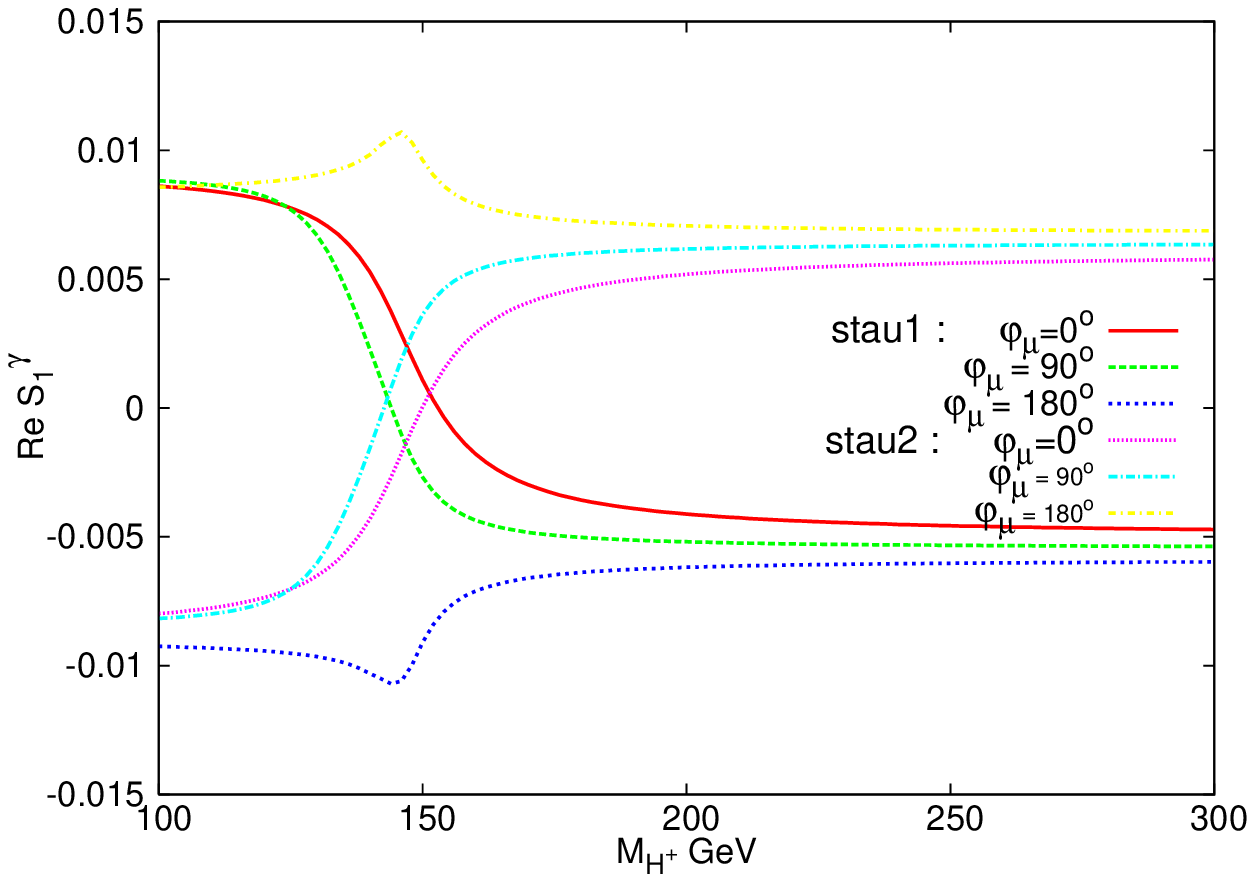}
\includegraphics[width=5cm]{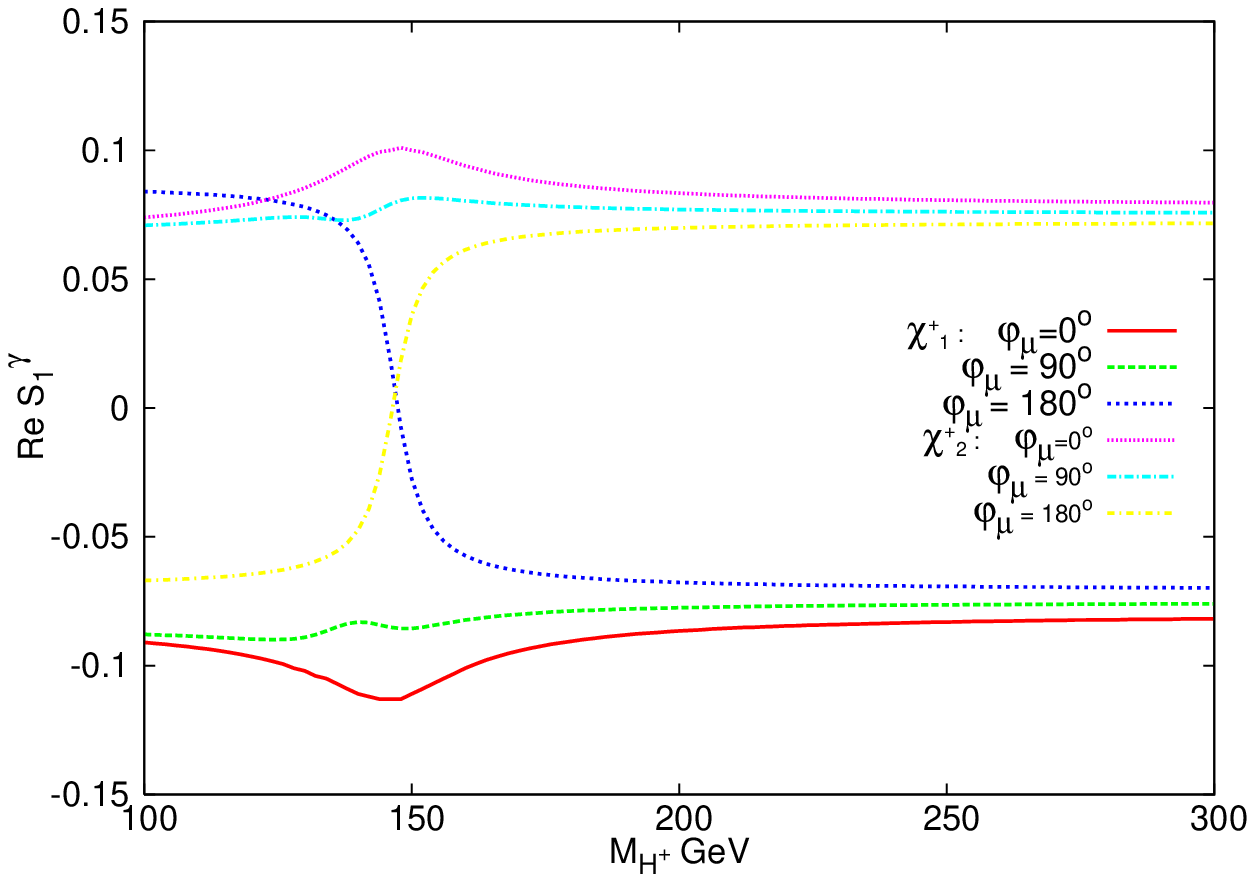}
\includegraphics[width=5cm]{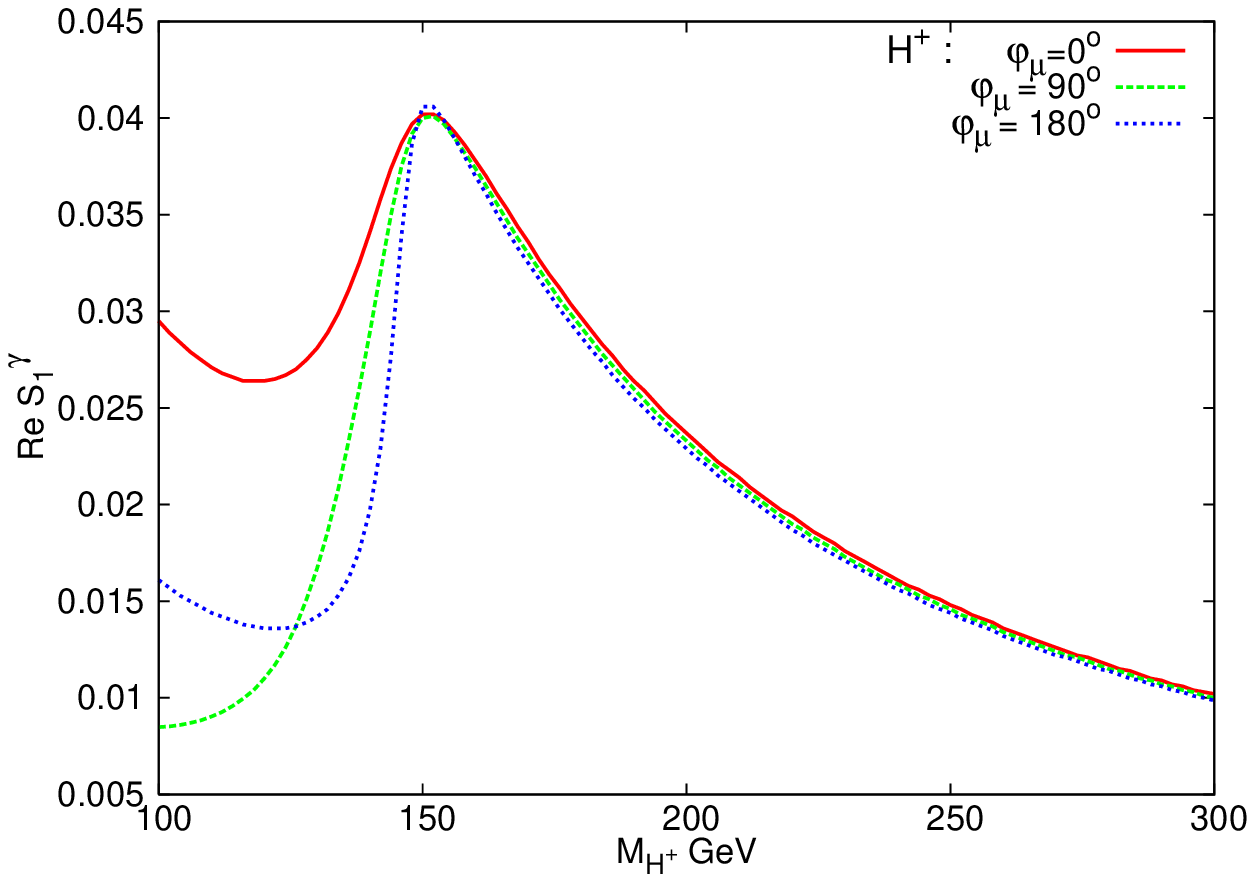}
\caption{\it 
Different contributions to Re $(S^{\gamma}_1)$ against the input parameter 
$M_{H^+}$ with SUSY parameters as in Fig. \ref{fig:mixO1} and 
$M_{\tilde U_3}=1$ TeV.
Top row: (from left) SM, $\tilde t_{1,2}$ and $\tilde b_{1,2}$.
Bottom row: (from left) 
$\tilde \tau_{1,2}$, $\tilde \chi^+_{1,2}$ and $H^+$.
}
\label{fig:SPR1}
\end{figure}

\begin{figure}
\includegraphics[width=5cm]{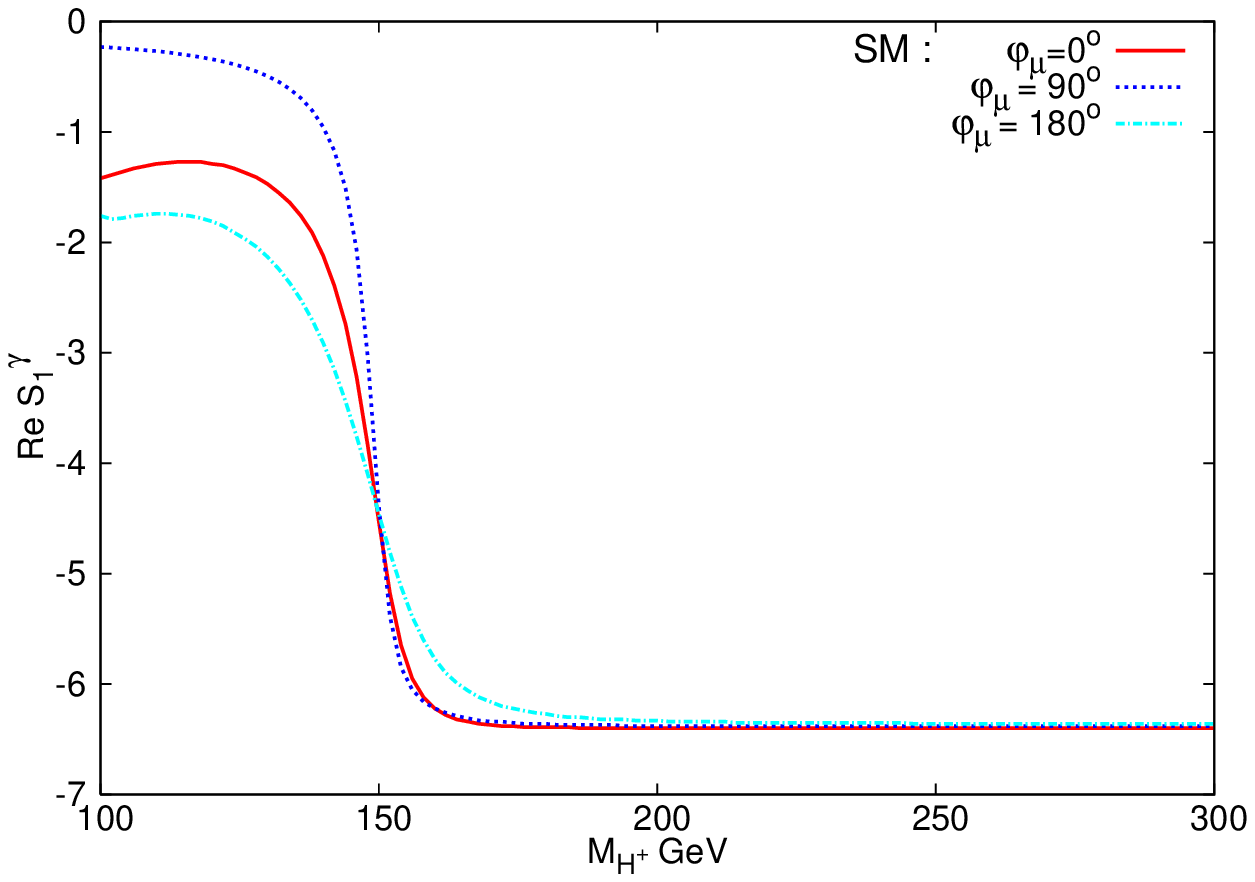}
\includegraphics[width=5cm]{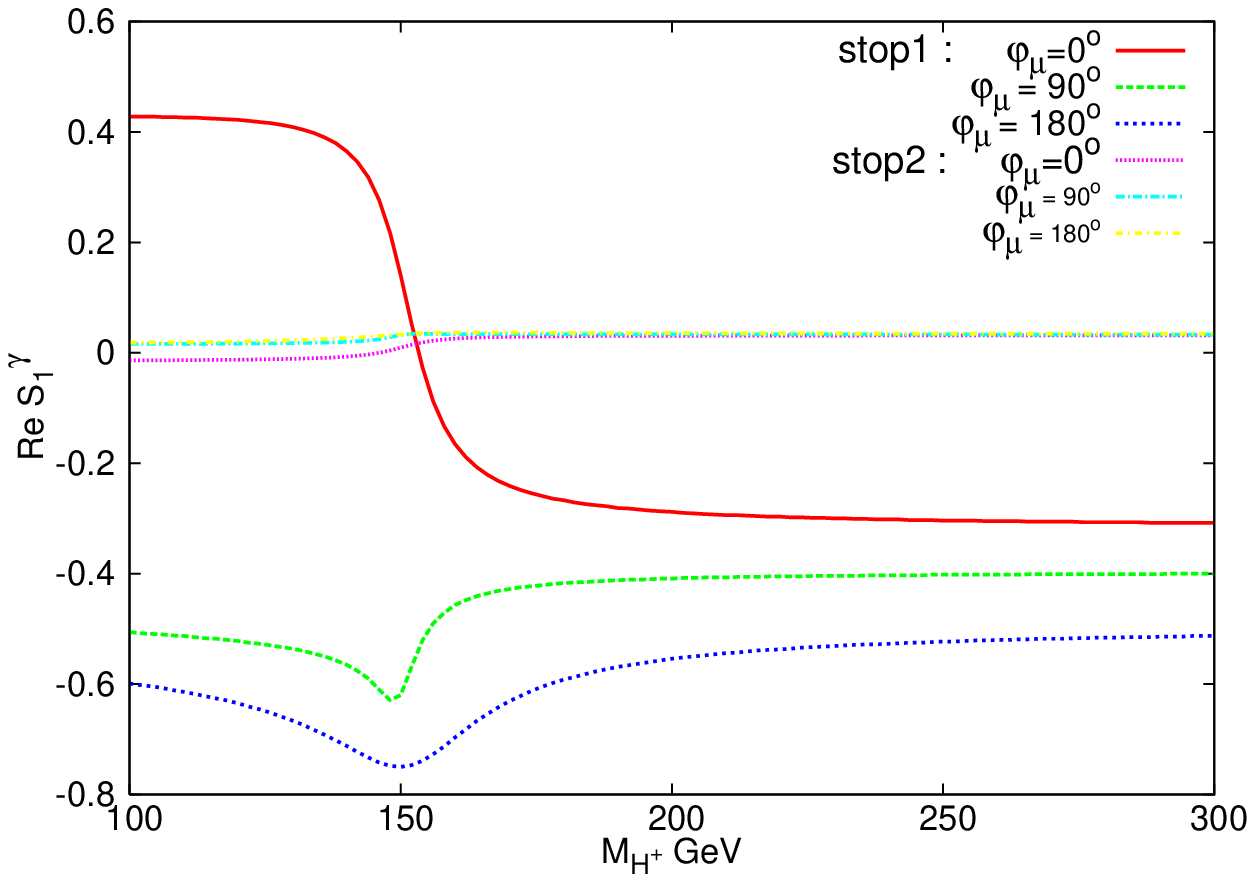}
\includegraphics[width=5cm]{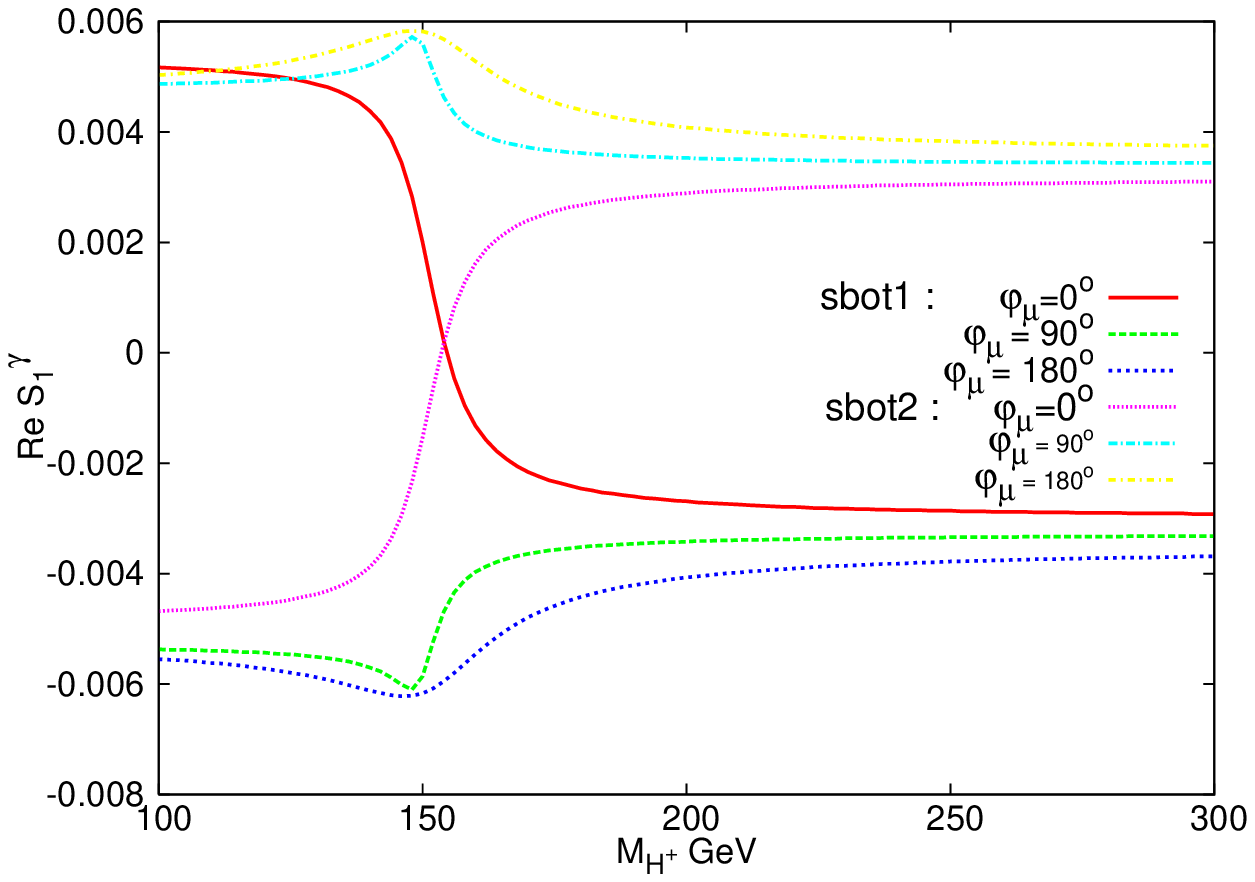}

\includegraphics[width=5cm]{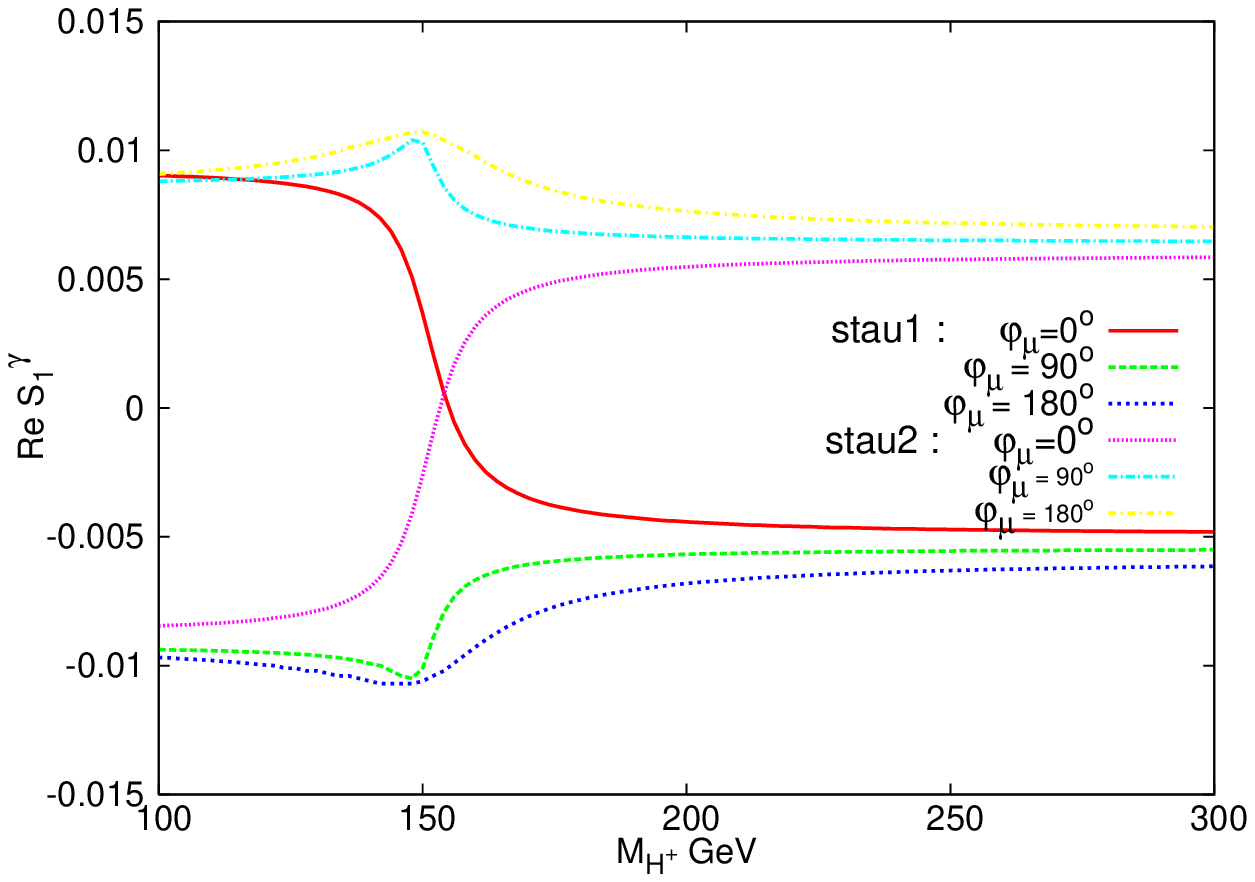}
\includegraphics[width=5cm]{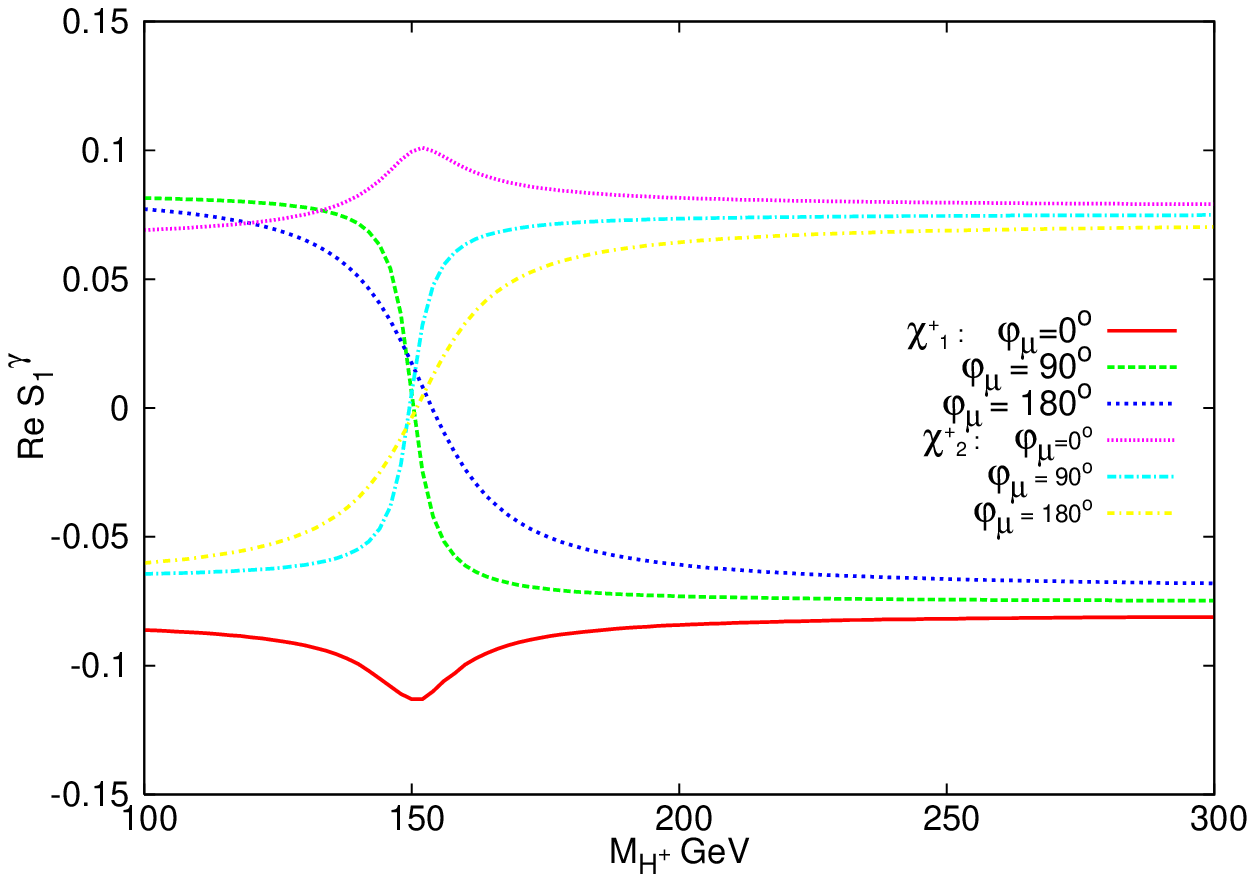}
\includegraphics[width=5cm]{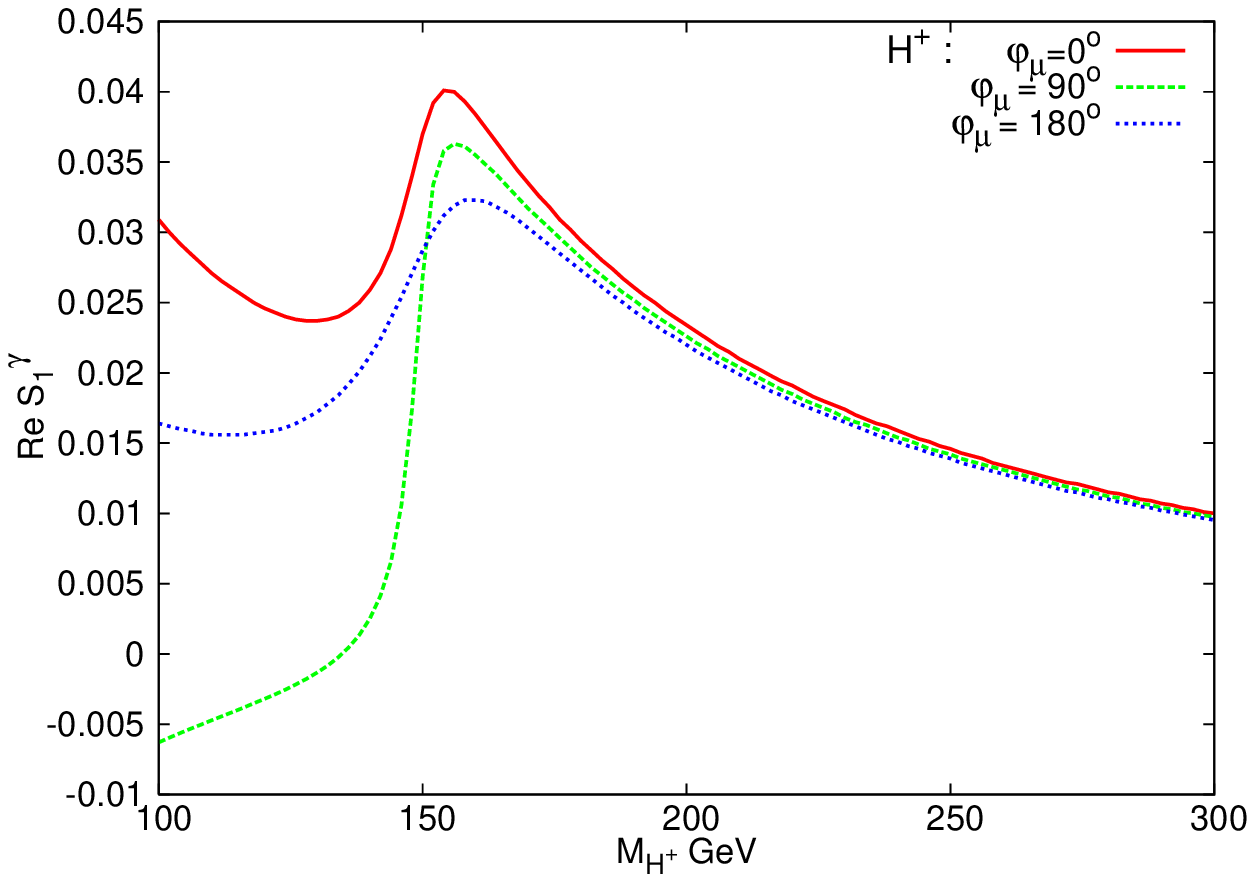}
\caption{\it 
Different contributions to Re $(S^{\gamma}_1)$ against the input parameter 
$M_{H^+}$ with SUSY parameters as in Fig. \ref{fig:mixO1} and 
$M_{\tilde U_3}=250$ GeV.
Top row: (from left) SM, $\tilde t_{1,2}$ and $\tilde b_{1,2}$.
Bottom row: (from left) 
$\tilde \tau_{1,2}$, $\tilde \chi^+_{1,2}$ and $H^+$.
}
\label{fig:SPR2}
\end{figure}

\begin{figure}
\includegraphics[width=5cm]{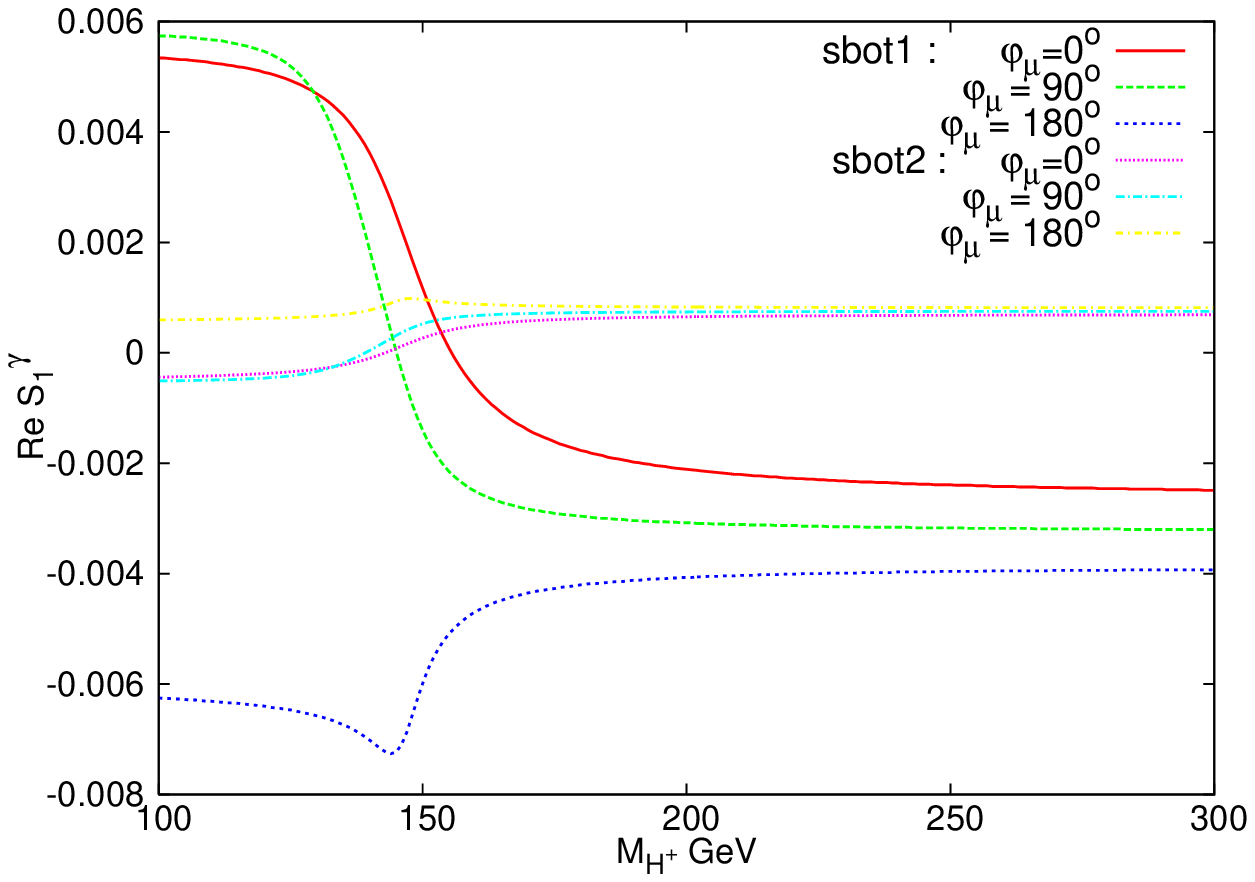}
\includegraphics[width=5cm]{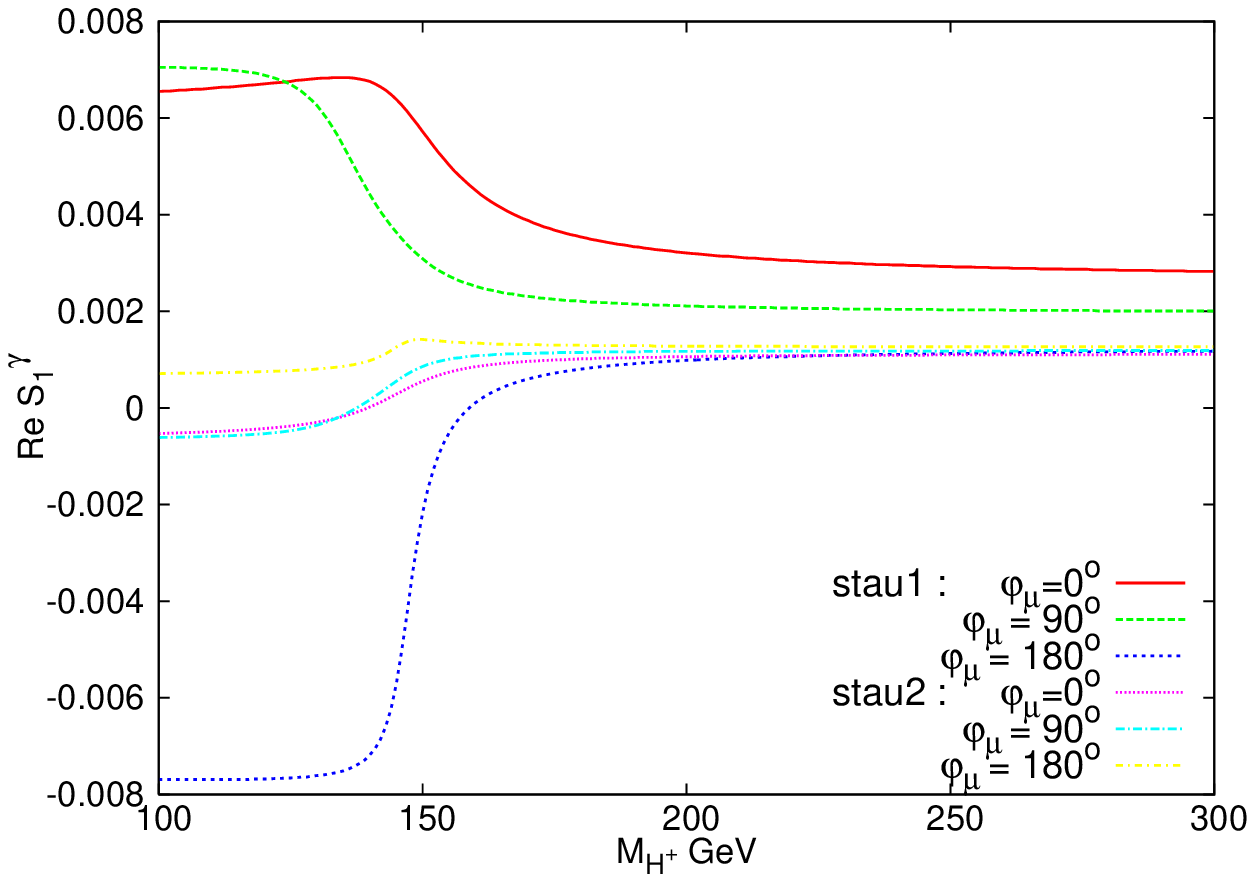}
\includegraphics[width=5cm]{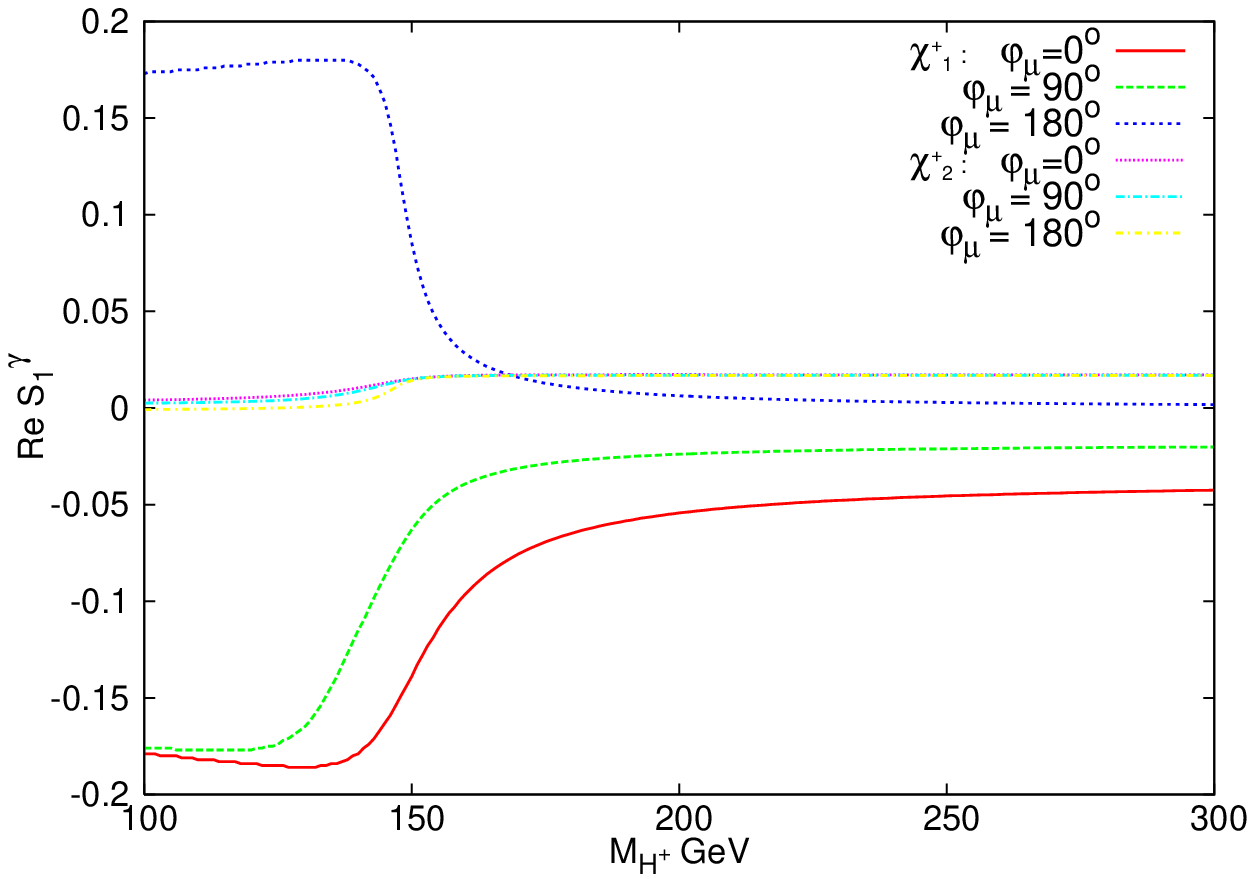}
\caption{\it 
Contributions to Re $(S^{\gamma}_1)$ when the respective sparticle is light.
Left plot is with $M_{\tilde D_3}=310$ GeV 
(or $m_{\tilde b_1}\sim 300$ GeV),
middle one is with $M_{\tilde E_3}=310$ GeV (or $M_{\tilde \tau_1}\sim 300$ GeV)
and the right one is with $M_2=100$ GeV (or $M_{\chi^\pm_1}\sim 100$ GeV). 
All other sparticles in the loop are of the order of TeV.
The SM contribution in each case remains more or less the same as in 
Fig.~\ref{fig:SPR1} (no light sparticle).
}
\label{fig:SPR3}
\end{figure}

\begin{figure}
\begin{center}
\includegraphics[width=7cm]{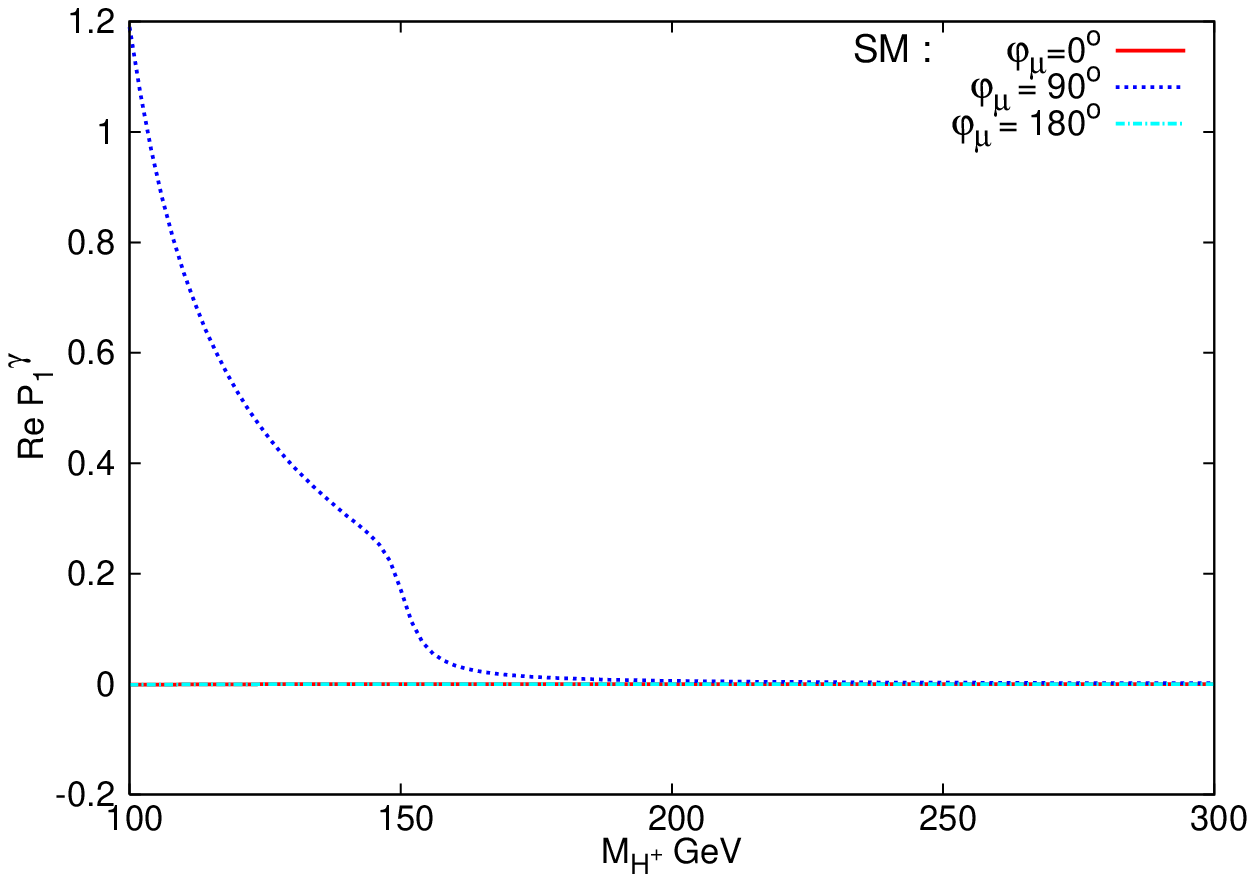}
\includegraphics[width=7cm]{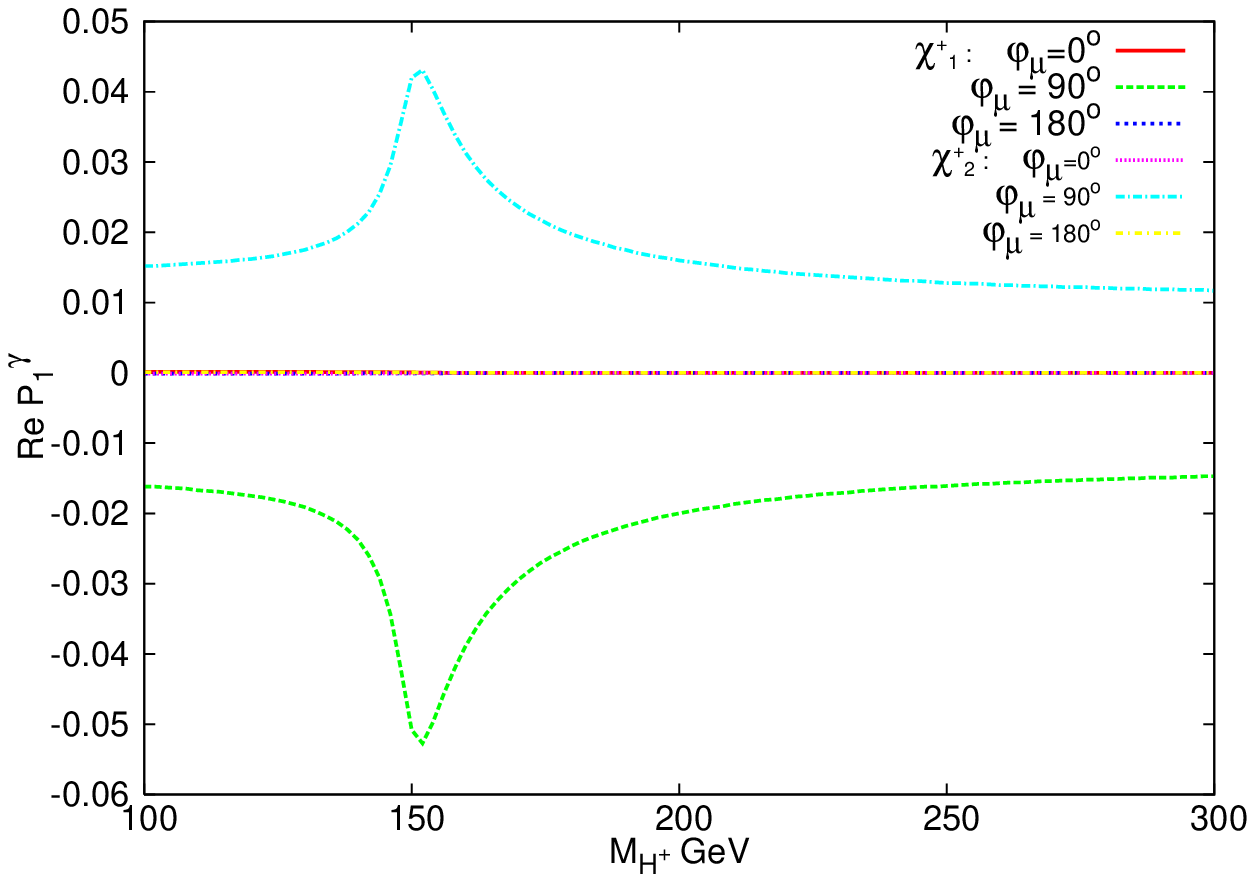}

\includegraphics[width=7cm]{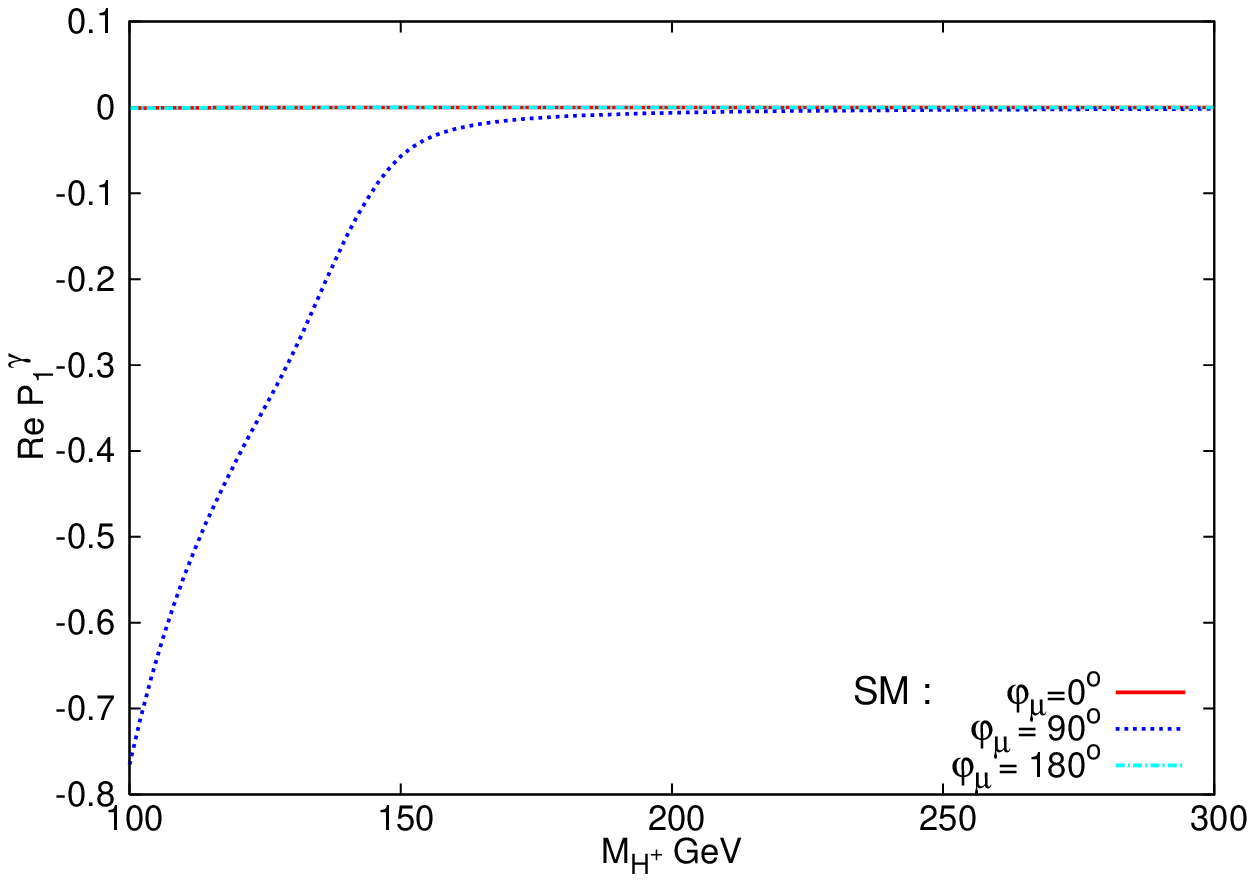}
\includegraphics[width=7cm]{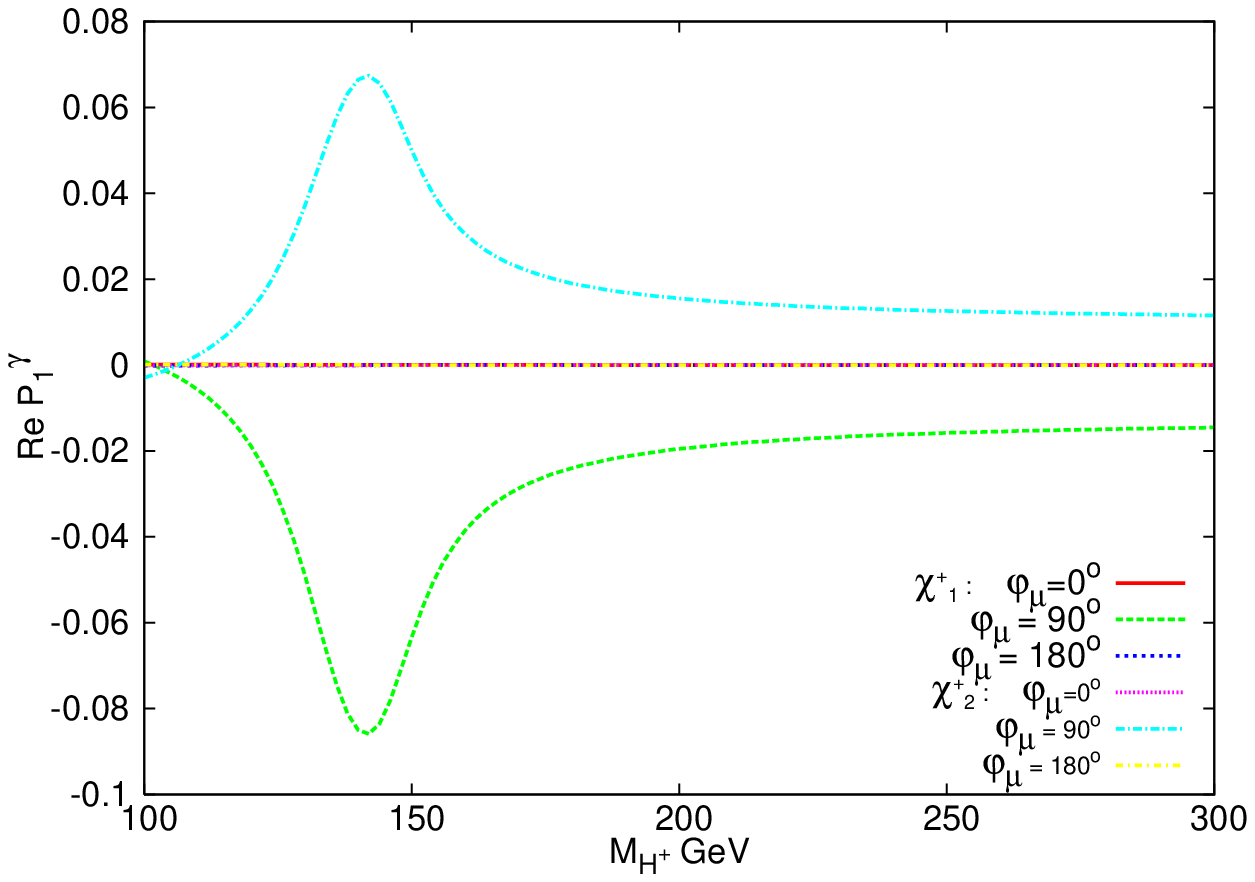}
\caption{\it 
SM and chargino contributions (as indicated) to Re $(P^{\gamma}_1)$. 
Top row corresponds to the case with $M_{\tilde U_3}=250$ GeV and 
bottom row with $M_{\tilde U_3}=1$ TeV.
The contributions in CP-conserving parameter points
($\phi_\mu = 0^\circ$ and $\phi_\mu = 180^\circ$) are all consistent
with zero.
}
\label{fig:PPR}
\end{center}
\end{figure}

\begin{figure}
\begin{center}
\includegraphics[width=7cm]{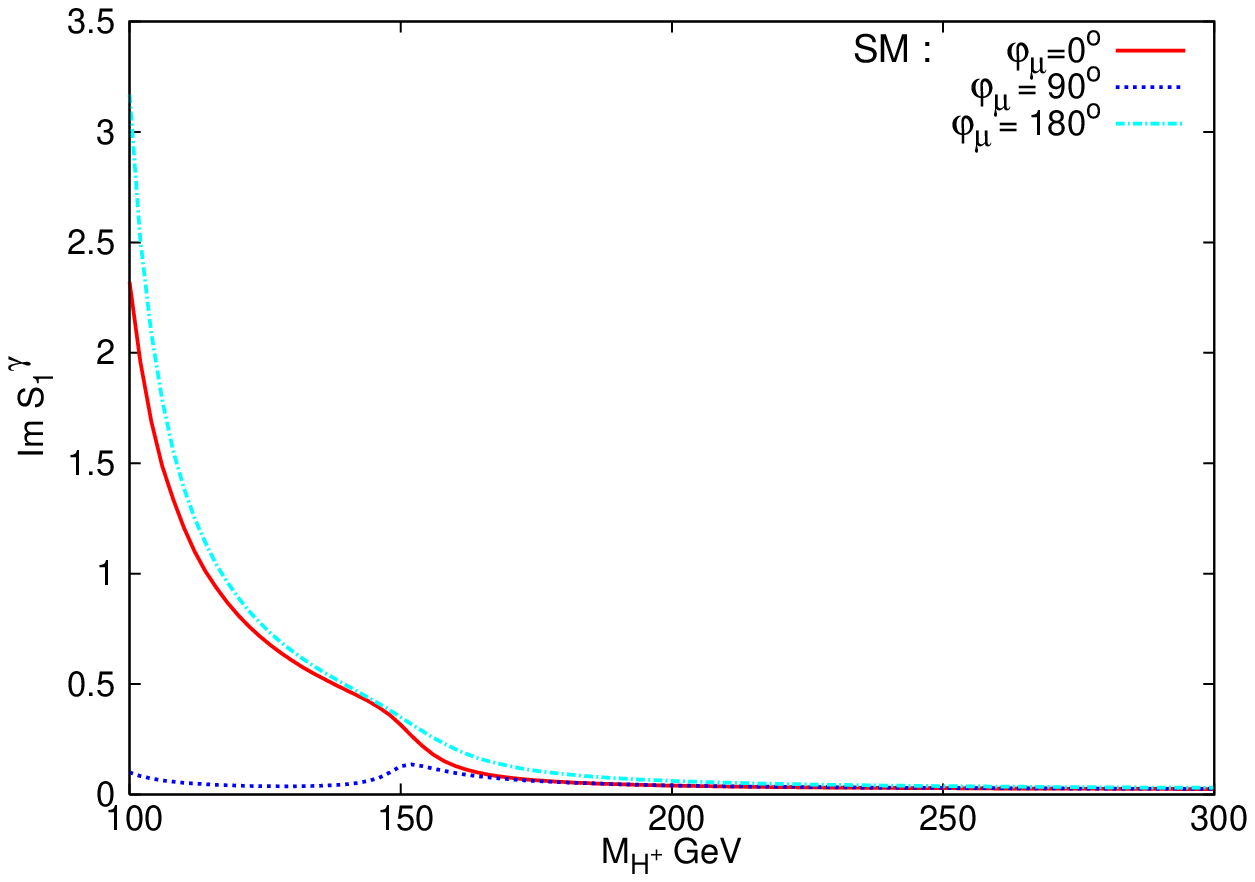}
\includegraphics[width=7cm]{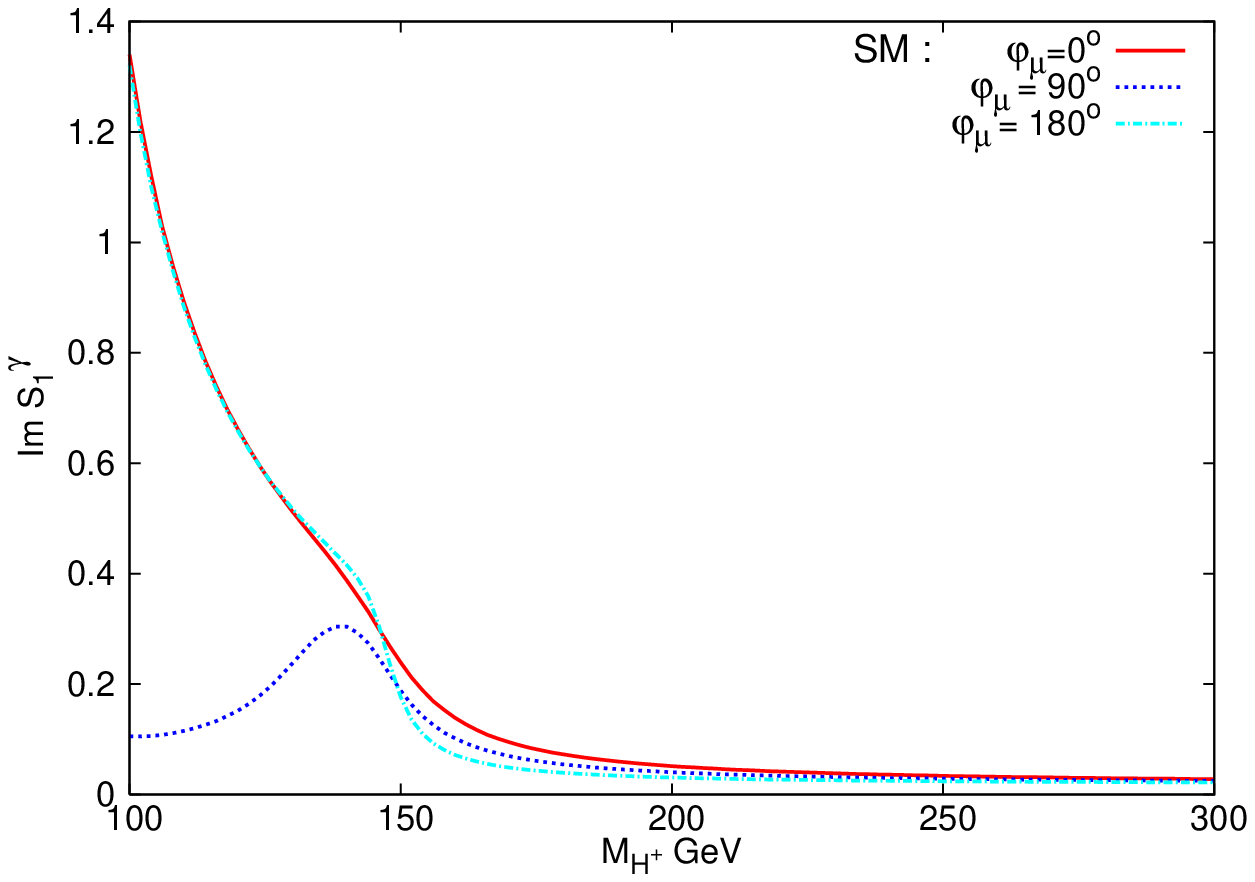}

\includegraphics[width=7cm]{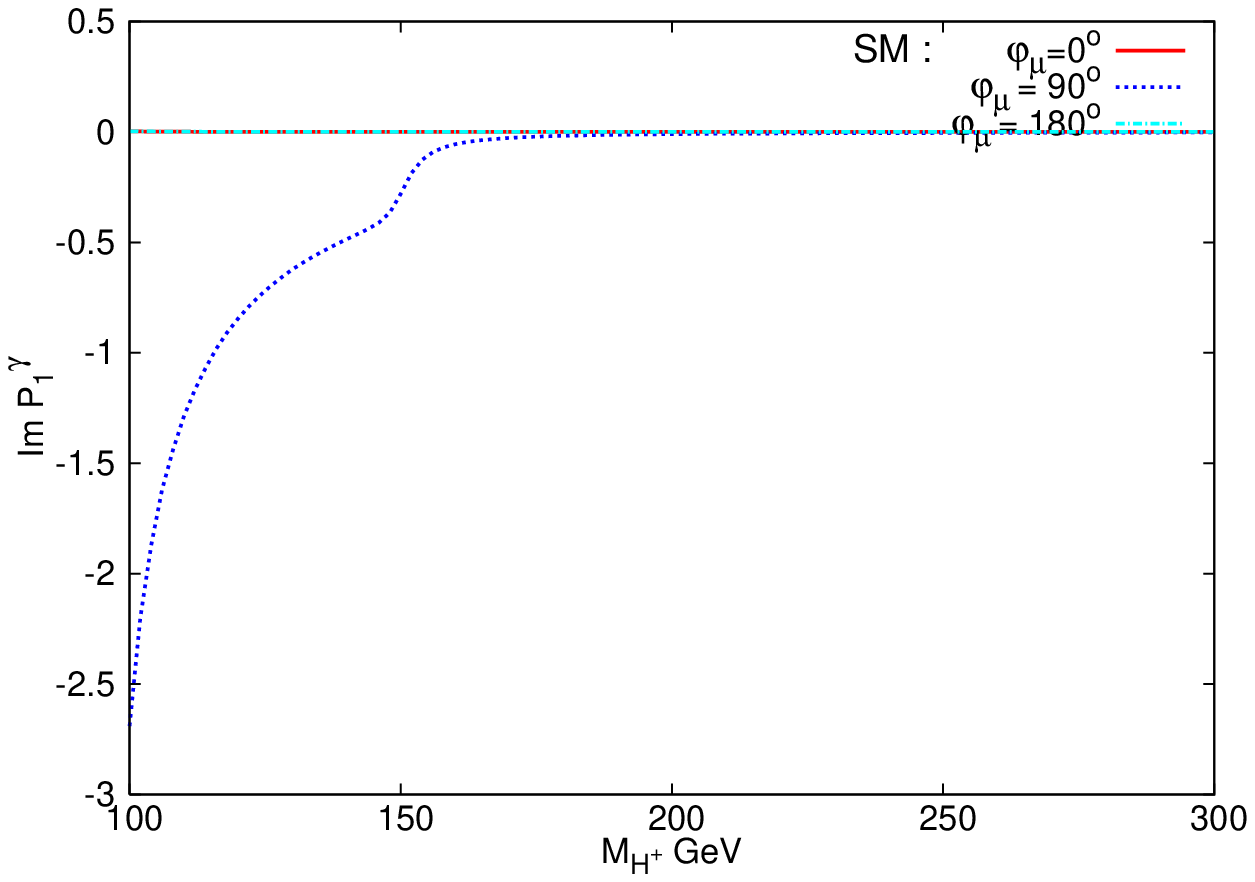}
\includegraphics[width=7cm]{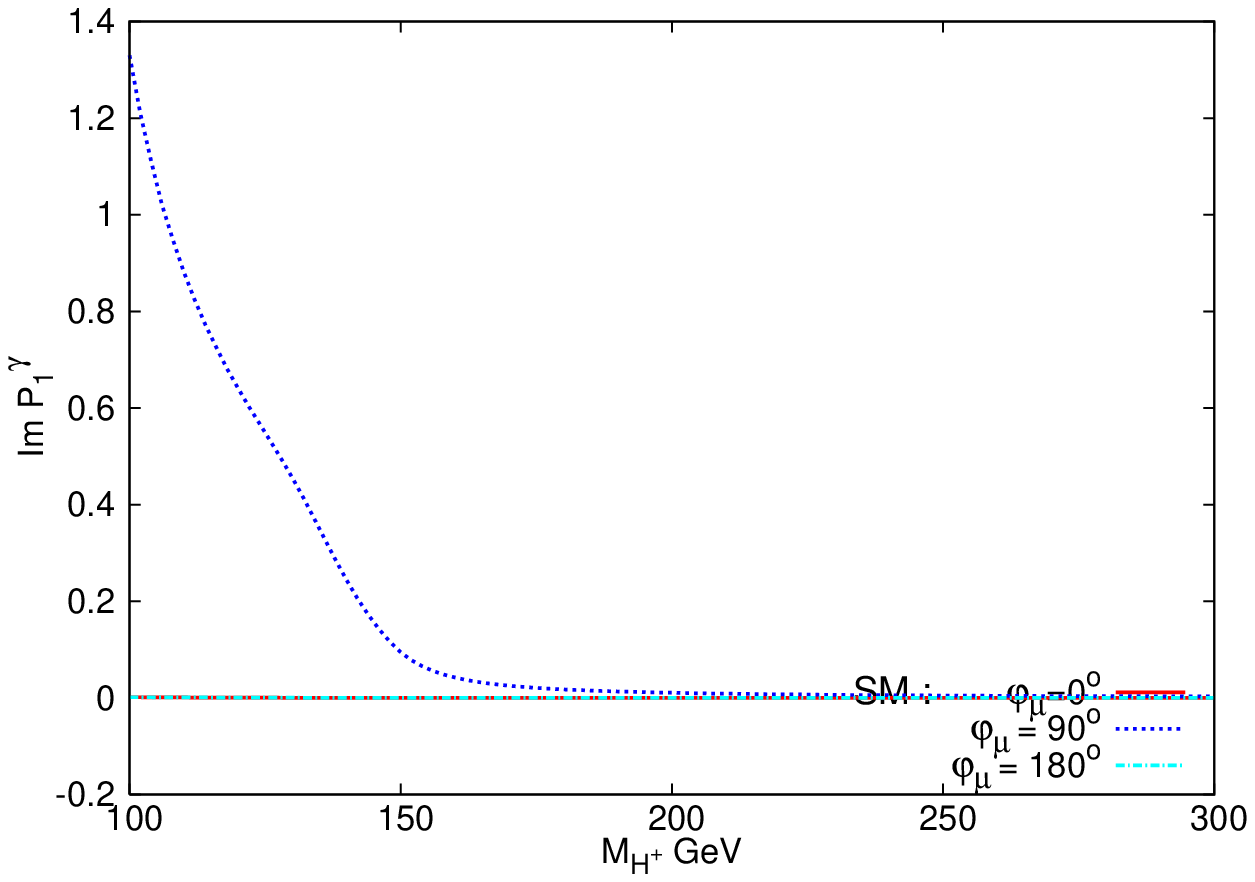}

\caption{\it 
SM contribution to Im $(S^{\gamma}_1)$ (first row) and Im $(P^\gamma_1)$ (second 
row) against the input parameter 
$M_{H^+}$, for the case with $M_{\tilde U_3}=250$ GeV (left column) and  with 
$M_{\tilde U_3}=1$ TeV (right column).
Contributions from superparticles are zero.
}
\label{fig:Im}
\end{center}
\end{figure}

\vspace*{0.15cm}
These CP-mixing effects feed into the decay amplitude of Eq.~(\ref{eq:Melement})
through couplings of the $H_i$'s to the SM and SUSY particles in the loop 
(see Fig.~\ref{fig:HiggsPho}) at one-loop and tree-level, respectively. In Figs.~\ref{fig:SPR1}--\ref{fig:Im} 
we show different contributions to the amplitude 
of  $H_1\rightarrow \gamma\gamma$. Clearly, the SM contribution is dominant 
in all cases.  Among the major contributions within the SM, 
that from the $W^\pm$ 
loop is about 5 times larger than the top quark contribution
for the whole range of $M_{H^+}$ (for the chosen set of SUSY parameters),
while the bottom quark and tau lepton contributions are about an order of 
magnitude smaller
over the lower range of $M_{H^+}$ (except around 100 GeV) and negligibly small for 
larger values.  Magnitudes of both the $W^\pm$ and top quark contributions grow 
with $M_{H^+}$, which is a reflection of the fact that these couplings are 
proportional to the mixing matrix element $O_{21}$.
When all SUSY states are heavy (Fig.~\ref{fig:SPR1}),  
all sparticle (and $H^+$) contributions to the real part of $S^\gamma_1$
are two or three orders of magnitude 
smaller than the SM term. The exception is the chargino contribution,
which is at the most 10\%. 
The effect of  a light $\tilde t_1$ (Fig.~\ref{fig:SPR2})
enters in two different ways. Firstly, it 
affects the SM couplings to $H_1$ through loop corrections
(compare top left plots in Figs.~\ref{fig:SPR1} and \ref{fig:SPR2}).
This is more 
prominent when $\phi_\mu = 180^\circ$. 
Effects due to change in $\phi_\mu$ are different from the case
with no light sparticle.  Secondly, the $\tilde t_1$ contribution
(top middle plot in Fig.~\ref{fig:SPR2})
is now comparable (about 40\% in the CP-conserving case) to that of the SM.
Contributions of other sparticles are not changed much going from larger
to smaller masses of the respective sparticle as shown in  
Fig.~\ref{fig:SPR3}.
In Fig. \ref{fig:PPR} contributions to the real part of $P^\gamma_1$ are 
plotted against $M_{H^+}$. These terms come through the pseudo-scalar component
in $H_1$. Since the squarks, sleptons and the charged Higgs boson do not 
couple to $a$, only SM objects and charginos contribute to $P^\gamma_1$.  
For the CP-conserving case it vanishes as expected. When $\phi_\mu$ is 
non-zero $H_1$ has an $a$ component and there is a non-vanishing $P^\gamma_1$, 
as illustrated by the curve corresponding to $\phi_\mu=90^\circ$ in 
Fig. \ref{fig:PPR}.
Figure \ref{fig:Im} shows the imaginary parts of $S^\gamma_1$ and $P^\gamma_1$ 
which are sensitive to the
value of $\phi_\mu$. For the $H_1$ mass range considered here
only the SM contribution is complex. 
Again, $P^\gamma_1$ 
being the contribution from $a$ coupling to the (s)particles, its 
imaginary part vanishes in the CP-conserving cases. The $H_1\rightarrow
\gamma\gamma$ width is not sensitive to the sign of Im $P^\gamma_1$, 
therefore the difference between the cases of light $\tilde t_1$ and
no light sparticle is not very dramatic.

\vspace*{0.15cm}
From our MSSM parameter space scans, the following 
generic features on the sensitivity of width and BR to the 
CP-violating
phase, $\phi_\mu$, have emerged. The effect is most pronounced around
the crossing region ($M_{H^+}\sim$ 150 GeV). This is expected since
the scalar/pseudo-scalar mixing in $H_1$ is largest here. For much higher
values of $M_{H^+}$, $H_1$ is purely a scalar. There can still be a CP-violating 
effect through its sfermion couplings though. The latter will 
be more visible 
when there is a light-sparticle in the loop
 (a stop, in particular). For $M_{H^+}< 150$ GeV there
is still sufficient mixing to have substantial difference in the BR.  
But in this region $M_{H_1}$ is also changed by about 10 to 15\% between
$\phi_\mu=0^\circ$ and $\phi_\mu=90^\circ$. We will however concentrate on the region
$M_{H^+}> 150$ GeV, where  $M_{H_1}>115$ GeV for the parameter sets 
considered. Moreover, the effect of $\phi_\mu \ne 0$ on $M_{H_1}$ in this 
region is within 1 GeV, which is less than the experimental uncertainty 
expected at the LHC. In particular, we have learnt that the width and BR of the decay
 $H_1\rightarrow
\gamma\gamma$ are very sensitive to the $\tilde t_1$ mass. 
The effect comes through both a modification of the $H_1W^+W^-$ 
couplings and the presence of a light
$\tilde t_1$ in the triangle loops of the decay amplitude.

\begin{figure}
\epsfysize=10cm \epsfxsize=8cm
\epsfbox{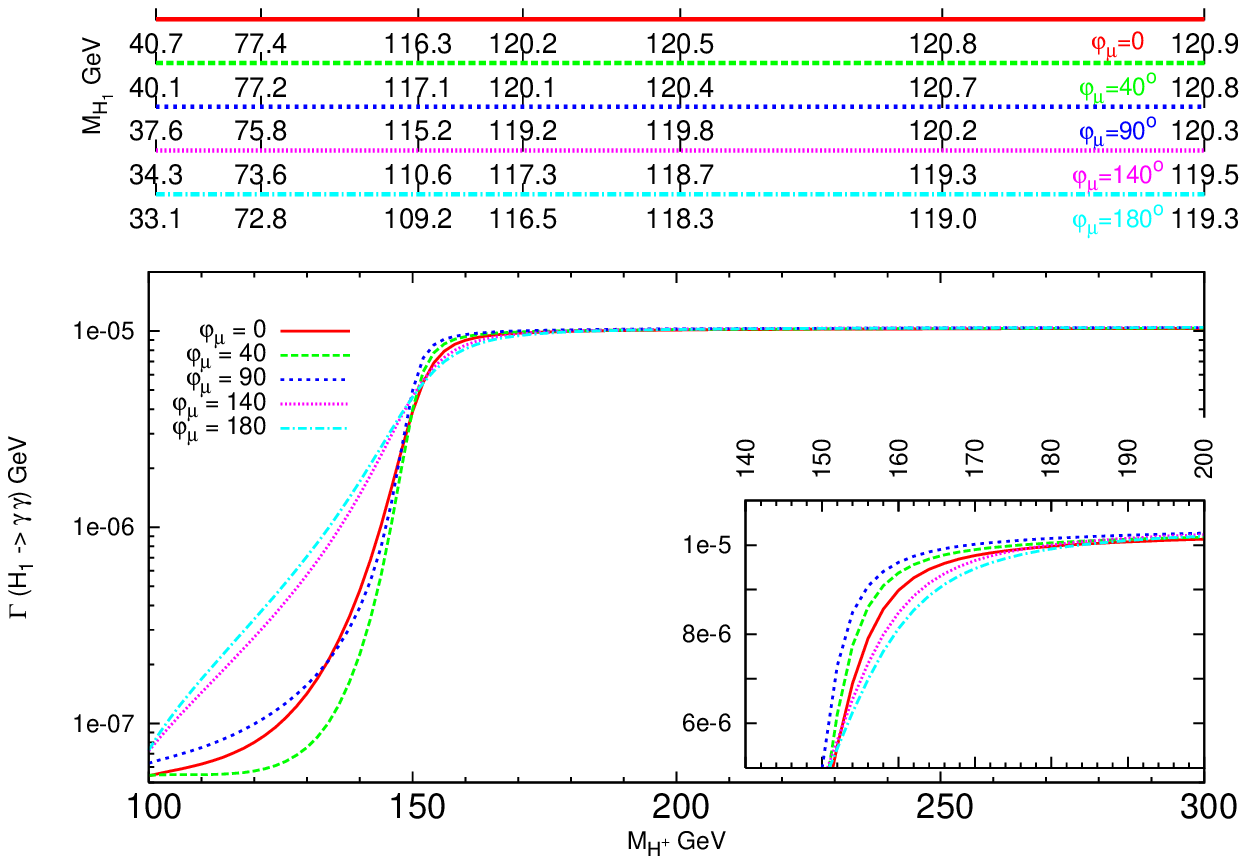}
\epsfysize=10cm \epsfxsize=8cm
\epsfbox{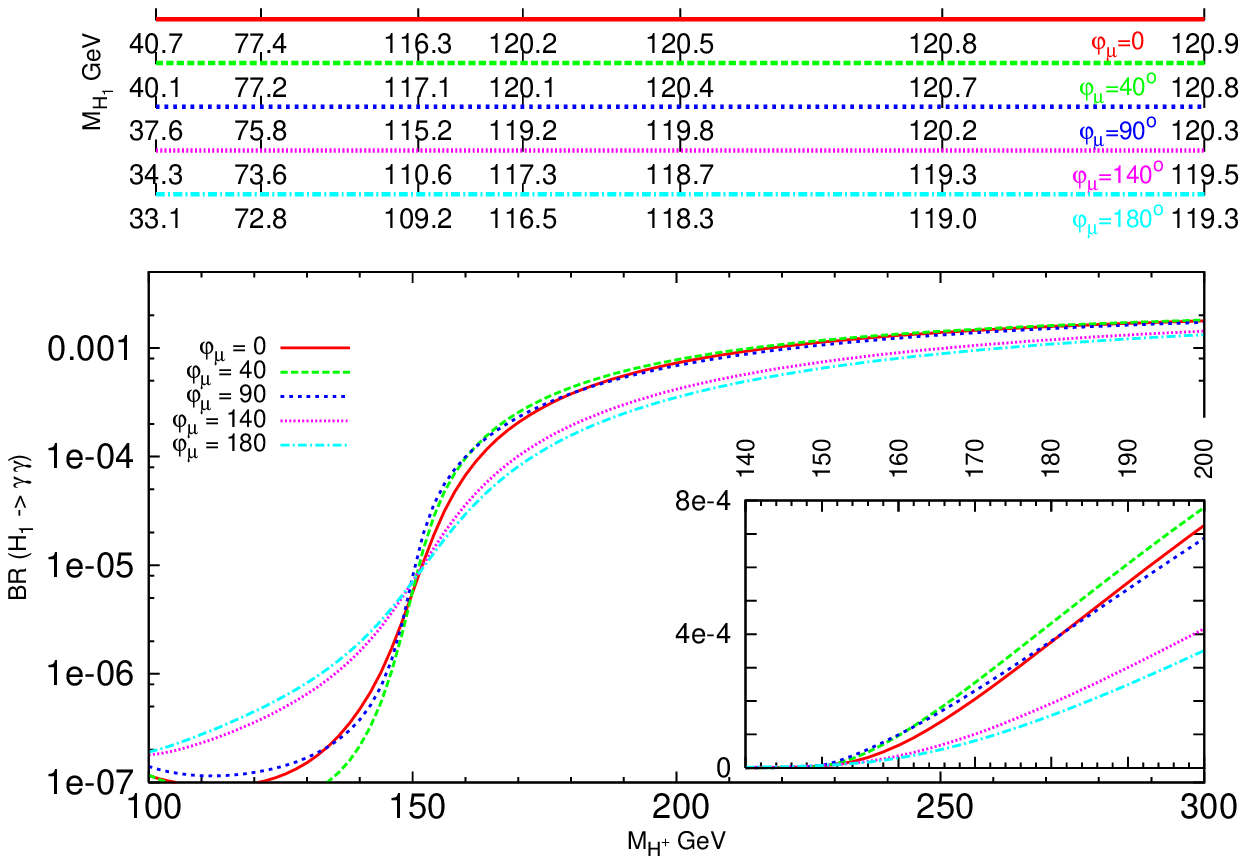}

\epsfysize=10cm \epsfxsize=8cm
\epsfbox{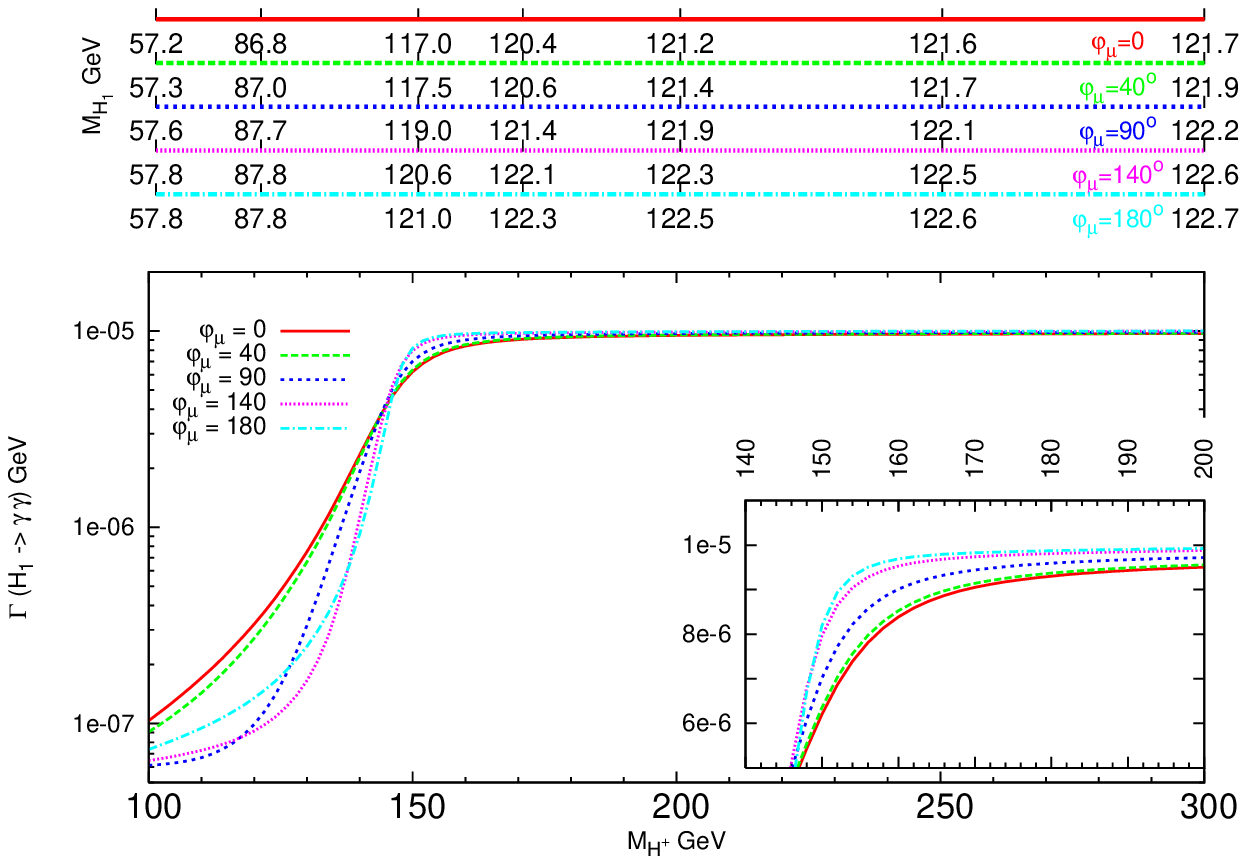}
\epsfysize=10cm \epsfxsize=8cm
\epsfbox{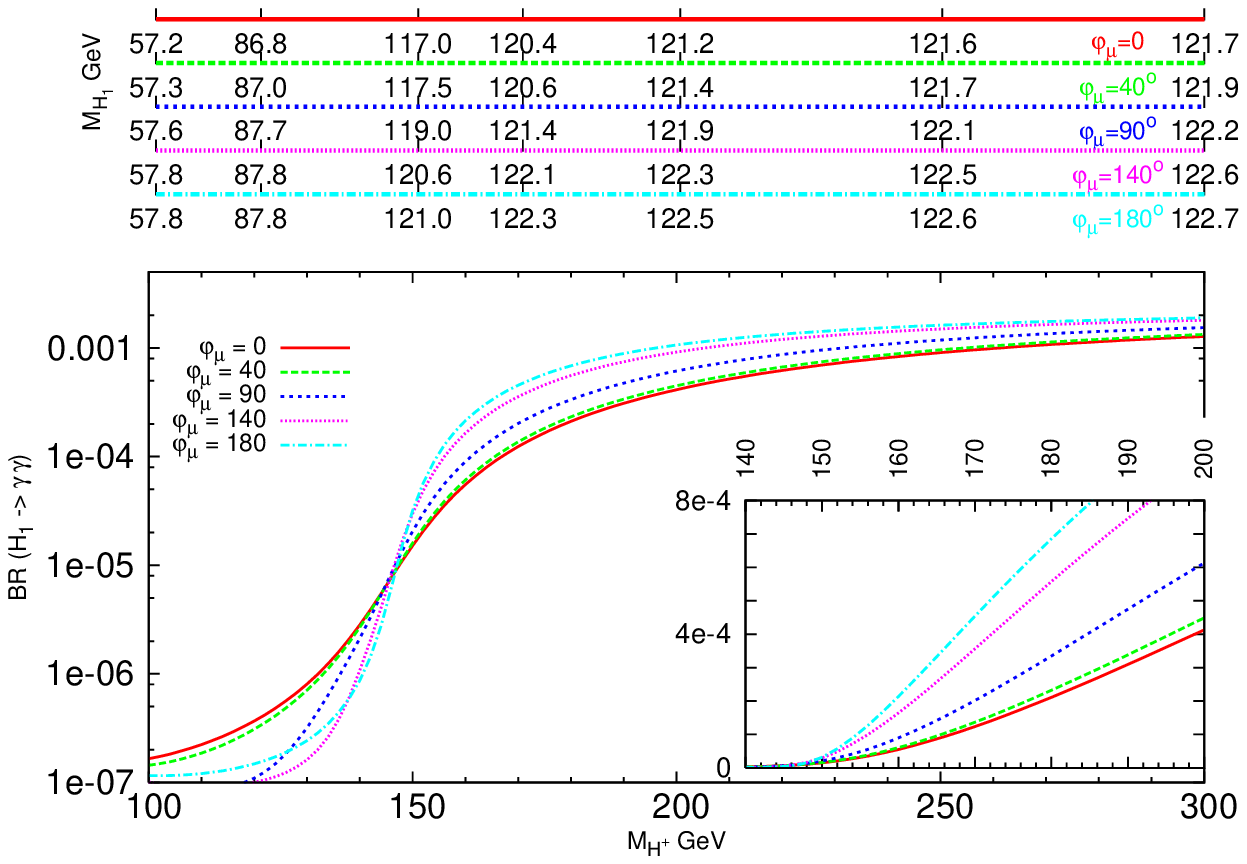}
\caption{\it 
Width (left column) and BR (right column)  of $H_1\rightarrow \gamma\gamma$ 
against the input parameter $M_{H^+}$ for $|A_f|=1.5$ TeV, $|\mu|=1$ TeV
and $\tan\beta=20$. 
Values of $M_{H_1}$ corresponding to representative points on $M_{H^+}$ axis 
are indicated on the horizontal lines above separately for the values of 
$\phi_\mu$ used.
Top row corresponds to the case with $M_{\tilde U_3}=250$ GeV,
while the bottom one corresponds to the case with $M_{\tilde U_3}=1$ TeV.
}
\label{fig:BR_1}
\end{figure}

\begin{figure}
\epsfysize=10cm \epsfxsize=8cm
\epsfbox{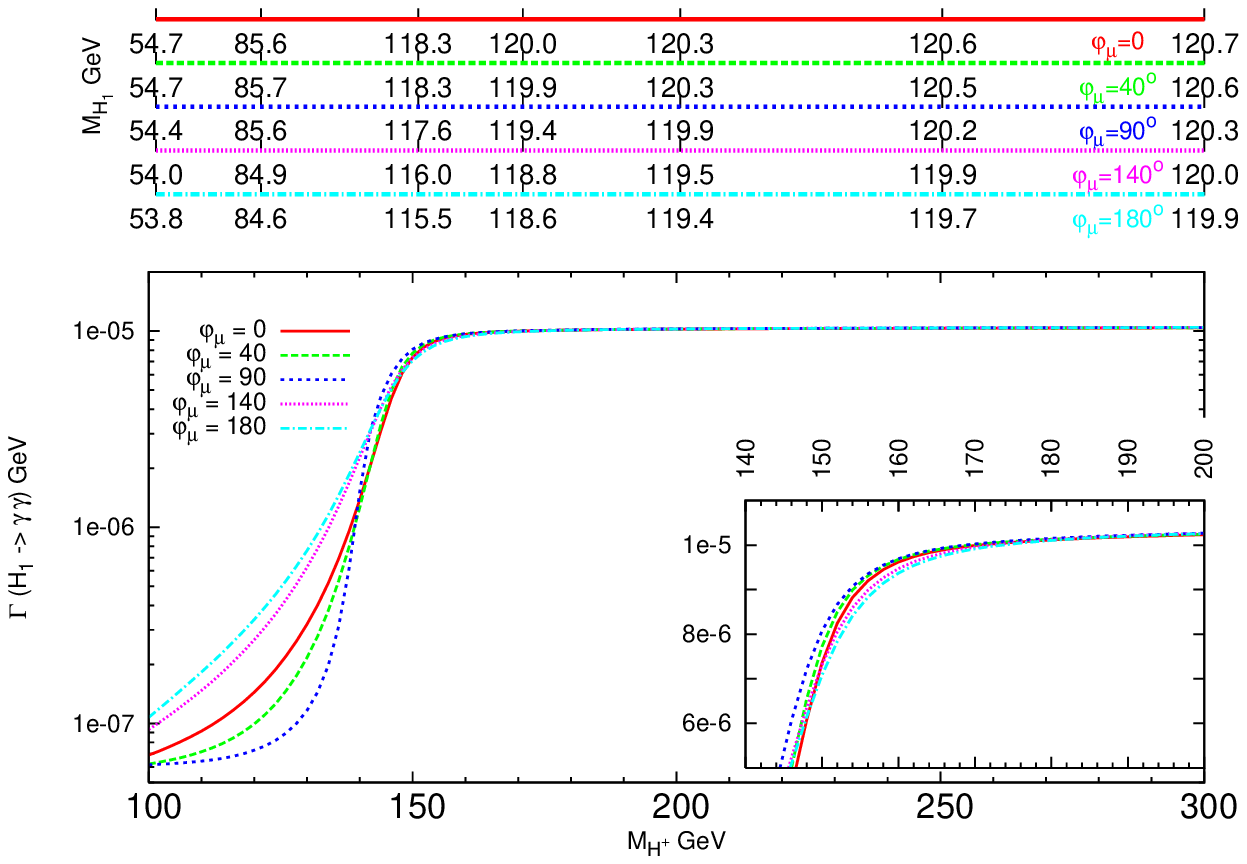}
\epsfysize=10cm \epsfxsize=8cm
\epsfbox{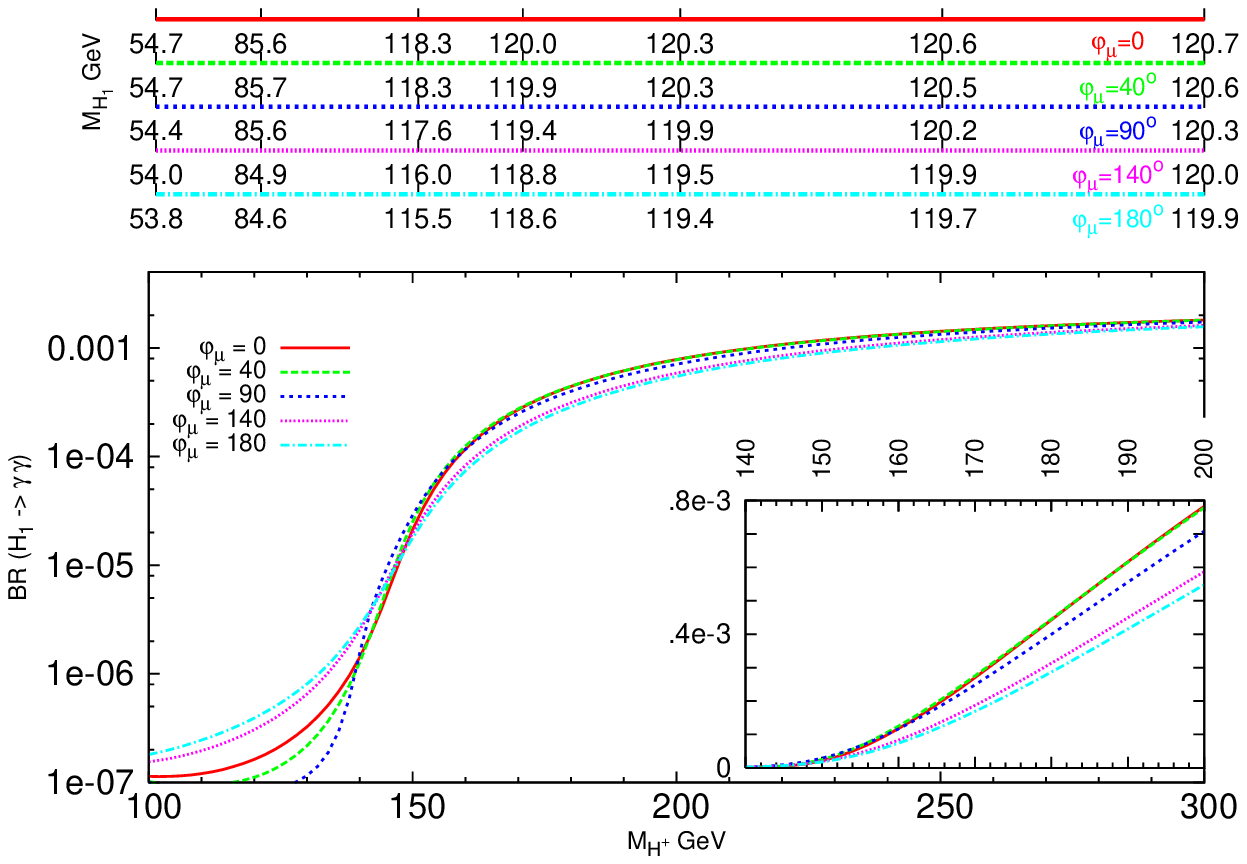}

\epsfysize=10cm \epsfxsize=8cm
\epsfbox{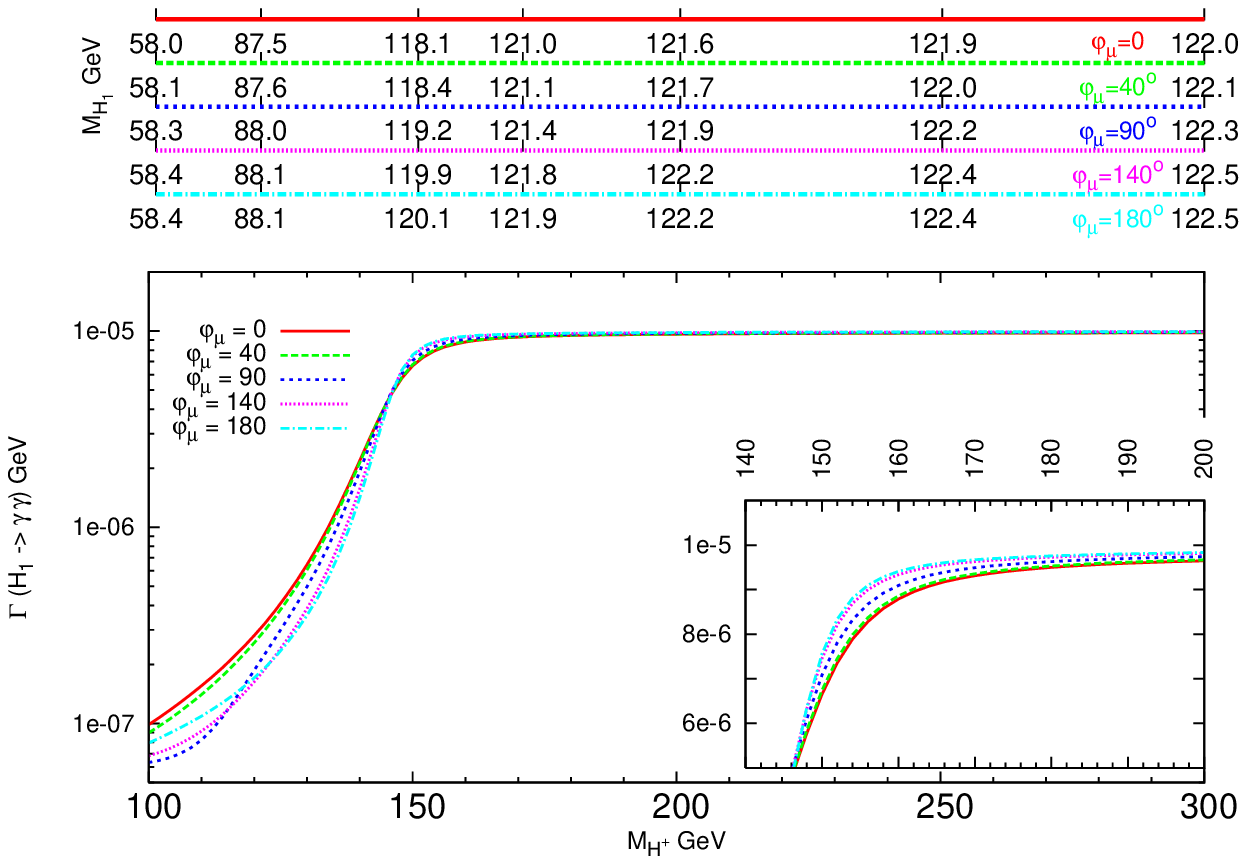}
\epsfysize=10cm \epsfxsize=8cm
\epsfbox{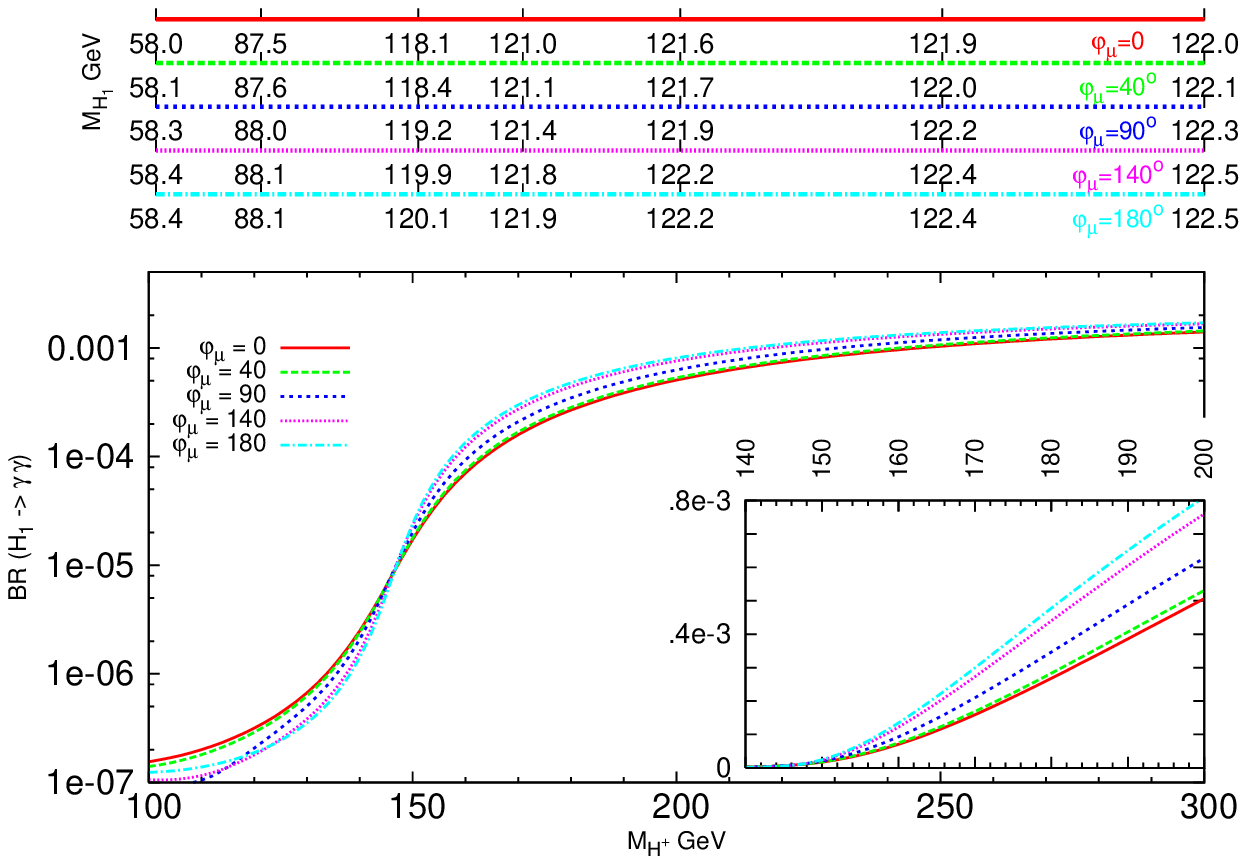}
\caption{\it 
Similar to Fig. \ref{fig:BR_1}, but with $|A_f|=1.5$ TeV, $|\mu|=0.5$ TeV 
and $\tan\beta=20$.
}
\label{fig:BR_3}
\end{figure}

\begin{figure}
\epsfysize=10cm \epsfxsize=8cm
\epsfbox{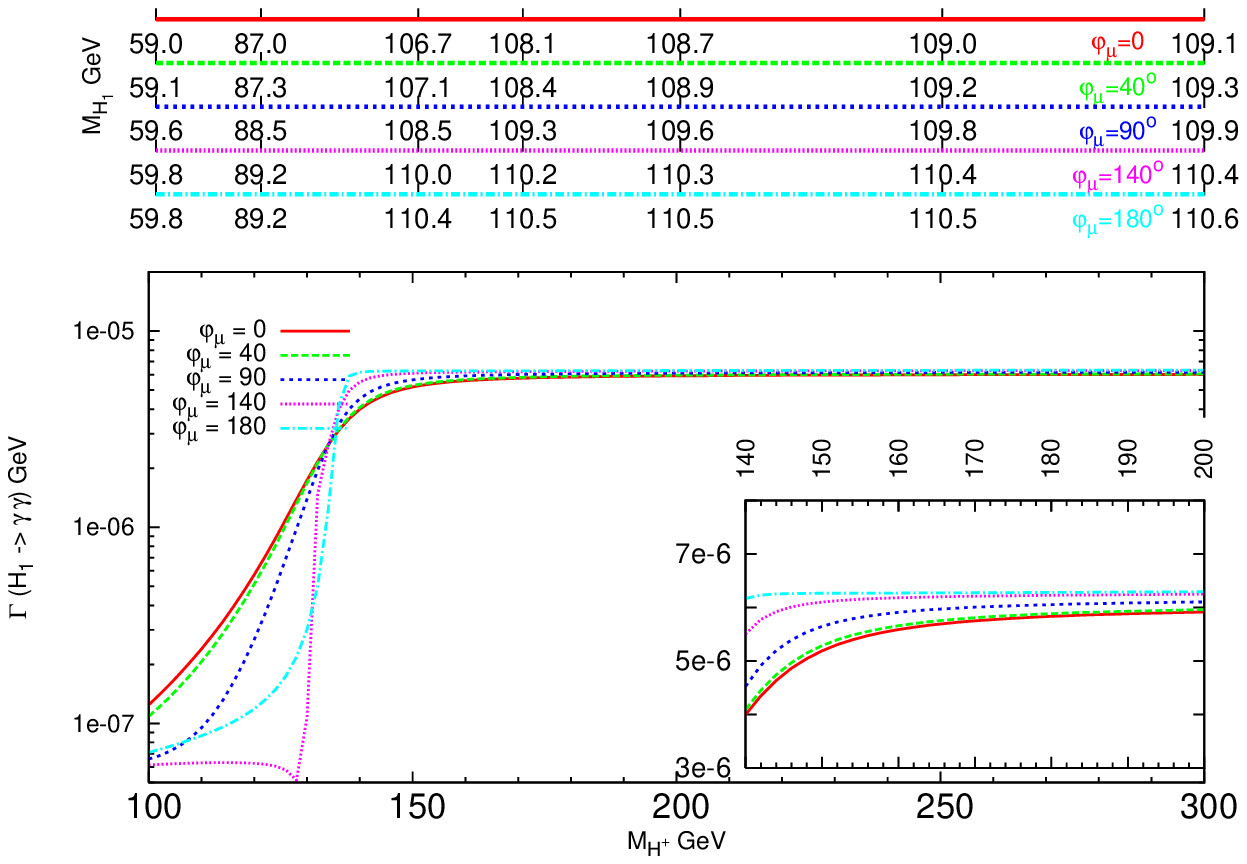}
\epsfysize=10cm \epsfxsize=8cm
\epsfbox{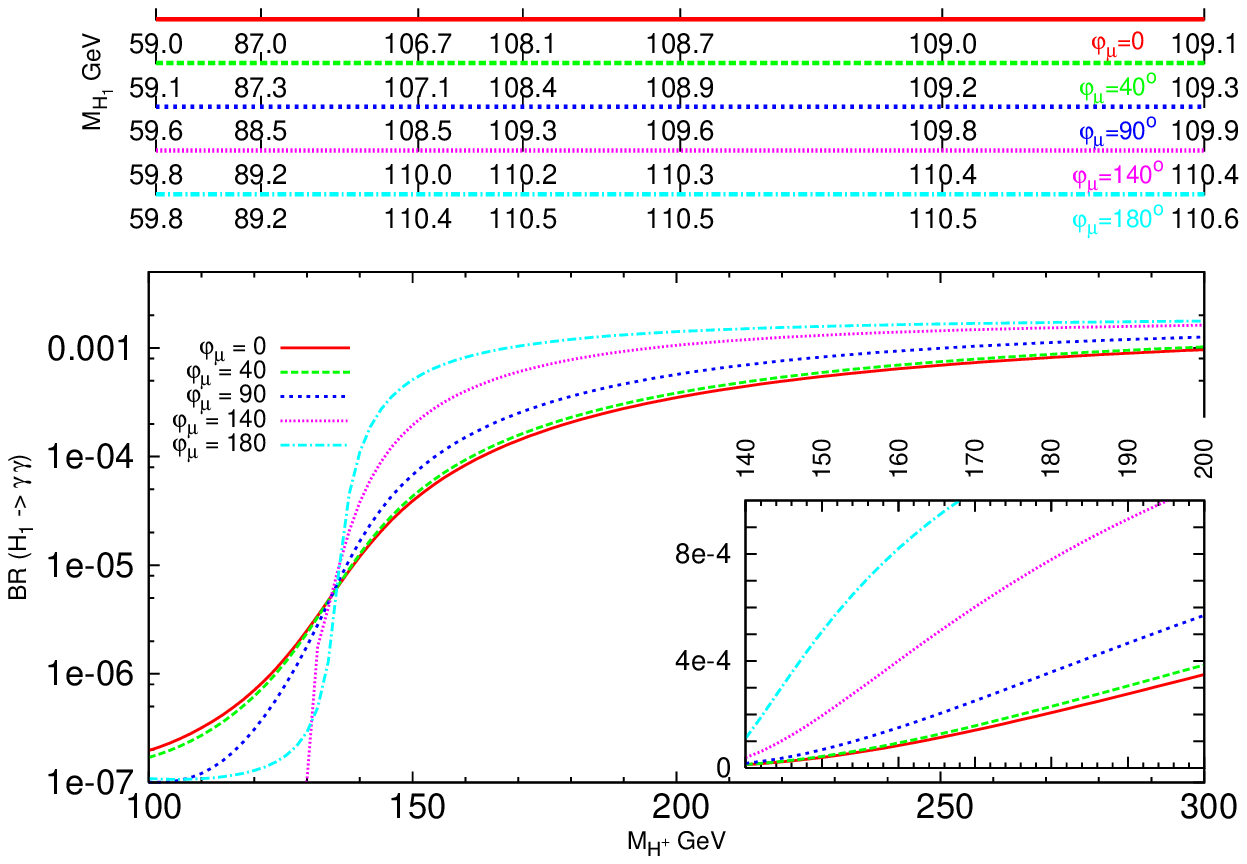}

\epsfysize=10cm \epsfxsize=8cm
\epsfbox{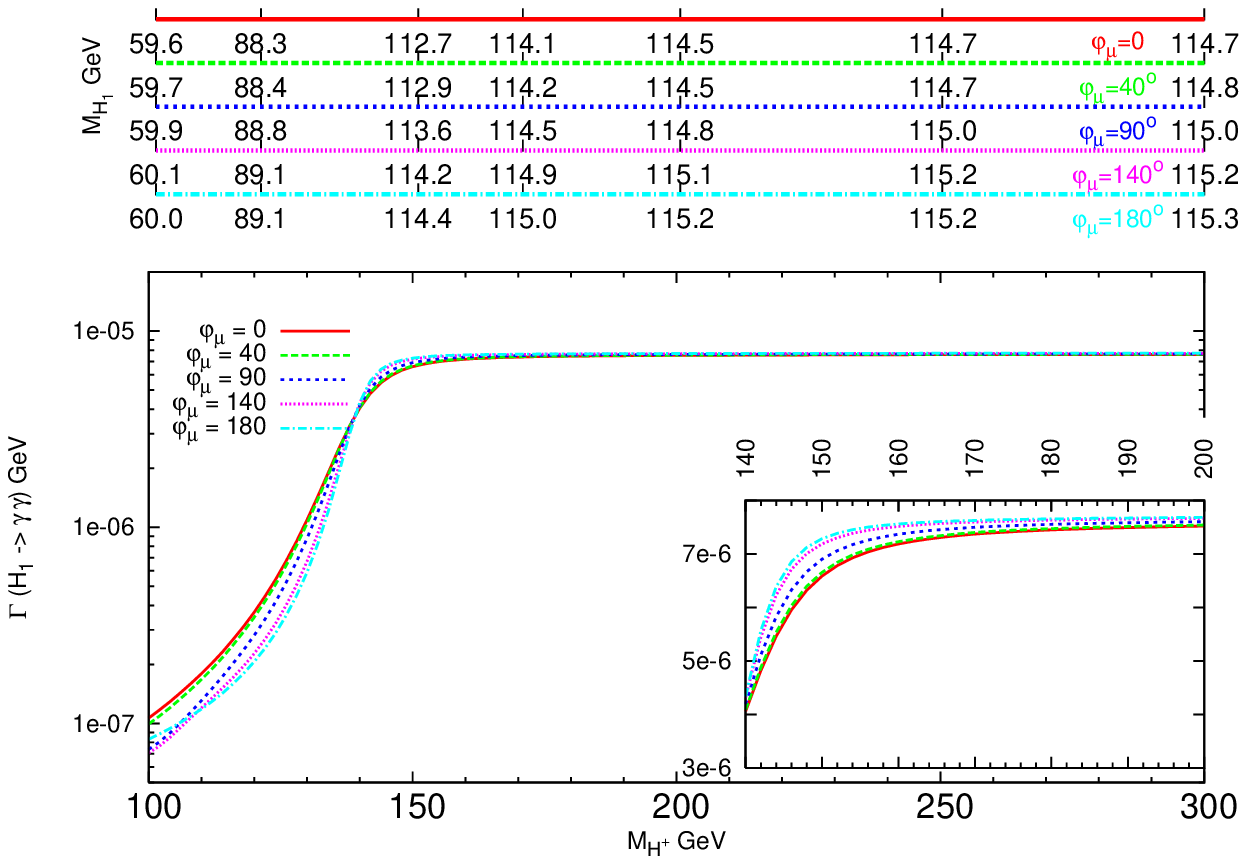}
\epsfysize=10cm \epsfxsize=8cm
\epsfbox{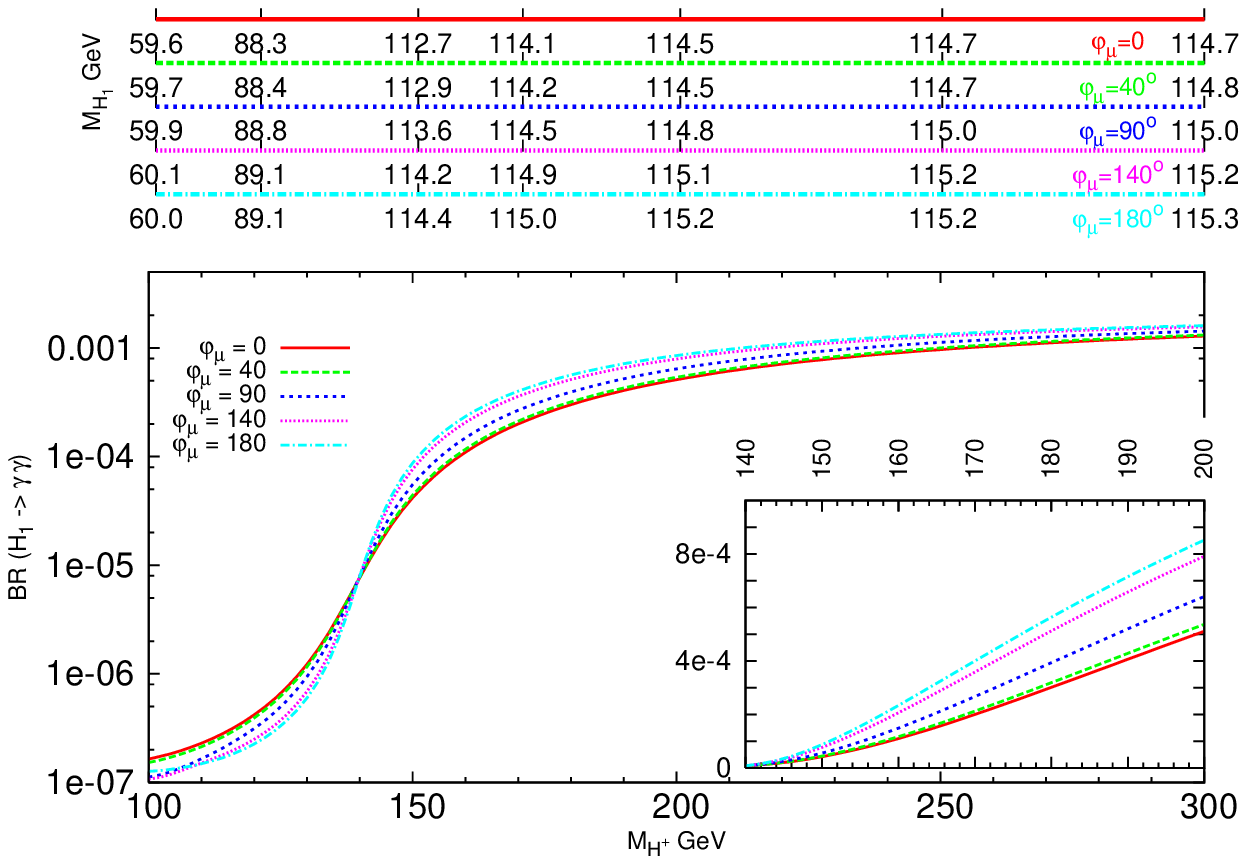}
\caption{\it 
Similar to Fig. \ref{fig:BR_1}, but with $|A_f|=0.5$ TeV, $|\mu|=1$ TeV 
and $\tan\beta=20$.
}
\label{fig:BR_2}
\end{figure}

\begin{figure}
\epsfysize=10cm \epsfxsize=8cm
\epsfbox{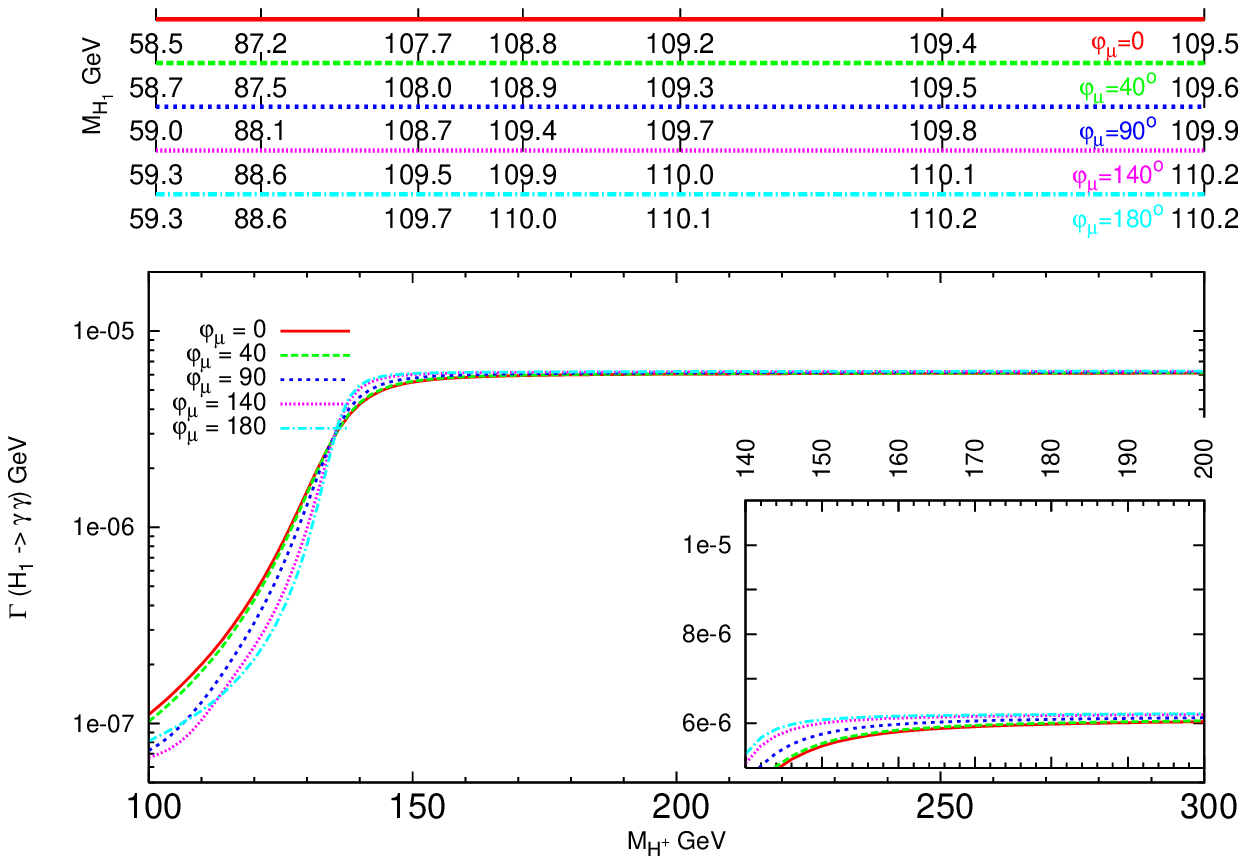}
\epsfysize=10cm \epsfxsize=8cm
\epsfbox{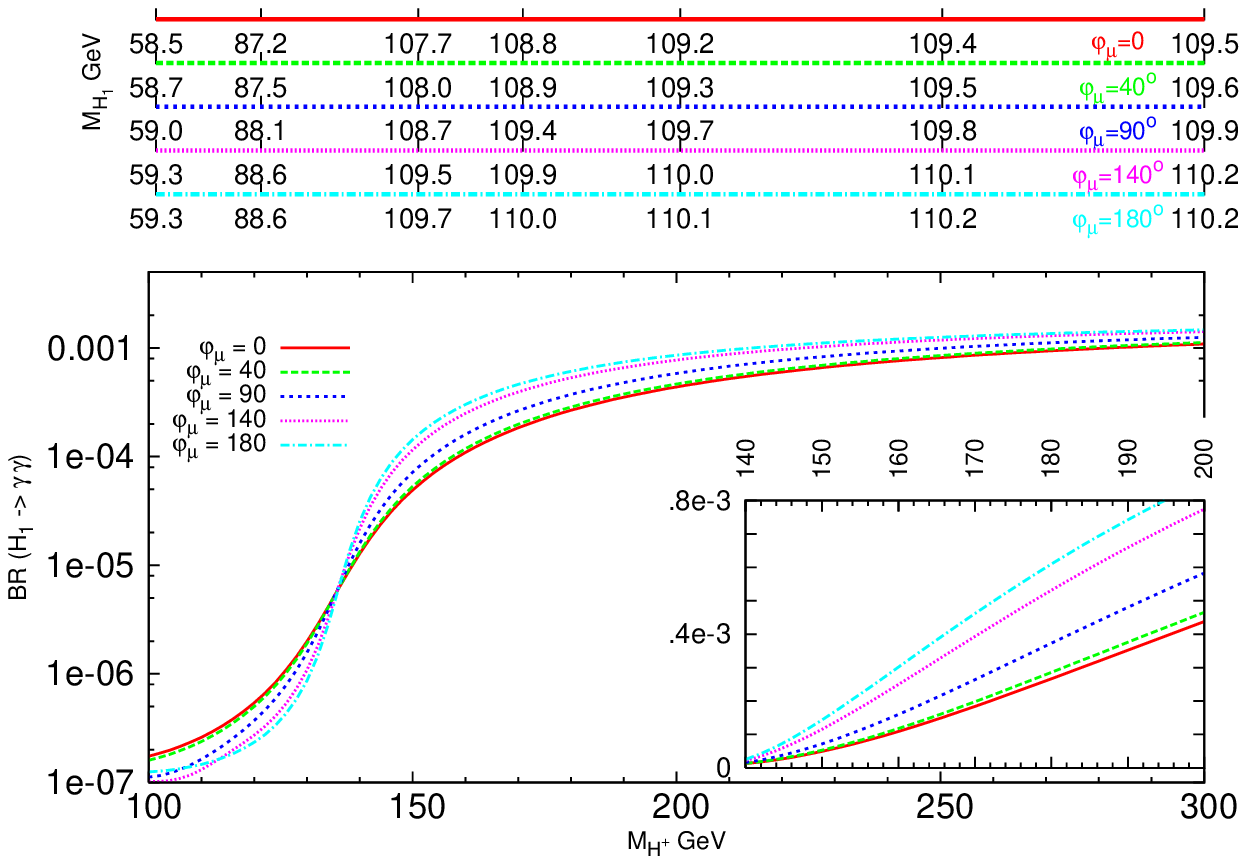}

\epsfysize=10cm \epsfxsize=8cm
\epsfbox{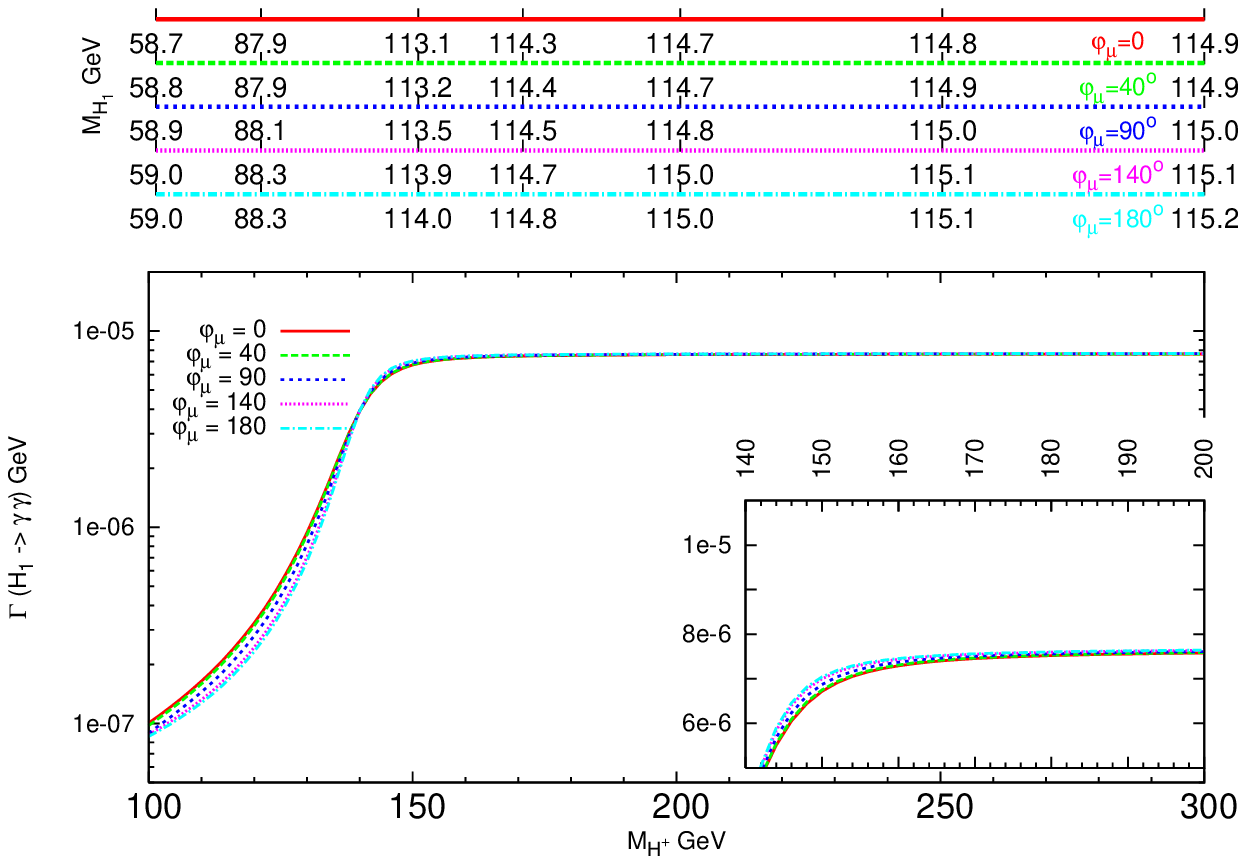}
\epsfysize=10cm \epsfxsize=8cm
\epsfbox{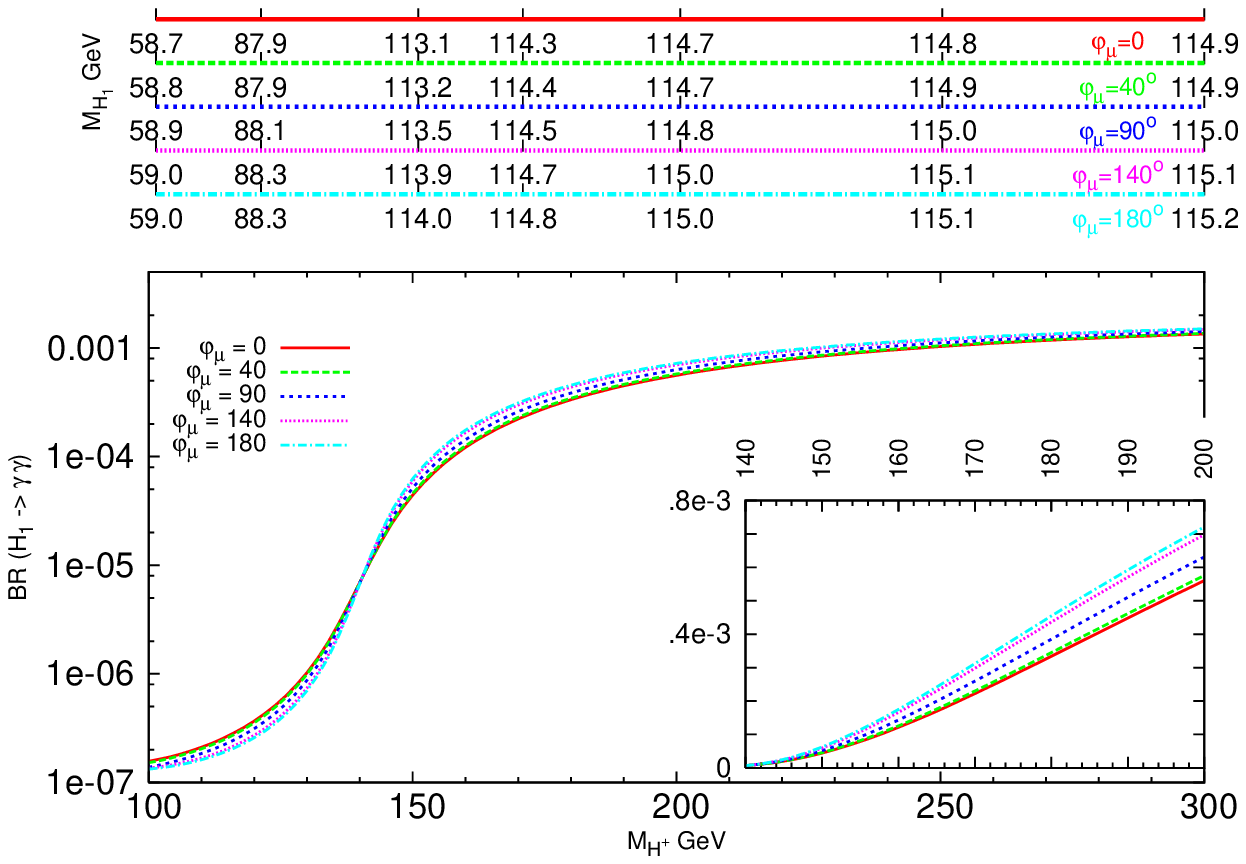}
\caption{\it 
Similar to Fig. \ref{fig:BR_1}, but with $|A_f|=0.5$ TeV, $|\mu|=0.5$ TeV 
and $\tan\beta=20$.
}
\label{fig:BR_4}
\end{figure}

\begin{figure}
\epsfysize=10cm \epsfxsize=8cm
\epsfbox{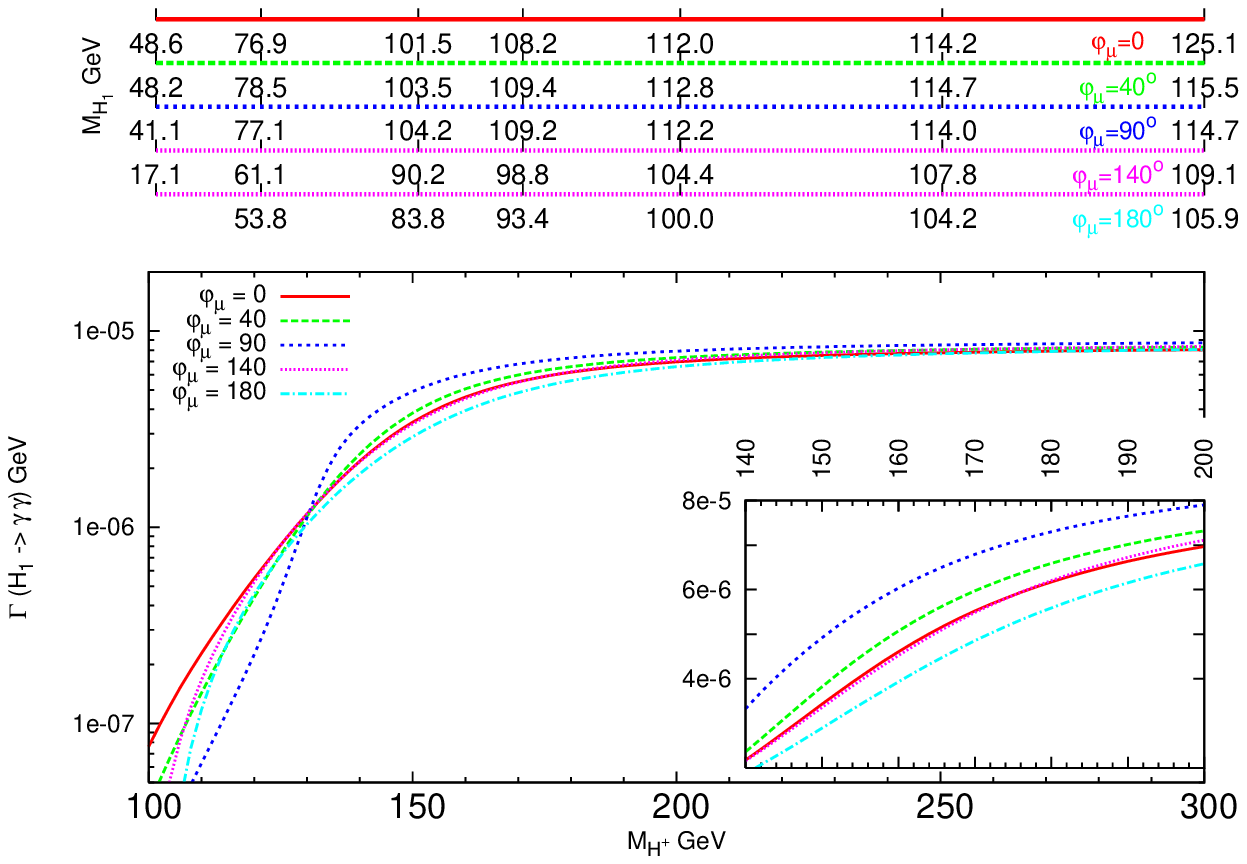}
\epsfysize=10cm \epsfxsize=8cm
\epsfbox{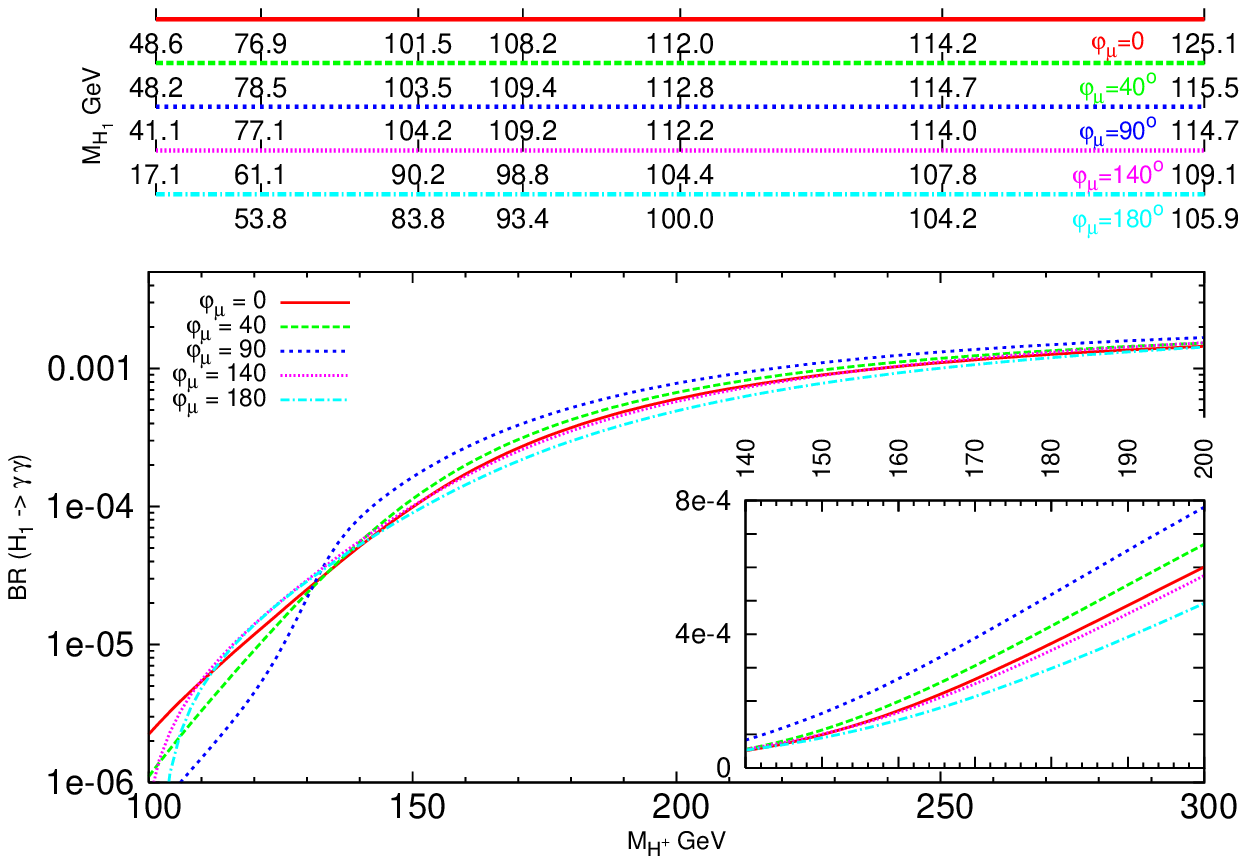}

\epsfysize=10cm \epsfxsize=8cm
\epsfbox{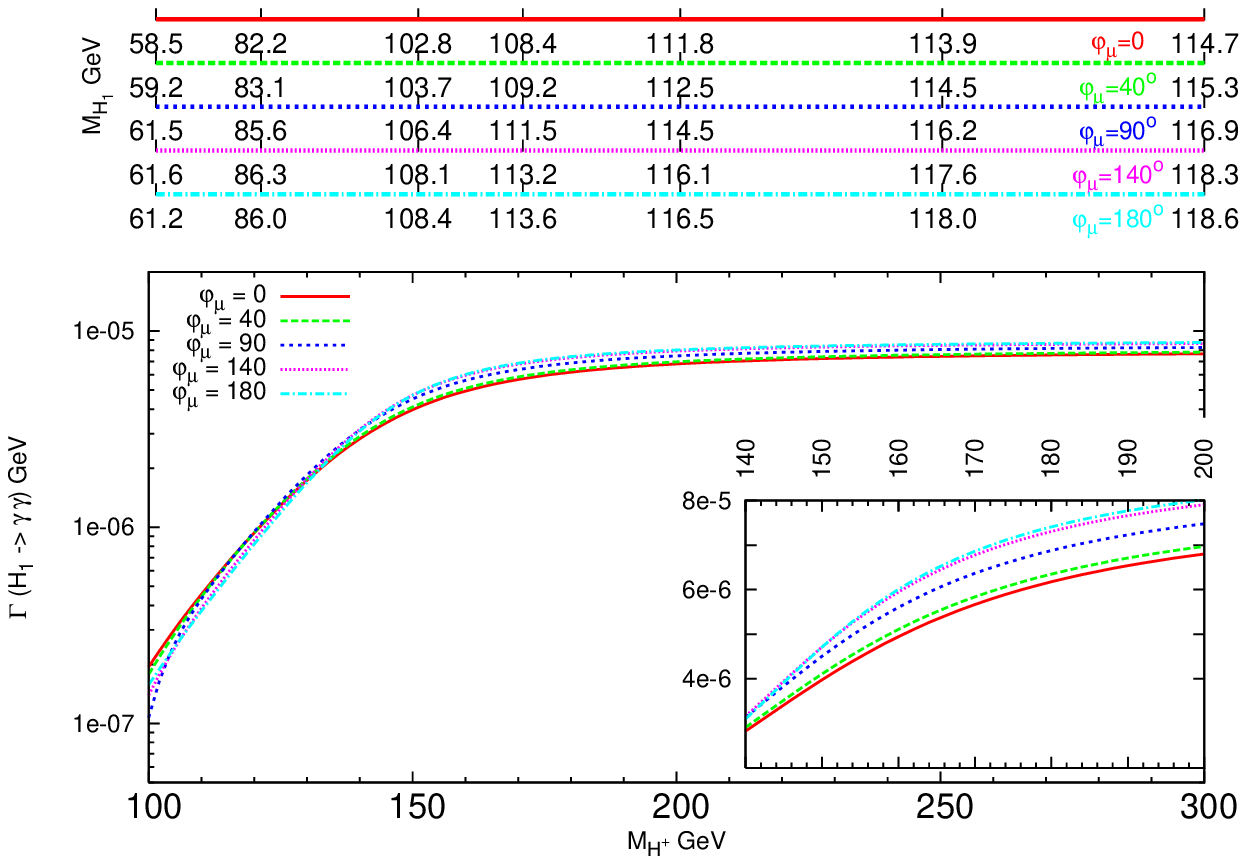}
\epsfysize=10cm \epsfxsize=8cm
\epsfbox{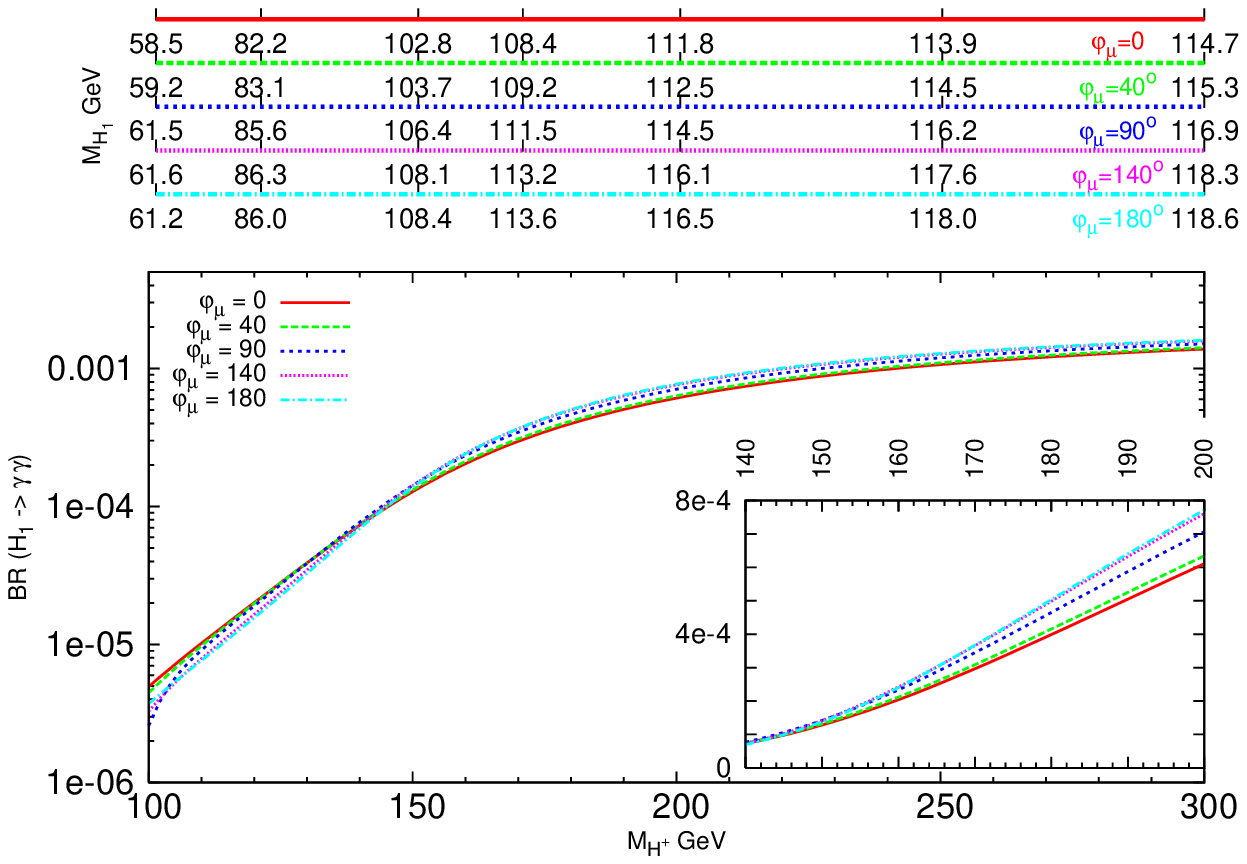}
\caption{\it 
Similar to Fig. \ref{fig:BR_1}, but with $|A_f|=1.5$ TeV, $|\mu|=1$ TeV 
and $\tan\beta=5$.
}
\label{fig:BR_tb05_1}
\end{figure}

\begin{figure}
\epsfysize=10cm \epsfxsize=8cm
\epsfbox{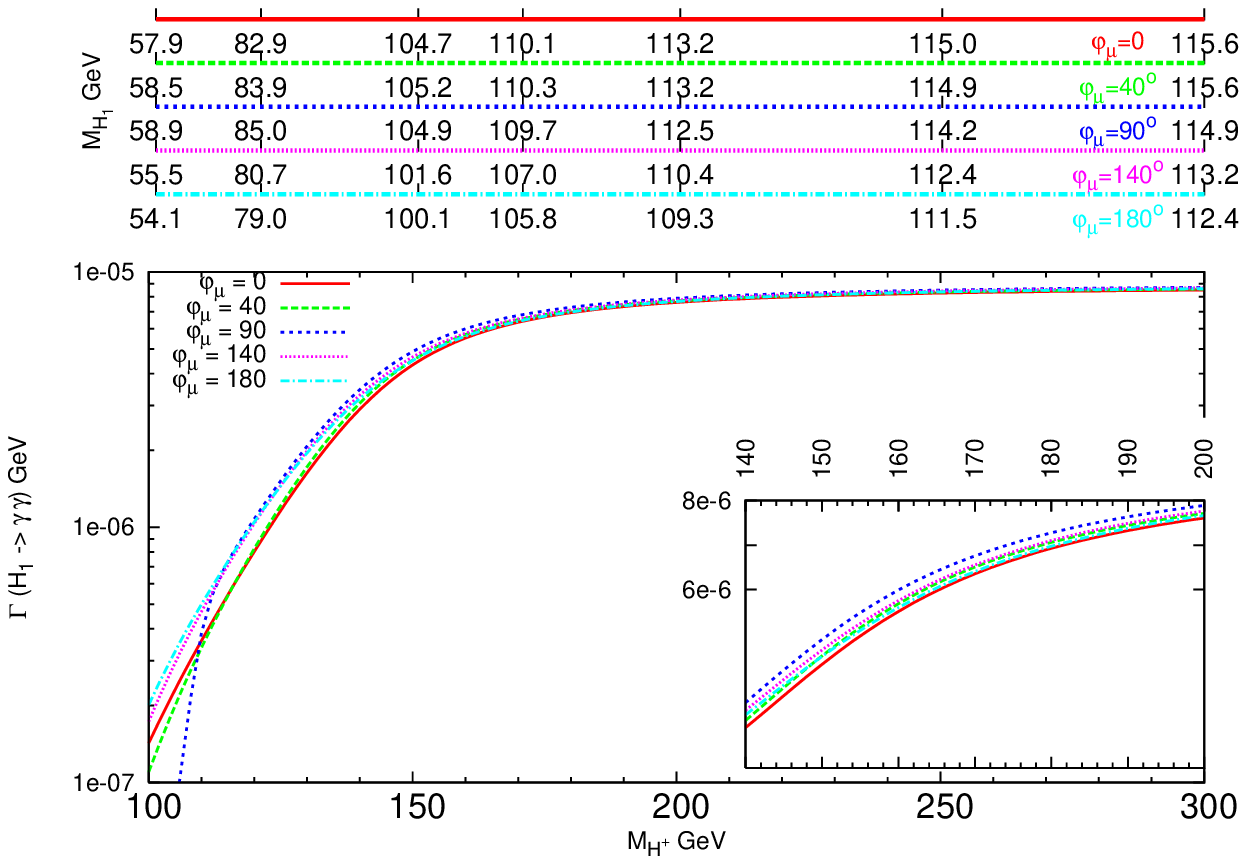}
\epsfysize=10cm \epsfxsize=8cm
\epsfbox{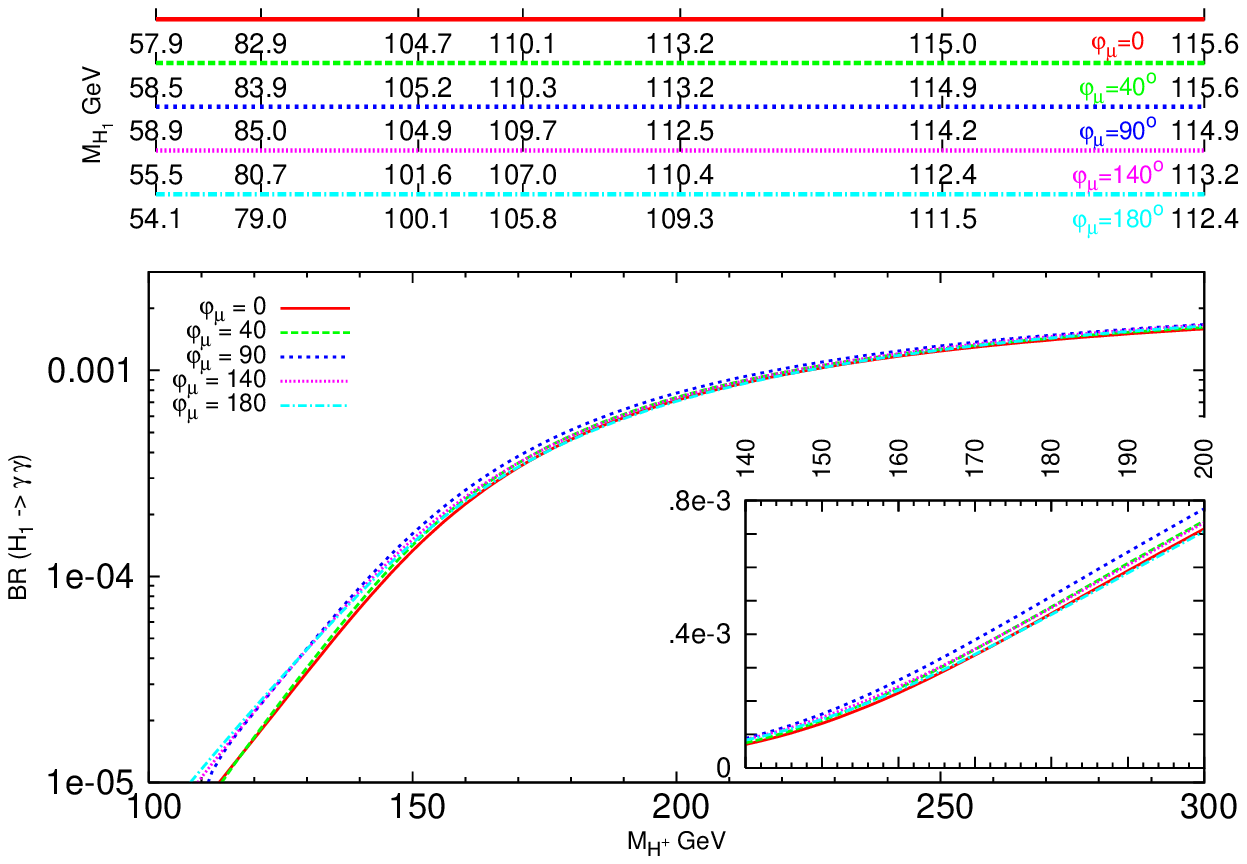}

\epsfysize=10cm \epsfxsize=8cm
\epsfbox{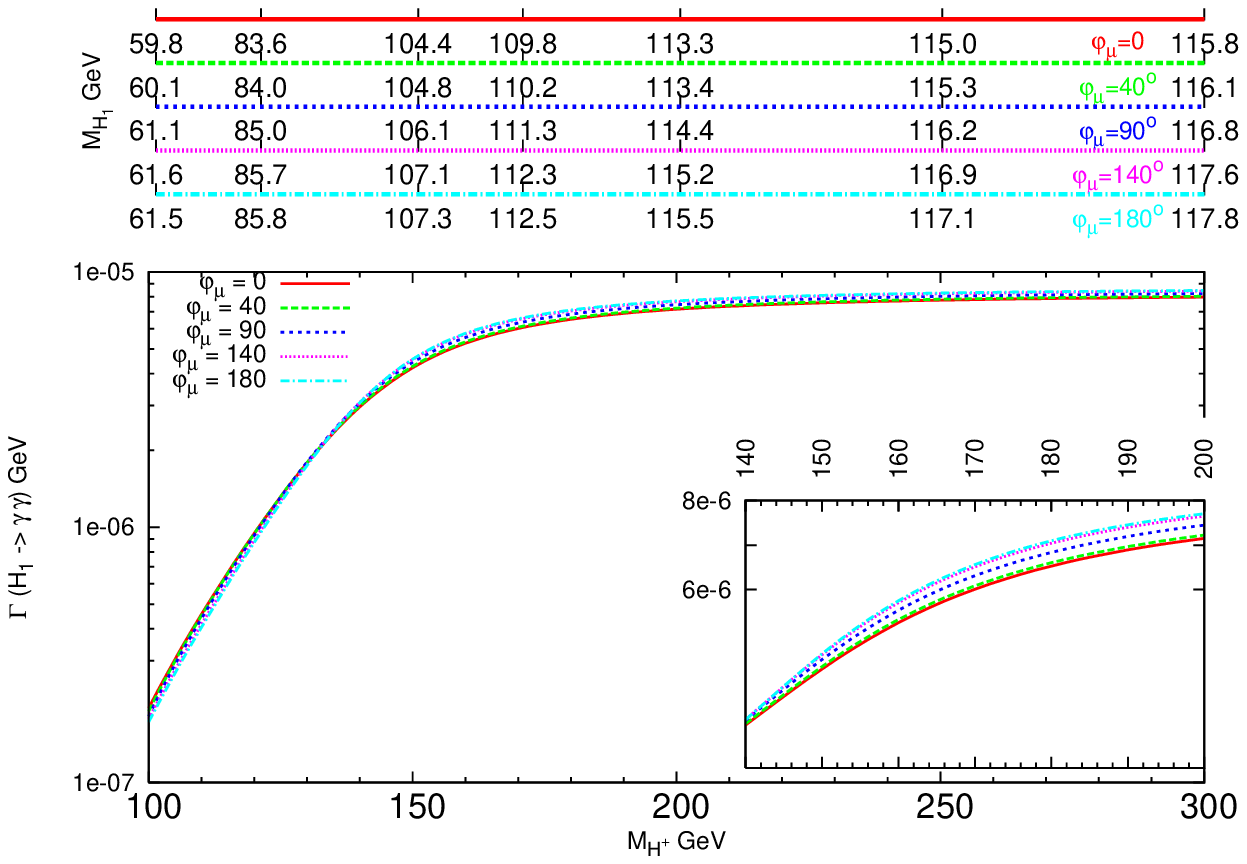}
\epsfysize=10cm \epsfxsize=8cm
\epsfbox{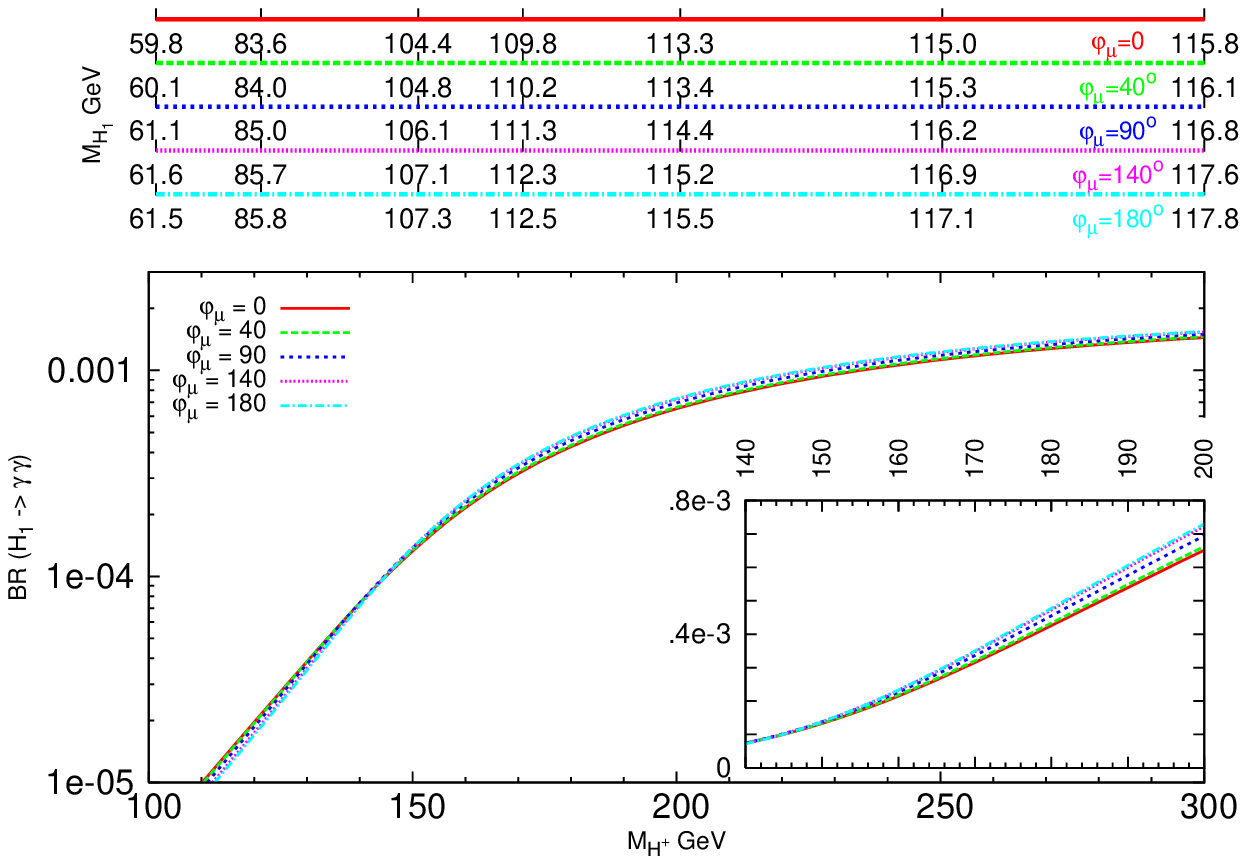}
\caption{\it 
Similar to Fig. \ref{fig:BR_1}, but with $|A_f|=1.5$ TeV, $|\mu|=0.5$ TeV 
and $\tan\beta=5$.
}
\label{fig:BR_tb05_3}
\end{figure}

\begin{figure}
\epsfysize=10cm \epsfxsize=8cm
\epsfbox{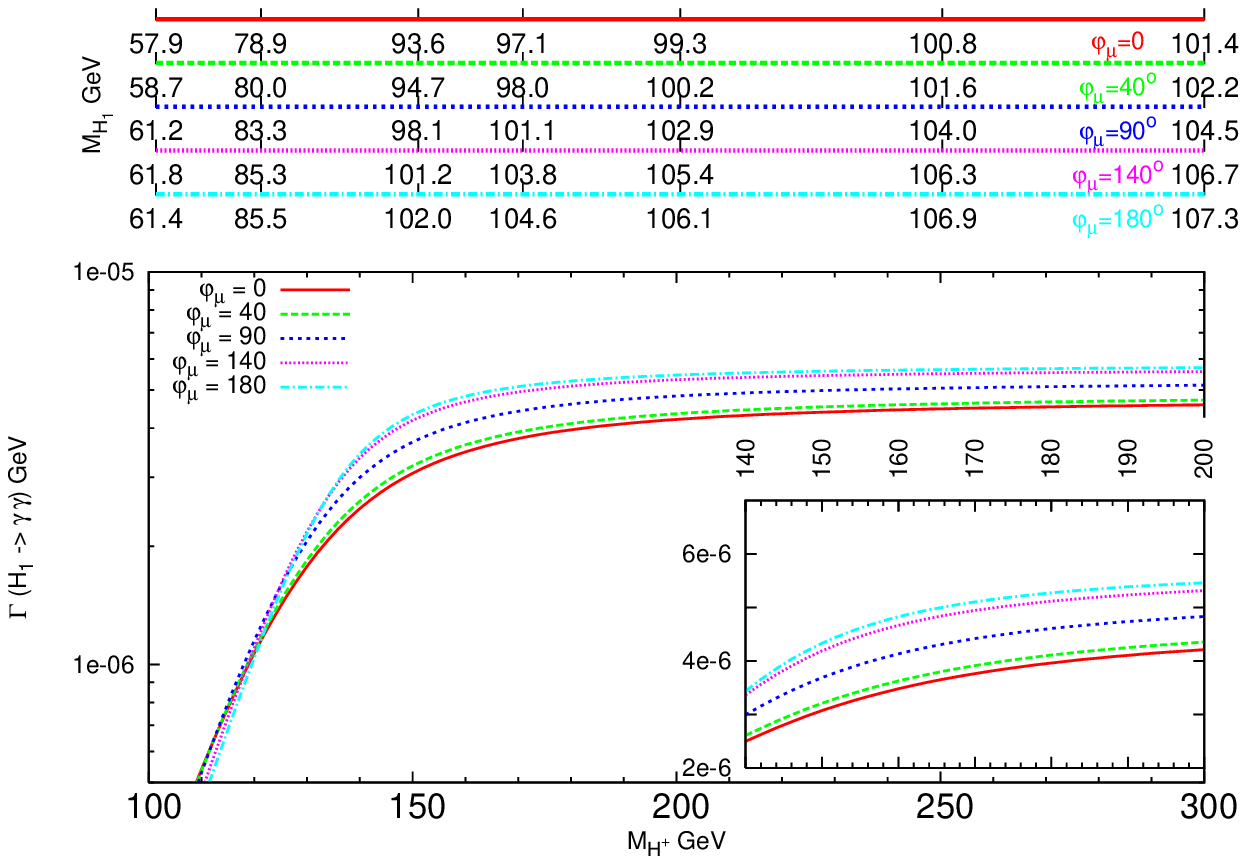}
\epsfysize=10cm \epsfxsize=8cm
\epsfbox{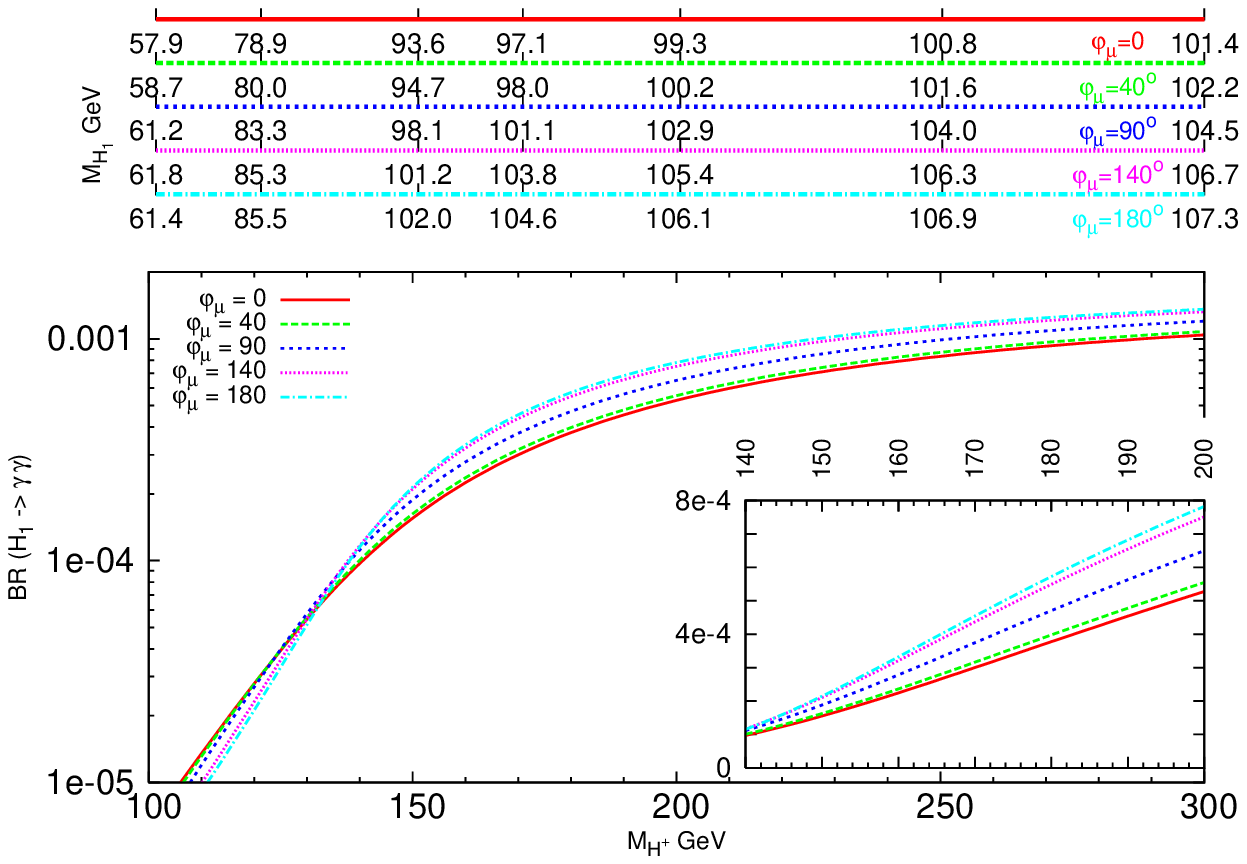}

\epsfysize=10cm \epsfxsize=8cm
\epsfbox{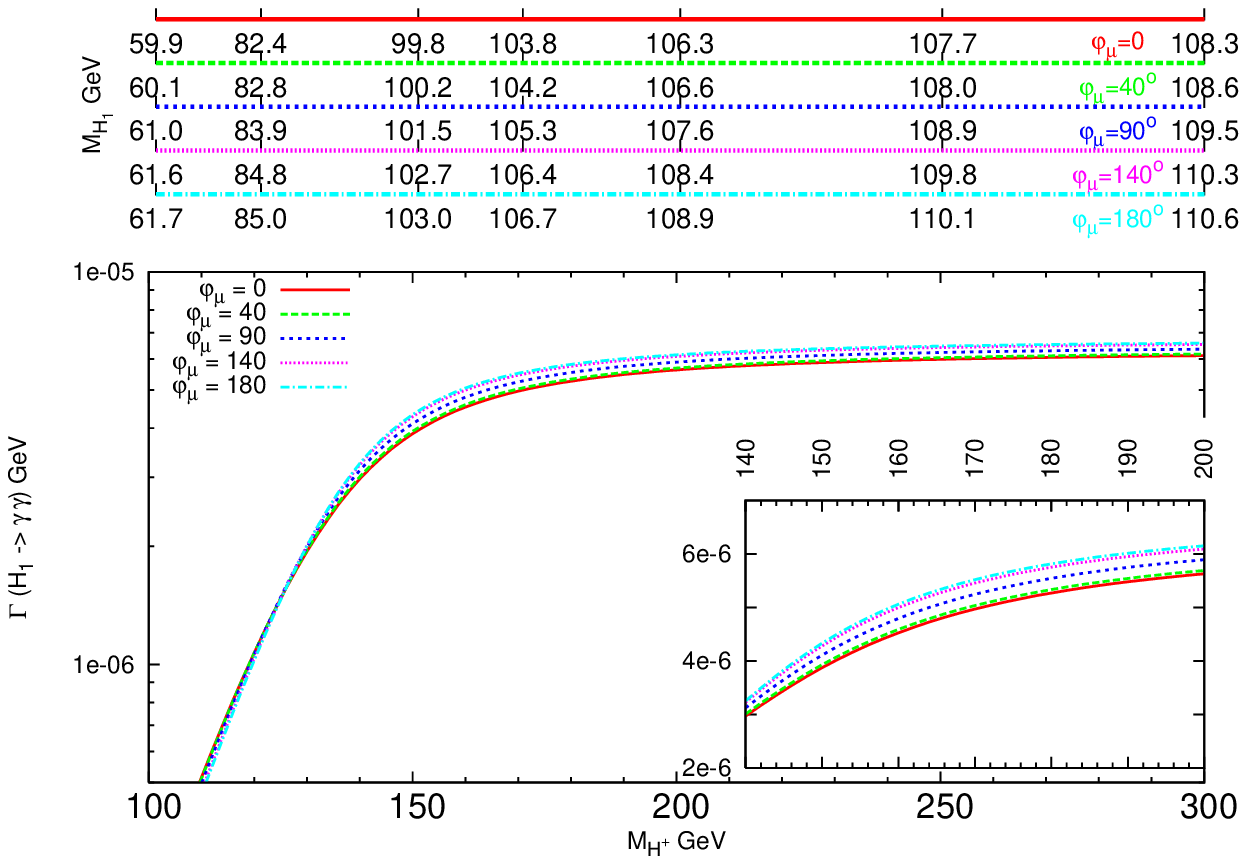}
\epsfysize=10cm \epsfxsize=8cm
\epsfbox{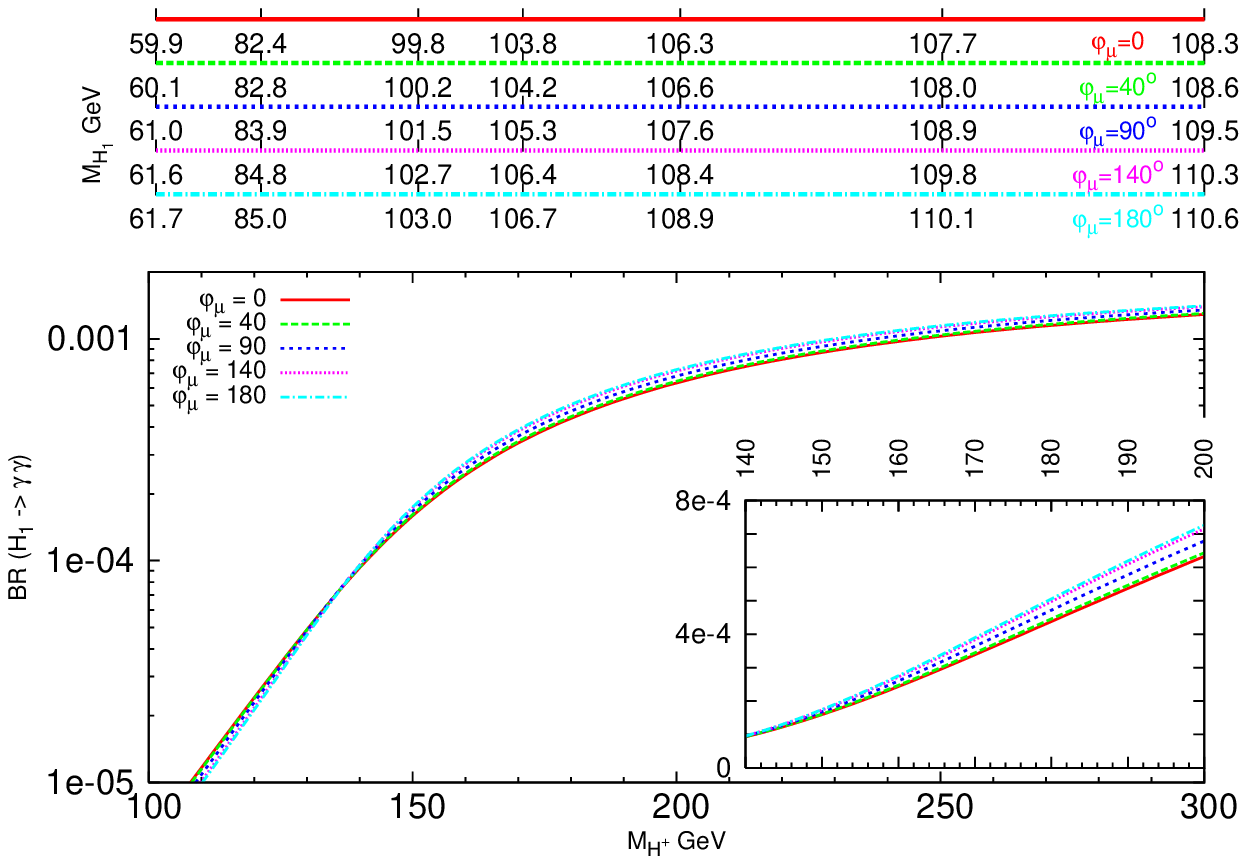}
\caption{\it 
Similar to Fig. \ref{fig:BR_1}, but with $|A_f|=0.5$ TeV, $|\mu|=1$ TeV 
and $\tan\beta=5$.
}
\label{fig:BR_tb05_2}
\end{figure}

\begin{figure}
\epsfysize=10cm \epsfxsize=8cm
\epsfbox{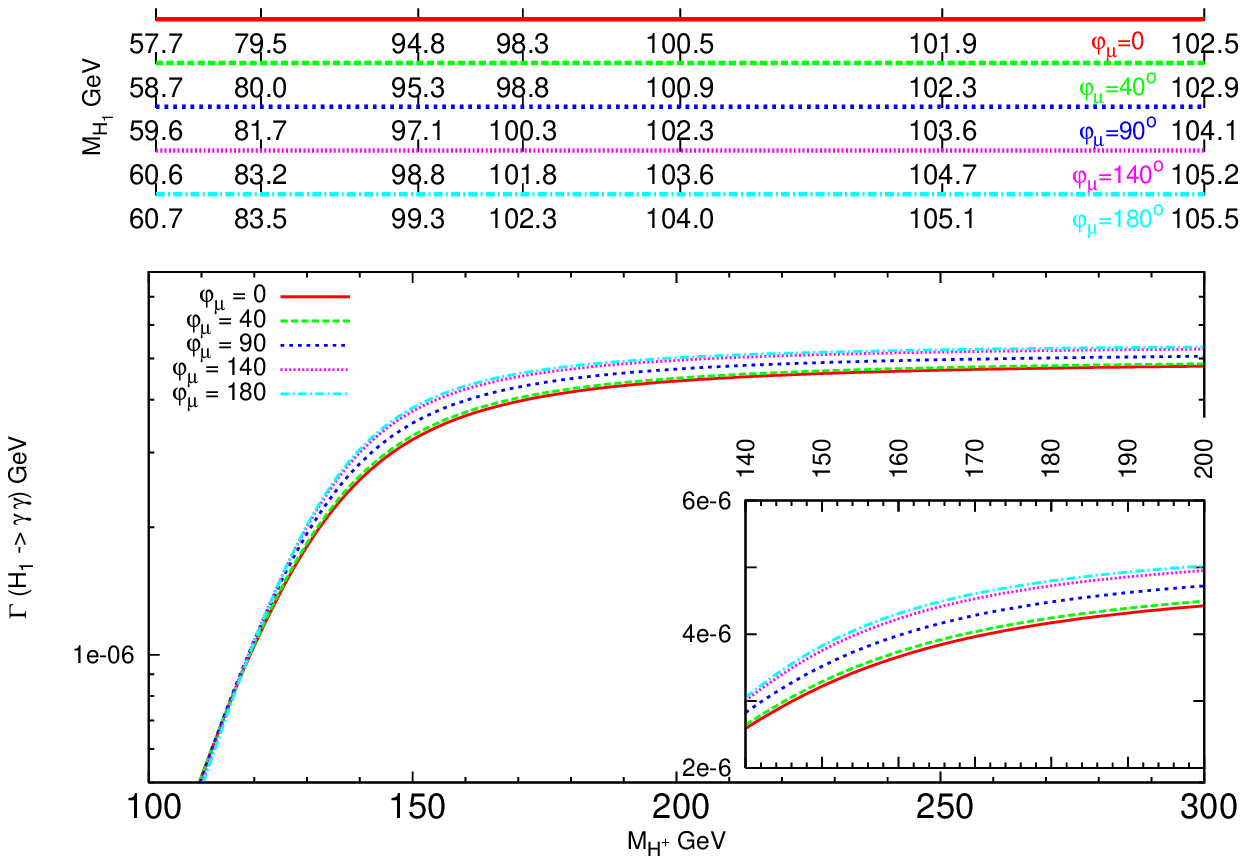}
\epsfysize=10cm \epsfxsize=8cm
\epsfbox{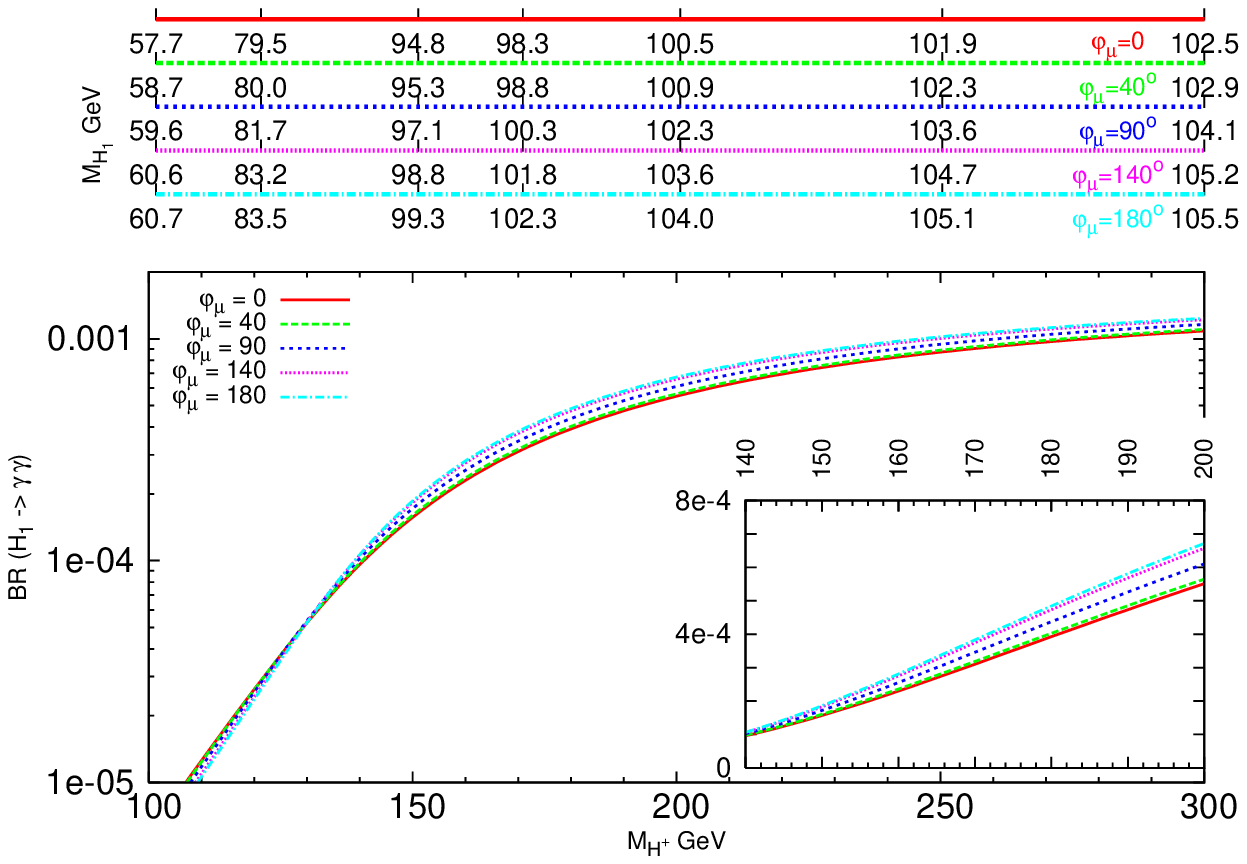}

\epsfysize=10cm \epsfxsize=8cm
\epsfbox{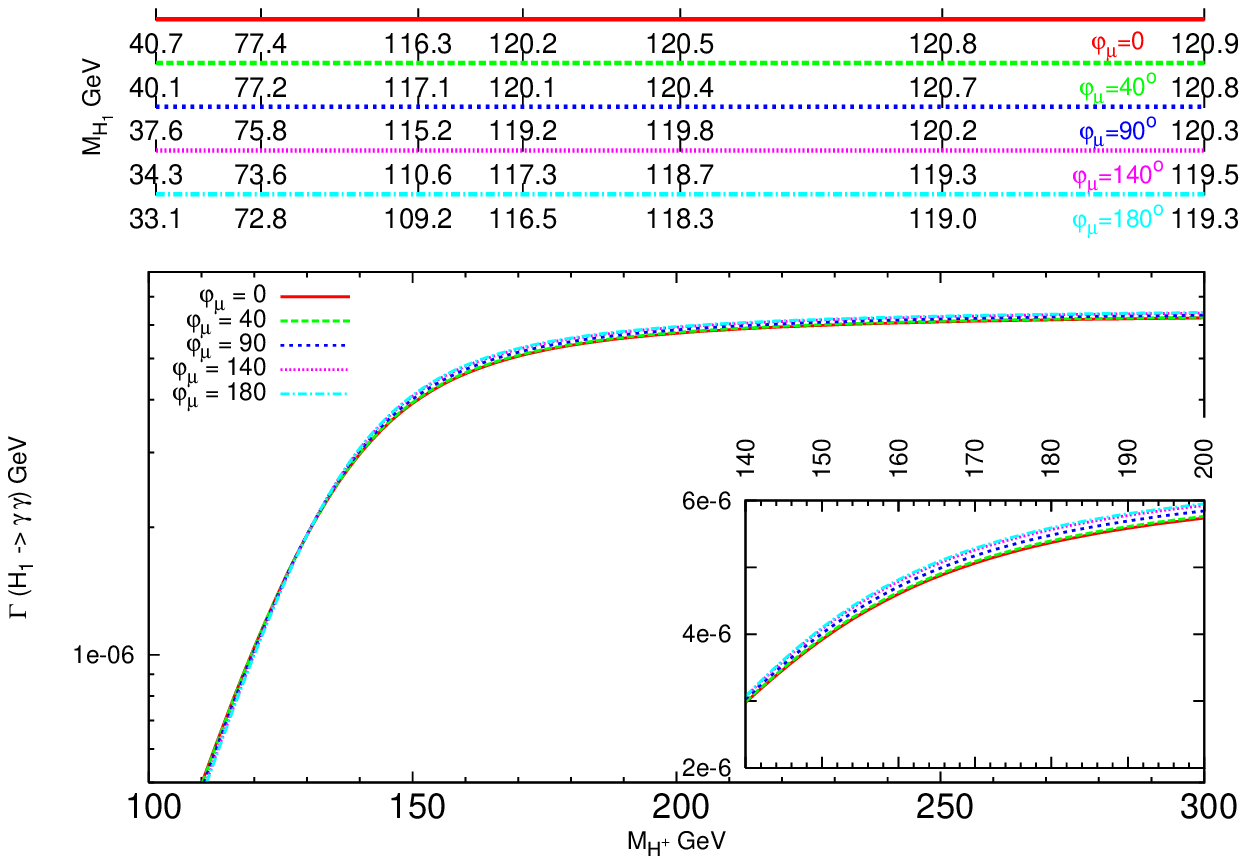}
\epsfysize=10cm \epsfxsize=8cm
\epsfbox{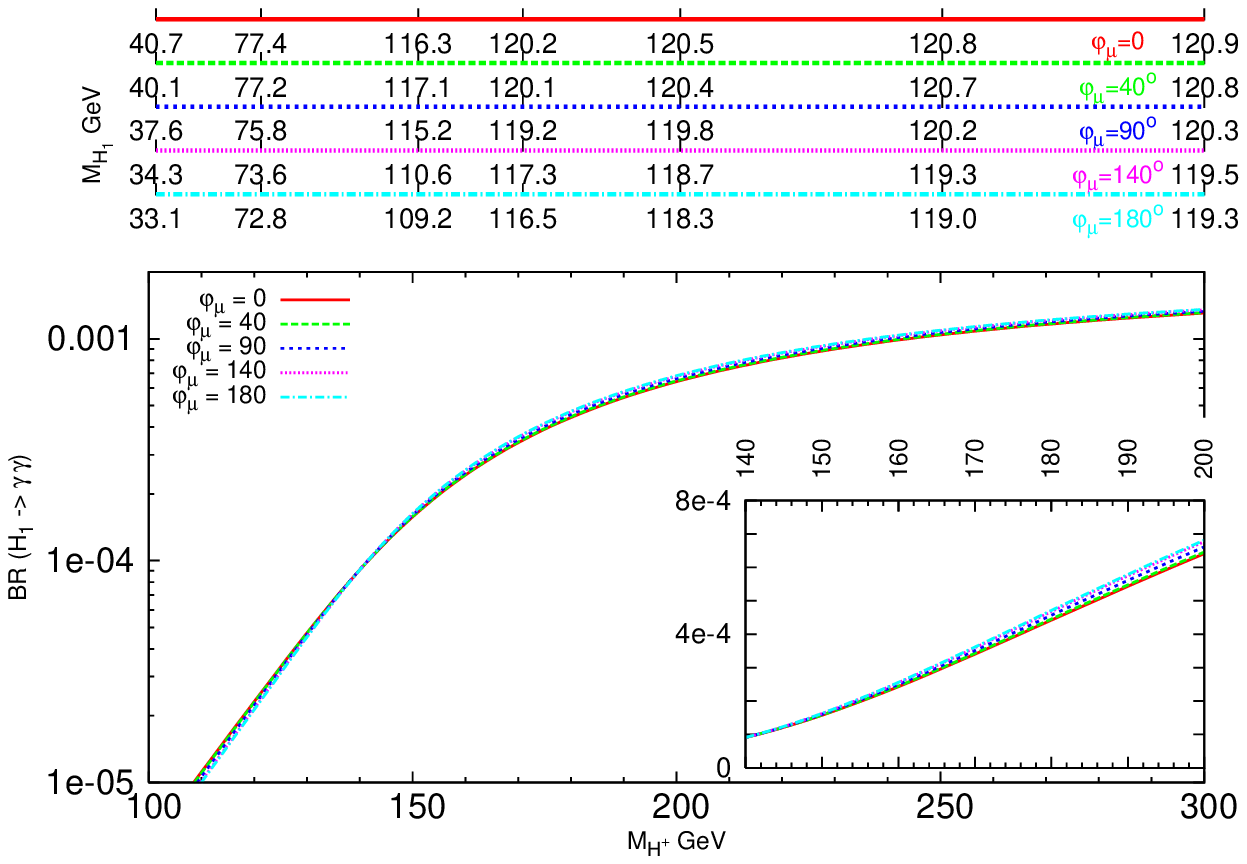}
\caption{\it 
Similar to Fig. \ref{fig:BR_1}, but with $|A_f|=0.5$ TeV, $|\mu|=0.5$ TeV 
and $\tan\beta=5$.}
\label{fig:BR_tb05_4}
\end{figure}

\vspace*{0.15cm}
As it is unfeasible to present all the results of our scan, we have picked out a few
different discrete choices of the relevant MSSM parameters and plotted 
width and BR against $M_{H^+}$ for the latter
in Figs.~\ref{fig:BR_1}--\ref{fig:BR_tb05_4}.
We choose five representative values between $0^\circ$ and $180^\circ$
for the phase of $\mu$%
\footnote{To aid the reader, through multiple x-axes and labels, 
 we have included the mass of $H_1$ in each plot corresponding to $M_{H^\pm}$ for the
given choice of other MSSM parameters. Besides, in each figure we have zoomed
near and above the aforementioned cross over point at $M_{H^\pm}\approx150$ GeV.
Finally, as the BR is the measurable quantity, we dwell on this while only plotting the
width for reference.}.
Fig.~\ref{fig:BR_1} is with $|A_f|=1.5$ TeV, $|\mu|=1$ TeV and $\tan\beta=20$. 
Comparing the two cases of $M_{\tilde U_3}=250$ GeV and  $M_{\tilde U_3}=1$ TeV 
there is a qualitative difference in the sensitivity to $\phi_\mu$. This is 
also true for Fig.~\ref{fig:BR_3}, which assumes
 $|A_f|=1.5$ TeV and $\tan\beta=20$, but a smaller $|\mu|=500$ GeV.
In the first case there is an increase in the BR over the 
region $M_{H^+}>150$ GeV (and decrease over the region $M_{H^+}<150$ GeV) as 
$\phi_\mu$ is switched on. This relative change with $\phi_\mu$ is
maximised for some value of $\phi_\mu$ 
around $40^\circ$, beyond which the change in BR decreases again to about 50\%
at $\phi_\mu = 180^\circ$.
In the second case there is no such a trend as
there is a 50\% increase in the BR for $\phi_\mu=90^\circ$ at  
$M_{H^+}\sim 200$ GeV and the effect grows larger for $\phi_\mu>90^\circ$.
Other general features are the following.
The dependence on $\phi_\mu$ decreases with lower values of $|\mu|$ as
seen from Fig.~\ref{fig:BR_3}.
The value of $\phi_\mu$ with maximum BR in the
presence of a light $\tilde t_1$ decreases compared to 
the case when $|\mu|=1$ TeV.
In contrast, a smaller value of $|A_f|=500$ GeV 
(Figs.~\ref{fig:BR_2} and \ref{fig:BR_4}) keeps the picture 
qualitatively the same for the two cases of light and heavy $\tilde t_1$:
e.g., in the region $M_{H^+}>150$ GeV, the BR
increases with increasing $\phi_\mu$. But, while in the first case 
($M_{\tilde U_3}=250$ GeV) there is a 50\% increase for $\phi_\mu=90^\circ$ 
at $M_{H^+}=200$ GeV, in the second case ($M_{\tilde U_3}=1$ TeV) there is a
reduction of less than 20\%. Again, the deviations can be substantially larger for 
$\phi_\mu>90^\circ$.
Sensitivity of BR($H_1\rightarrow \gamma\gamma$) to $\phi_\mu$ 
is however reduced considerably for lower values of 
$\tan\beta$, while the qualitative features remain the same,
as is  illustrated in
Figs.~\ref{fig:BR_tb05_1}--\ref{fig:BR_tb05_4}
for $\tan\beta=5$.


\section{Conclusion}

In this paper, we have demonstrated that the decay channel $H_1\rightarrow \gamma\gamma$
is particularly suitable to probe the possible presence of CP-violating effects
in the MSSM. This mode is in fact not only
very sensitive to variations of the coupling $H_1W^+W^-$ pertaining to the dominant SM loop
-- induced from mixing amongst
Higgs states via one-loop effects (as
shown in previous literature) -- but also to contributions of a 
light $\tilde t_1$ in the triangle-loop defining the decay process --
which is in fact a tree-level effect induced by a complex  $\mu$ parameter
(while the trilinear coupling $A_t$ is taken real). 

In particular, our detailed analyses indicate that studies of the 
di-photon channel of a light Higgs boson (with mass below 130 GeV or so) found at the LHC
may eventually enable one to disentangle the
CP-violating case from the CP-conserving one, so long that the relevant SUSY parameters 
entering  $H_1\rightarrow \gamma\gamma$
are measured elsewhere. This is not phenomenologically unconceivable, as the $H_1\to\gamma\gamma$ detection
mode requires a very high luminosity, unlike the discovery of those sparticles (and the measurement of
their masses and couplings) that enter the Higgs process studied here. Furthermore, while explicit
CP violation could affect the mass of the lightest Higgs state of the underlying SUSY model, we have
restricted ourselves to regions of parameter space where -- for an identical choice of all 
SUSY inputs but $\phi_\mu$, the only relevant phase in the scenarios
considered here
-- the difference between the $M_{H_1}$ values in the CP-violating case and CP-conserving
one are below the experimental uncertainty on the determination of such a quantity, so that it would not
be possible to confirm or disprove the existence of complex parameters in the SUSY Lagrangian by solely
isolating a $H_1$ resonance.  

A complete analysis
will eventually require to fold the decay process studied here in Narrow Width Approximation
with propagator effects and the appropriate
production mode (gluon-gluon fusion and Higgs-strahlung in this case), where
similar CP-violating
effects may enter. This will be done in a future publication
\cite{preparation}. Finally, as
argued in the Introduction, for the scenario we considered here, with very
heavy squarks and vanishing
trilinear couplings for the first and second generations, we can evade the
EDM constraints on the
CP-violating phases. A detailed analysis of this is also deferred to future
work.

\section*{Acknowledgments}

PP's research is supported by the Framework Programme 6 via a Marie Curie 
International Incoming Fellowship, contract number MIF1-CT-2004-002989. 
This research has been partially financed by the NATO Collaborative Linkage 
Grant no. PST.CLG.980066. SM thanks the RTN European Programme
MRTN-CT-2006-035505 (HEPTOOLS, Tools and Precision Calculations
for Physics Discoveries at Colliders) for partial financial support.


\vskip 5mm
\noindent
\begin{thebibliography}{99}

\bibitem{Higgs-hunter} For reviews, see: A. Djouadi, hep-ph/0503172, hep-ph/0503173; J. F. Gunion, H. E. Haber, G. Kane, S. Dawson, {\em The Higgs Hunter's Guide} (Addison-Wesley, Reading, MA, 1990).

\bibitem{Maekawa} N. Maekawa, Phys. Lett. B {\bf282}, 387 (1992).

\bibitem{Pilaftsis-PLB} A. Pilaftsis, Phys. Lett. B \textbf{435}, 88 (1998).

\bibitem{Georgi-Pais} H. Georgi, A. Pais, Phys. Rev. D \textbf{10}, 1246 (1974).
\bibitem{Pamoral} A. Pomarol, Phys. Lett. B \textbf{ 287}, 331 (1992); N. Haba, Phys. Lett. B \textbf{398}, 305 (1997);  O. C. W. Kong, F. L. Lin, Phys. Lett. B \textbf{418}, 217 (1998).

\bibitem{PlWagner} A. Pilaftsis, C. E. M. Wagner, Nucl. Phys. B \textbf{553}, 3 (1999).

\bibitem{Dugan} M. Dugan, B. Grinstein, L. Hall, Nucl. Phys. B \textbf{255}, 413 (1985).

\bibitem{dchang} D. Chang, W. Y. Keung, A. Pilaftsis, Phys. Rev. Lett. \textbf{82}, 900 (1999).

\bibitem{EDM1} P. Nath, Phys. Rev. Lett. B \textbf{66}, 2565 (1991), Y. Kuzikuri, N. Oshimo, Phys. Rev. D \textbf{45}, 1806 (1992);
{\it ibid.} \textbf{46},  3025 (1992);
S. Abel, S.~Khalil, O.~Lebedev, Nucl.\ Phys.\  B {\bf 606}, 151 (2001);
K.~A.~Olive, M.~Pospelov, A.~Ritz, Y.~Santoso, Phys.\ Rev.\  D {\bf 72}, 075001 (2005);
S. Abel, O.~Lebedev, JHEP, {\bf 0601}, 133 (2006).

\bibitem{EDM3} T. Ibrahim, P. Nath, Phys. Lett. B \textbf{418}, 98 (1998); Phys. Rev. D \textbf{57}, 478 (1998); {\it ibid.} \textbf{58}, 019901 (1998); T. Falk, K. A. Olive, Phys. Lett. B \textbf{439}, 71 (1998); M. Brhlik, G. J. Good, G. L. Kane, Phys. Rev. D \textbf{59}, 115004 (1999); S. Pokorski, J. Rosiek, C. A. Savoy, Nucl. Phys. B \textbf{570}, 81 (2000).

\bibitem{YaserAyazi:2006zw}
  S.~Yaser Ayazi, Y.~Farzan,
  Phys.\ Rev.\  D {\bf 74}, 055008 (2006).

\bibitem{ap-edm} A. Pilaftsis, Nucl. Phys. B {\bf 644}, 263 (2002).

\bibitem{EDM2} S. Dimopoulos, G. F. Giudice, Phys. Lett. B \textbf{357}, 573 (1995); A. Cohen, D. B. Kaplan, A. E. Nelson, Phys. Lett. B \textbf{388}, 588 (1996); A. Pamarol, D. Tommasini, Nucl. Phys. B \textbf{488}, 3 (1996).

\bibitem{Bartl:2003ju}
  A.~Bartl, W.~Majerotto, W.~Porod, D.~Wyler,
  Phys.\ Rev.\  D {\bf 68}, 053005 (2003).

\bibitem{ChoiSik}S. Y. Choi, J. S. Lee, Phys. Rev. D \textbf{61}, 015003 (2000); {\it ibid.}  \textbf{61}, 115002 (2000); {\it ibid.} \textbf{62}, 036005 (2000).

\bibitem{ChoiSik2}S. Y. Choi, J. S. Lee, Phys. Rev. D \textbf{61}, 115002 (2000).

\bibitem{ChoiKao}S. Y. Choi, K. Hagiwara, J. S. Lee, Phys. Rev. D \textbf{64}, 032004 (2001).

\bibitem{ChoiDrees} S. Y. Choi, M. Drees, J. S. Lee, J. Song, Eur. Phys. J. C \textbf{25}, 307 (2002).

\bibitem{CarEllis1a} M. Carena, J. Ellis, A. Pilaftsis,
  C. E. M. Wagner, Phys. Lett. B \textbf{495}, 155 (2000);
M. Carena, J. Ellis, S. Mrenna, A. Pilaftsis, C. E. M. Wagner, Nucl. Phys. B \textbf{659}, 145 (2003).

\bibitem{EllisPLee} J. Ellis, J. S. Lee, A. Pilaftsis, Phys. Rev. D \textbf{70}  075010 (2004); Mod. Phys. Lett. A \textbf{21}, 1405 (2006).

\bibitem{Choi:2004rf}
  S.~Y.~Choi, M.~Drees, B.~Gaissmaier,
  Phys.\ Rev.\  D {\bf 70}, 014010 (2004).

\bibitem{Bartl:2003pd}
  A.~Bartl, S.~Hesselbach, K.~Hidaka, T.~Kernreiter, W.~Porod,
  Phys.\ Lett.\  B {\bf 573}, 153 (2003);
  Phys.\ Rev.\  D {\bf 70}, 035003 (2004).

\bibitem{ToddILC}
A.~Bartl, H.~Fraas, S.~Hesselbach, K.~Hohenwarter-Sodek,
G.~A.~Moortgat-Pick, JHEP {\bf 0408}, 038 (2004);
A.~Bartl, H.~Fraas, S.~Hesselbach, K.~Hohenwarter-Sodek, T.~Kernreiter
and G.~Moortgat-Pick, JHEP {\bf 0601}, 170 (2006);
Eur.\ Phys.\ J.\  C {\bf 51}, 149 (2007).

\bibitem{ATLASTDR2} ATLAS Collaboration, {\em ATLAS Detector and Physics Performance: Technical Design Report, 2}, CERN-LHCC-99-015 (1999). 

\bibitem{CMS} CMS Collaboration, {\em CMS Physics: Technical Design Report v.2: Physics Performance}, CERN-LHCC-2006-021 (2006).

\bibitem{dedes} A. Dedes, S. Moretti, Phys. Rev. Lett. \textbf{84}, 22 (2000); Nucl. Phys. B \textbf{576}, 29 (2000).

\bibitem{CarEllis1} M. Carena, J. Ellis, A. Pilaftsis, C. E. M. Wagner, Nucl. Phys. B \textbf{586}, 92 (2000).

\bibitem{ChoiKaoJae} S. Y. Choi. K. Hagiwara, J. S. Lee,
Phys. Lett. B \textbf{529}, 212 (2002).
\bibitem{dilipVH} A. Arhrib, D.K. Ghosh, O.C.W. Kong,
Phys. Lett. B \textbf{537}, 217 (2002).

\bibitem{ChoiDrees04} S. Y. Choi, M. Drees, B. Gaissmaier, Phys. Rev. D, \textbf{70}, 014010 (2004); D. A. Demir, Phys. Rev. D, {\bf 60}, 055006 (1999).

\bibitem{dilip} D.K. Ghosh, S. Moretti, Eur. Phys. J. C, {\bf 42}, 341 (2005);
D.K. Ghosh, R.M. Godbole, D.P. Roy, Phys. Lett. B {\bf 628}, 131 (2005). 

\bibitem{4leptons} 
R. Godbole, D.J. Miller, S. Moretti, M. Muhlleitner, in
hep-ph/0608079 and hep-ph/0602198; A. Skjold, P. Osland,
Phys. Lett. B {\bf 329}, 305 (1994).

\bibitem{paper1}S. Moretti, S. Munir, P. Poulose, Phys. Lett. B {\bf 649}, 
206 (2007).

\bibitem{preparation} S. Hesselbach, S. Moretti, S. Munir,  P. Poulose, in preparation.

\bibitem{heinmeyer} S.~Heinemeyer, Int.\ J.\ Mod.\ Phys.\ A {\bf 21}, 2659 (2006);~M.~Frank, T.~Hahn, S.~Heinemeyer, W.~Hollik, H.~Rzehak, G.~Weiglein,
hep-ph/0611326.

\bibitem{CPSuperH} J. S. Lee \textit{et al.}, Comput. Phys. Commun. \textbf{156}, 283 (2004).

\bibitem{CarEllis2} M. Carena, J. R. Ellis, A. Pilaftsis, C. E. M. Wagner, Nucl. Phys. B \textbf{625}, 345 (2002). 

\bibitem{Hemph} B. Ananthanarayan, G. Lazarides, Q. Shafi, Phys. Rev. D {\bf 44},1613 (1991); 
  R. Hemphling, Phys. Rev. D \textbf{49}, 6168 (1994); L. Hall, R. Rattazzi, U. Sarid, Phys. Rev D \textbf{50}, 7048 (1994); M. Carena, M. Olechowski, S. Pokorski, C. E. M. Wagner, Nucl. Phys. B \textbf{426}, 269 (1994); D. Pierce, J. Bagger, K. Matchev, R. Zhang, Nucl. Phys. B \textbf{491}, 3 (1997).

\bibitem{Coarasa} J. A Coarasa, R. A. Jimenez, J. Sola, Phys. Lett. B \textbf{389}, 312 (1996); R. A. Jimenez, J. Sola, Phys. Lett. B \textbf{389}, 53 (1996); K. T. Matchev, D. M. Pierce, Phys. Lett. B \textbf{445}, 331 (1999); P. H. Chankowski, J. Ellis, M. Olechowski, S. Pokorski, Nucl. Phys. B \textbf{544}, 39 (1999); K. S. Babu, C. Kolda, Phys. Lett. B \textbf{451}, 77 (1999).

\end{thebibliography}
\end{document}